\documentclass[12pt]{article}

\renewcommand{\arraystretch}{1.2} 

\usepackage[pdftex]{graphicx}
\usepackage{array}
\usepackage[makeroom]{cancel}
\usepackage{cite}
\usepackage{pdflscape}
\usepackage{amsmath}
\usepackage{amssymb}
\usepackage{hyperref}
\usepackage[utf8]{inputenc}
\usepackage{listings}
\usepackage[dvipsnames]{xcolor}
\usepackage{xspace}
\usepackage{braket}
\usepackage{siunitx}
\usepackage{verbatim}
\usepackage{subfigure}
\usepackage[normalem]{ulem}
\usepackage{slashed}
\usepackage{multirow}
\usepackage[thinlines]{easytable}
\usepackage{url}
\usepackage{longtable}
\usepackage{booktabs}
\usepackage{bbm}
\usepackage[dvipsnames]{xcolor}
\usepackage{placeins}
\definecolor{tit}{rgb}{0.1,0.2,0.4}
\numberwithin{equation}{section}
\usepackage{caption}
\renewcommand{\Re}{\operatorname{Re}}

\newcommand{\Disc}{\operatorname{Disc}_{b \bar s}}

\newcommand{\OPE}{\text{OPE}}
\newcommand{\QCD}{\text{QCD}}
\newcommand{\QED}{\text{QED}}

\newcommand{\lamkin}{\lambda_{\text{kin}}}
\newcommand{\EOS}{\texttt{EOS}\xspace}
\newcommand{\N}{\mathcal{N}}
\newcommand{\A}{\mathcal{A}}
\newcommand{\F}{\mathcal{F}}
\newcommand{\T}{\mathcal{T}}
\renewcommand{\L}{\mathcal{L}}
\renewcommand{\P}{\mathcal{P}}
\newcommand{\cO}{\mathcal{O}}

\newcommand{\eq}[1]{\begin{equation} #1 \end{equation}}
\newcommand{\eqa}[1]{\begin{eqnarray} #1 \end{eqnarray}}
\newcommand{\av}[1]{\langle #1 \rangle}

\newcommand{\qsmax}{q_{\rm max}^2}

\newcommand{\SP}[2]{\ensuremath{\mathcal{S}_{#1}^{#2}}}
\newcommand{\FP}[2][P]{\ensuremath{\mathcal{F}^{B\to #1}_{#2}}}
\newcommand{\HP}[2][P]{\ensuremath{\mathcal{H}^{B\to #1}_{#2}}}
\newcommand{\hHP}[2][P]{\ensuremath{\hat{\mathcal{H}}^{B\to #1}_{#2}}}

\newcommand{\SV}[2]{\ensuremath{\mathcal{S}_{#1}^{#2}}}
\newcommand{\FV}[2][B\to V]{\ensuremath{\mathcal{F}^{#1}_{#2}}}
\newcommand{\HV}[2][B\to V]{\ensuremath{\mathcal{H}^{#1}_{#2}}}

\newcommand{\hHV}[2][B\to V]{\ensuremath{\hat{\mathcal{H}}^{#1}_{#2}}}

\newcommand{\FM}[2][B\to M]{\ensuremath{\mathcal{F}^{#1}_{#2}}}
\newcommand{\HM}[2][B\to M]{\ensuremath{\mathcal{H}^{#1}_{#2}}}
\newcommand{\hHM}[2][B\to M]{\ensuremath{\hat{\mathcal{H}}^{#1}_{#2}}}

\newcommand{\outerF}[2][B\to M]{\ensuremath{\phi^{#1}}_{#2}}

\newcommand{\tV}[2][B\to M]{\ensuremath{\tilde{\mathcal{V}}^{#1}_{#2}}}

\newcommand{\para}{\parallel}
\newcommand{\eps}{\varepsilon}
\newcommand{\order}[1]{\mathcal{O}\left(#1\right)}

\newcommand{\BR}{\ensuremath{\mathcal{B}}}

\newcommand{\GeV}{\,\text{GeV}}

\newcommand{\Eq}[1]{Eq.~(\ref{#1})}
\newcommand{\Eqs}[2]{Eqs.~(\ref{#1})-(\ref{#2})}
\newcommand{\Reff}[1]{Ref.~\cite{#1}}
\newcommand{\Sec}[1]{Section~\ref{#1}}

\newcommand{\App}[1]{Appendix~\ref{#1}}
\newcommand{\Tab}[1]{Table~\ref{#1}}
\newcommand{\Tabs}[2]{Tables~\ref{#1}-\ref{#2}}
\newcommand{\Fig}[1]{Figure~\ref{#1}}

\newcounter{TODO}

\makeatletter

\def\dvd{\@ifstar\@@dvd\@dvd}
\newcommand{\@dvd}[1]{\textcolor{Magenta}{[\textbf{DvD:} #1]}}
\newcommand{\@@dvd}[1]{\textcolor{Magenta}{#1}}

\def\jv{\@ifstar\@@jv\@jv}
\newcommand{\@jv}[1]{\textcolor{Purple}{[\textbf{JV:} #1]}}
\newcommand{\@@jv}[1]{\textcolor{Purple}{#1}}

\def\mr{\@ifstar\@@mr\@mr}
\newcommand{\@mr}[1]{\textcolor{CornflowerBlue}{[\textbf{MR:} #1]}}
\newcommand{\@@mr}[1]{\textcolor{CornflowerBlue}{#1}}

\def\ng{\@ifstar\@@ng\@ng}
\newcommand{\@ng}[1]{\textcolor{ForestGreen}{[\textbf{NG:} #1]}}
\newcommand{\@@ng}[1]{\textcolor{ForestGreen}{#1}}
\makeatother

\lstdefinestyle{customcpp}{
  backgroundcolor=\color{lightgray},
  basicstyle=\ttfamily\footnotesize,
  belowcaptionskip=1\baselineskip,
  breaklines=true,
  frame=L,
  xleftmargin=\parindent,
  language=C++,
  showstringspaces=false,
  basicstyle=\footnotesize\ttfamily,
  keywordstyle=\bfseries\color{green!40!black},
  commentstyle=\itshape\color{purple!40!black},
  identifierstyle=\color{blue},
  stringstyle=\color{orange},
}
\lstnewenvironment{cppsource}[1][]
{%
    \lstset{
        language=C++,
        style=customcpp,
        #1
    }
}
{%
}

\newcommand*\justify{%
  \fontdimen2\font=0.4em
  \fontdimen3\font=0.2em
  \fontdimen4\font=0.1em
  \fontdimen7\font=0.1em
  \hyphenchar\font=`\-
}

\begin{document}

\begin{titlepage}

\begin{flushright}
SI-HEP-2022-12, P3H-22-059, TUM-HEP-1401/22, EOS-2022-02
\end{flushright}

$\ $
\vspace{-2mm}
\begin{center}
\fontsize{16}{20}\selectfont
\bf 
Improved Theory Predictions and Global Analysis\\[2mm]
of Exclusive $\boldsymbol{b\to s\mu^+\mu^-}$ Processes 
\end{center}

\vspace{2mm}

\begin{center}
{Nico Gubernari$^{\,a}$, M\'eril Reboud$^{\,b}$, Danny van Dyk$^{\,b}$, Javier Virto$^{\,c,d}$}\\[5mm]
{\it\small
{$^{\, a}$} 
Universit\"at Siegen, Naturwissenschaftliche Fakult\"at,\\
Walter-Flex-Stra\ss{}e 3, 57068 Siegen, Germany
\\[2mm]
{$^{\, b}$} 
Technische Universit\"at M\"unchen, Physik Department,\\
James-Franck-Stra\ss{}e 1, 85758 Garching, Germany
\\[2mm]
{$^{\, c}$}
Departament de Física Quàntica i Astrofísica,
Universitat de Barcelona,\\
Martí i Franqués 1, 08028 Barcelona, Catalunya
\\[2mm]
{$^{\, d}$}
Institut de Ciències del Cosmos (ICCUB), Universitat de Barcelona,\\
Martí i Franqués 1, 08028 Barcelona, Catalunya
\\[2mm]
}
\end{center}

\vspace{1mm}
\begin{abstract}\noindent
\vspace{-5mm}

We provide improved Standard Model theory predictions for the exclusive rare semi\-muonic processes
$B\to K^{(*)}\mu^+\mu^-$ and $B_s\to\phi\mu^+\mu^-$. Our results are based on
a novel parametrization of the non-local form factors, which
manifestly respects a recently developed dispersive bound. We critically compare our predictions
to those obtained in the framework of QCD factorization.
Our predictions provide, for the first time, parametric estimates of the
systematic uncertainties due to non-local contributions.
Comparing our predictions within the Standard Model to available experimental data,
we find a large tension for $B\to K\mu^+\mu^-$. A simple model-independent analysis
of potential effects beyond the Standard Model yields results compatible with
other approaches, albeit with larger uncertainties for the $B\to K^*\mu^+\mu^-$
and $B_s\to \phi\mu^+\mu^-$ decays. 
Our approach yields systematically improvable predictions,
and we look forward to its application in further analyses beyond the Standard Model.

\end{abstract}

\end{titlepage}

\newpage

\setcounter{tocdepth}{2}
\tableofcontents


\section{Introduction}
\label{sec:intro}
\setcounter{equation}{0}

One the most impressive achievements of the flavor physics program in the 21st century is the thorough study of rare $b\to s\ell^+\ell^-$ transitions,
which has resulted in strong constraints on physics beyond the Standard Model (BSM).
These constraints are now one order of magnitude stronger compared to the situation in $\sim 2010$~\cite{Altmannshofer:2011gn,Beaujean:2012uj,Descotes-Genon:2012isb}, forcing the theory community to revisit in depth the various issues in theory predictions and the source and size of their hadronic uncertainties.

There are two categories of $b\to s\ell^+\ell^-$ observables that are gathering special interest: those probing
lepton-flavor universal (LFU) physics, and those probing lepton-flavor {\it non}-universality~(LFNU). From the point of view of theoretical predictions, the ones that can be predicted with best accuracy are the LFNU ratios, for instance $R_{K}$ or $R_{K^*}$, whose Standard Model (SM) predictions have negligible uncertainties~\cite{Hiller:2003js,Bordone:2016gaq}.
Nevertheless, $b\to s\ell^+\ell^-$ observables require in general the calculation of local and non-local form factors (FFs).
LFNU ratios are thus good ``smoking guns'' of BSM physics, {\it if} the corresponding theory turns out to violate LFU significantly. But in order to infer more precisely a potential new physics pattern (e.g., the pattern of BSM contributions to the Wilson coefficients of the EFT at low energies), it is much more advantageous to use all available data. Not to mention that in the presence of significant LFNU, the hadronic uncertainties in ratios such as $R_K$ do not exhibit the level of cancellations present in the SM~\cite{Boucenna:2016wpr,Capdevila:2016ivx,Serra:2016ivr}.

The combination of theoretical and experimental work in the last decade has eventually led to the establishment
of the so-called ``$B$ anomalies''.
The $B$ anomalies have been in the focus point since the $P'_5$~\cite{Descotes-Genon:2012isb} measurement in 2013~\cite{LHCb:2013ghj,Descotes-Genon:2013wba,Altmannshofer:2013foa,Beaujean:2013soa,Horgan:2013pva}.
They are of interest to the particle physics community as a whole, as new measurements seem to confirm once and again
the patterns of the early analyses.
Global fits to all available data already pointed to the current pattern of BSM physics even before the $R_K$ measurement in 2014~\cite{LHCb:2014vgu}.
Since then, global fits include also separate analyses with only LFNU observables~\cite{Altmannshofer:2014rta,Descotes-Genon:2015uva,DAmico:2017mtc,Geng:2017svp,Capdevila:2017bsm,Hurth:2017hxg,Ciuchini:2017mik}.
\\

Exclusive $b\to s\ell^+\ell^-$ observables in $B\to K^{(*)}\ell^+\ell^-$ decays were first systematically discussed
by Beneke, Feldmann and Seidel (BFS) in 2001~\cite{Beneke:2001at} using QCD-factorization (QCDF) in the heavy-quark limit~\cite{Beneke:1999br,Beneke:2000ry}. 
The QCDF approach is characterized by two elements.
First, it exploits the large-recoil relations for local FFs~\cite{Charles:1998dr,Beneke:2000wa,Bauer:2000yr}.
Second, it incorporates the contributions to the non-local FFs where a hard-collinear interaction with the $B$-meson spectator quark is present.
The main contribution to the non-local FFs is  given by the perturbative contribution from $(\bar s b)(\bar c c)$ operators,
which can be framed as a local operator-product expansion (OPE) and leads to a contribution to the amplitude proportional to the local FFs.

With new more refined calculations of the complete set of local FFs, both within the light-cone sum rules (LCSR)~\cite{Khodjamirian:2017fxg,Bharucha:2015bzk,Gubernari:2018wyi,Descotes-Genon:2019bud}
and lattice QCD (LQCD)~\cite{Bouchard:2013eph,Bailey:2015dka,Horgan:2013hoa,Horgan:2015vla}, the practice of using the large-recoil FF relations for the ``factorizable'' part of the
amplitude has been mostly abandoned. The theory determination of the full set of local FFs is now performed by simultaneously fitting
a parametrization based on the $z$-expansion to
the LCSR determinations at low $q^2$ and the LQCD determinations at large $q^2$  (see, e.g., Refs.~\cite{Bharucha:2015bzk,Gubernari:2018wyi}).
This procedure provides a very accurate (and systematic) determination of the $q^2$ dependence in the whole semileptonic region.

Concerning the non-local FFs, a comprehensive discussion was given by Khodjamirian, Mannel, Pivovarov and Wang (KMPW) in 2010~\cite{Khodjamirian:2010vf}, explaining how to extend the BFS formalism in the low-$q^2$ region and calculating the first correction to the perturbative contribution from the $(\bar s b)(\bar c c)$ operators, which is the next-to-leading term in a light-cone OPE (LCOPE).
In that paper, a dispersion relation has also been used to account for the presence of the narrow charmonium resonances
$J/\psi$ and $\psi(2S)$ in the dilepton spectrum.

Following the same spirit of the $z$-expansion for the local FFs, a similar approach was proposed for the non-local FFs in~\Reff{Bobeth:2017vxj}.
The analytic structure in this case is much more complicated, as can be understood directly from the two-loop perturbative corrections~\cite{Asatrian:2019kbk}.
Nevertheless, the $z$-expansion allows to use data on the non-leptonic decays $B\to M \psi$ to fix the residues of several poles in the non-local FFs,
providing direct constraints on the coefficients of the $z$-expansion.
Also in line with the case of local FFs, a dispersive bound that constrains the coefficients of the
$z$-expansion has been formulated recently~\cite{Gubernari:2020eft}, thereby providing a handle on the issue of convergence and truncation.
The theory calculations of the non-local FFs at negative $q^2$ (where the convergence of the LCOPE is fast) can also be used directly
in the fit to the coefficients of the $z$-expansion.
In this regard, an updated calculation of the next-to-leading contribution in the LCOPE has been presented in~\Reff{Gubernari:2020eft},
pointing to a much smaller effect than the previous calculation in Ref.~\cite{Khodjamirian:2010vf}. 
This change of magnitude is well understood, and can be traced back to missing hadronic matrix elements and updated input parameters.

Other developments include the calculation of local and non-local $B_s\to\phi$ FFs from LCSRs with $B_s$-meson distribution amplitudes~\cite{Gubernari:2020eft}, and the finite-width and non-resonant effects in $B\to K^*$ FFs~\cite{Descotes-Genon:2019bud}.\\

In the present paper we take the challenge of
providing state-of-the-art theory predictions for a variety of exclusive $b\to s\mu^+\mu^-$ observables,
and performing a global analysis of the available $b\to s\mu^+\mu^-$ data using these improved theory predictions.
Since we focus on $b\to s\mu^+\mu^-$ data, we do not include the LFNU ratios $R_K$ and $R_{K^*}$ in our analysis.

The most important results of our work are:
\begin{itemize}
    \item We provide the necessary framework to produce theory predictions both within and beyond the
    SM in presence of the dispersive bound for the non-local FFs. We go beyond the framework of QCDF in
    a way that is consistent with fundamental principles of QCD.
    Thus, our uncertainties are larger but better understood than those given in the literature.
    \item We confirm the substantial tension between the SM predictions and the presently available
    measurements in the three exclusive $b\to s\mu^+\mu^-$ processes considered.
    Our predictions within the SM are somewhat more compatible with the data than other QCDF-based
    predictions in the literature. However, a strong preference for BSM effects remains, with
    BSM-induced shifts to the Wilson coefficients that are compatible with those found in other works.
    
    \item We find that the \emph{local} FFs, rather than the non-local FFs, are now driving the bulk
    of the theory uncertainties. We look forward to new and updated LQCD analyses of the local FFs.
\end{itemize}

\bigskip

In~\Sec{sec:theory} we briefly review the theoretical framework together with the various definitions and conventions.
In~\Sec{sec:th-pred} we describe the details and the strategy followed in our analysis and provide our SM predictions.
\Sec{sec:confrontation} gathers the comparison of our results to experimental measurements and a model-independent BSM analysis. 
We discuss the conclusions that follow from our analysis and give an outlook into future prospects in~\Sec{sec:conclusions}.
The various appendices contain supplemental information concerning both analysis results and relevant formulas.

\section{Theoretical Framework}
\label{sec:theory}
\setcounter{equation}{0}

In this section we review the theoretical framework that we use to study $B\to K^{(*)}\ell^+\ell^-$ and $B_s \to \phi \ell^+\ell^-$ decays, which we collectively denote as $B\to M\ell^+\ell^-$.
The first step consists in factorizing the short-distance (perturbative) and the long-distance (non-perturbative) contributions using an effective field theory.
While the short-distance contributions --- encoded in the Wilson coefficients --- are well known and have been calculated in the literature with high accuracy, the long-distance contributions --- encoded in the hadronic matrix elements --- are the main obstacle to obtaining precise predictions for the observables of interest.
These matrix elements can be classified into local and non-local matrix elements that are usually decomposed in terms of local and non-local FFs, respectively.
We use the available calculations of these FFs and fit them to the parametrizations proposed in Ref.~\cite{Bharucha:2015bzk} (for the local FFs) and Ref.~\cite{Gubernari:2020eft} (for the non-local FFs). 
In this way we obtain SM predictions for $B\to M\ell^+\ell^-$ observables in the region of the momentum transfer  $0 < q^2 < M_{J/\psi}^2$.
Using the same framework and allowing a new physics contribution in certain Wilson coefficients, we also perform a model-independent global analysis of the experimental data in $b\to s \mu^+\mu^-$ transitions.

\subsection{Effective  Theory}
\label{sec:eft}

It is convenient to describe the $B$-meson decays within an effective field theory called the Weak Effective Theory (WET) or the Low-Energy Effective Theory (LEFT)~\cite{Buchalla:1995vs,Aebischer:2017gaw,Jenkins:2017jig}, where degrees of freedom at or above the electroweak scale have been integrated out. In the case of $b\to s\ell^+\ell^-$ transitions, the relevant set of effective operators up to mass-dimension six which is closed under renormalization contains 114 operators~\cite{Aebischer:2017gaw}.  
The WET Lagrangian for these transitions reads
\begin{align}
    \L_{\rm WET}^{sb \ell\ell} = 
    \L_{\QED} + \L_{\QCD} + \L_{D=6}^{sb \ell\ell}
    \,.
\end{align}
Here, $\L_{\QED}$ and $\L_{\QCD}$ are the Lagrangians describing electromagnetic and strong interactions among leptons and the five active quark flavors, and $\L_{D=6}^{sb \ell\ell}$ contains the dimension six WET operators:
\begin{align}
\label{eq:LD6}
\L_{D=6}^{sb \ell\ell} =
\frac{4 G_F}{\sqrt{2}} 
\left[
\lambda_t
\left(
\sum_{i=1}^2 C_i \cO_i^c
+
\sum_{i=3}^{10} C_i \cO_i
\right)
+ \lambda_u \left(
\sum_{i=1}^2 C_i 
\left(
    \cO_i^c - \cO_i^u
\right)
\right)
\right]
+\text{h.c.}
\,,
\end{align}
with $\lambda_q=V_{qb}V_{qs}^*$.
Here, we have included in $\L_{D=6}^{sb \ell\ell}$ only the operators that are relevant in the SM.
We use the operator basis defined in Ref.\cite{Chetyrkin:1996vx}:
\begin{align}
\cO_1^q  & = (\bar{s}_L \gamma_\mu T^a q_L) (\bar{q}_L \gamma^\mu T^a
b_L)
\,, &
\cO_2^q & = (\bar{s}_L \gamma_\mu q_L) (\bar{q}_L \gamma^\mu b_L) \,,
\nonumber\\
\cO_3 & = (\bar{s}_L \gamma_\mu b_L) \sum_p (\bar{p} \gamma^\mu
p) \, ,  &
\cO_4 & = (\bar{s}_L \gamma_\mu T^a b_L) \sum_p (\bar{p}
\gamma^\mu T^a p) \, ,  
\nonumber\\
\cO_5 & = (\bar{s}_L \gamma_\mu \gamma_\nu \gamma_\rho b_L)
\sum_p (\bar{p} \gamma^\mu \gamma^\nu \gamma^\rho p) \, ,
 &
\cO_6 & = (\bar{s}_L \gamma_\mu \gamma_\nu \gamma_\rho T^a b_L)
\sum_p (\bar{p} \gamma^\mu \gamma^\nu \gamma^\rho T^a p) \, ,
\\
\cO_7 & = \frac{e}{16 \pi^2}
m_b (\bar{s}_L \sigma^{\mu\nu} b_R) F_{\mu\nu}
\,,  &
\cO_8 & = \frac{g_s}{16 \pi^2}
m_b (\bar{s}_L \sigma^{\mu\nu} T^a b_R) G_{\mu\nu}^a
 \,,
\nonumber\\
\cO_9^\ell & = \frac{e^2}{16 \pi^2}
(\bar{s}_L \gamma_\mu b_L) (\bar{\ell} \gamma^\mu \ell)
\,, &
\cO_{10}^\ell & = \frac{e^2}{16 \pi^2}
(\bar{s}_L \gamma_\mu b_L) (\bar{\ell} \gamma^\mu \gamma_5 \ell)
\,,
\nonumber
\end{align}
with $q=u,c$ and $\ell = e,\mu,\tau$ while the sum runs over all five active quark flavors, i.e., $p=u,d,s,c,b$.
We use the following conventions:
$P_{R,L}\equiv(1\pm\gamma_5)/2$,
$\sigma_{\mu\nu}\equiv \frac{i}{2} [\gamma_\mu,\gamma_\nu]$,
the covariant derivative is given by $D_\mu q \equiv (\partial_\mu + i e Q_q A_\mu + i g_s T^A G^A_\mu)q$, and $m_b \equiv m_b(\mu)$ denotes the $\overline {\rm MS}$ $b$-quark mass.
Throughout this work we neglect the terms in \Eq{eq:LD6} proportional to $\lambda_u$, since they are CKM suppressed.
The numerical values for the Wilson coefficients used in this work are collected in \Tab{tab:WCs}.
\\

\begin{table}
\renewcommand{\arraystretch}{1.4}
\small
\centering
\begin{tabular}{@{}cccccccccc}
\toprule
$C_1(\mu_b)\!$ &   $C_2(\mu_b)\!$ &  $C_3(\mu_b)\!$ &  $C_4(\mu_b)\!$
& $C_5(\mu_b)\!$ & $C_6(\mu_b)\!$ & $C_7(\mu_b)\!$ & $C_8(\mu_b)\!$
& $C_9(\mu_b)\!$ & $C_{10}(\mu_b)\!$ \\[1mm]
\hline
$-0.2906$ & $1.010$ & $-0.0062$ & $-0.0873$ & $0.0004$ &
$0.0011$ &  $-0.3373$ & $-0.1829$ & $4.2734$ & $-4.1661$\\
\bottomrule
\end{tabular}
\caption{NNLO Wilson coefficients at the scale $\mu_b=4.2\GeV$~\cite{Bobeth:1999mk,Gorbahn:2004my}.
}
\label{tab:WCs}
\end{table}

In this framework, the decay amplitude for $\bar{B}\to  \bar{K}^{(*)}\ell^+\ell^-$ and $\bar{B}_s\to  \phi \ell^+\ell^-$ decays to the leading non-trivial order in QED can be written as
\begin{align}
    \A^{M\ell\ell}
    &\equiv 
        \frac{G_F\, \alpha_e\, V_{tb}^{} V_{ts}^*}{\sqrt{2} \pi}
        \bigg\{ (C_9 \,L^\mu_{V} + C_{10} \,L^\mu_{A})\ \! \FM[B\to M]{\mu}
        \!
        -  \frac{L^\mu_{V}}{q^2} \Big[  2 i m_b C_7\,\FM[B\to M]{T,\mu}  
        \! + 16\pi^2 \HM[B\to M]{\mu} \Big]   \bigg\}
        \,.
    \label{eq:amplitude}
\end{align}
Here, $q^2$ is the invariant squared mass of the lepton pair and
$L_{V(A)}^\mu \equiv \bar u_\ell(q_1) \gamma^\mu(\gamma_5) v_\ell(q_2)$ are leptonic currents.
We emphasize that the decay amplitude \eqref{eq:amplitude} depends on both the local matrix elements $\FM[B\to M]{(T),\mu}$ and the non-local matrix elements $\HM[B\to M]{\mu}$, defined as 
\begin{align}
    \label{eq:def-Fmu}
    \FM{\mu}(k, q) &\equiv \langle  M(k)|\bar{s}\gamma_\mu P_L\, b|\bar{B}(q+k)\rangle
    \ , \\
    \label{eq:def-FTmu}
    \FM{T,\mu}(k, q) &\equiv \langle  M(k)|\bar{s}\sigma_{\mu\nu} q^\nu P_R\, b|\bar{B}(q+k)\rangle
    \ , \\
    \HM{\mu}(k, q) 
    & \equiv
    \sum_p \HM{p,\mu} (k, q)
    \qquad (p=u,d,s,c,b)
    \,,
\end{align}
where
\begin{equation}
\begin{aligned}
    \!\!\!\!\HM{p,\mu}(k, q)  & \equiv
    i Q_p\!  \int\!  d^4x\, e^{i q\cdot x}\,
    \\
    &\times \av{ M(k) | \T\!\left\{ \bar p \gamma_\mu p(x) ,
    \left(
    C_1 \cO_1^c +
    C_2 \cO_2^c
    + 
    \sum_{i=3}^{6} C_i \cO_i 
    +
    C_8 \cO_8
    \right)\!\!(0)\!
    \right\}\!  | \bar B(q+k)}
    .
    \label{eq:Hpmu}
\end{aligned}
\end{equation}
For convenience, we have decomposed the non-local matrix element $\HM{\mu}$ into its different contributions $\HM{p,\mu}$ from the electromagnetic current.
The numerically dominant contributions to $\HM{\mu}$ come from  the operators $\cO_1^c$ and $\cO_2^c$ in $\HM{c,\mu}$.
We study the contributions to $\HM{c,\mu}$ using the parametrization and the corresponding dispersive bound derived for the first time in Ref.~\cite{Gubernari:2020eft}.
For completeness, we also include the contribution from the penguin operators in $\HM{c,\mu}$, which are suppressed by small Wilson coefficients. 
The operator $\cO_8$ does not contribute to $\HM{c,\mu}$ at the precision we are working at.

For our numerical analysis, we also take into account $\HM{s,\mu}$ and $\HM{b,\mu}$ in the framework of QCD factorization~\cite{Beneke:2001at,Beneke:2004dp}.
The relevant formulas are given in~\App{app:Hsb}.
We neglect $\HM{u,\mu}$ and $\HM{d,\mu}$, since they do not receive contributions from the four-quark operators $\cO_1^c$ and $\cO_2^c$ and hence they are numerically suppressed.
\\

The amplitude in~\Eq{eq:amplitude} is independent of the renormalization scale, with the scale dependence of the Wilson coefficients $C_i$ cancelling the scale dependence of the non-local matrix elements~$\HM{\mu}$. In practice, there is a remaining residual scale dependence due to the fact that the Wilson coefficients are only known up to NNLO while the short-distance part of~$\HM{\mu}$ is only known to NLO. In our framework, the determination of~$\HM{\mu}$ is based on the OPE calculation at negative $q^2$ --- which has the explicit scale dependence needed to cancel the scale dependence of the Wilson coefficients $C_{7,9}$ up to NLO ---, and the experimental data on the charmonium poles. This latter input to the determination is an observable and thus is scale independent, but this is consistent because, while~$\HM{\mu}$ is scale dependent, the residues at the charmonium poles are not. The conventional thing to do would be to vary the renormalization scale simultaneously in the Wilson coefficients and~$\HM{\mu}$ in order to estimate the effect of missing NNLO contributions. However, since this effect will be much smaller than the uncertainties introduced by the local FF determinations, we choose to fix the scale to $\mu_b=4.2\GeV$.

\subsection{Local Form Factors}
\label{sec:localFFth}

The matrix elements of local currents $\FM[B\to M]{\mu}$ and $\FM[B\to M]{T,\mu}$ can be decomposed in terms of the local FFs $\FM{(T),\lambda}$: 
\begin{align}
    \FM{(T),\mu}(k, q) \propto \sum_\lambda  \FM{(T),\lambda}(q^2)\, \SP{\mu}{\lambda}(k, q)\ ,
\end{align}
where the explicit form of this decomposition and of the Lorentz structures $\SP{\mu}{\lambda}$ is given in \App{app:FFsdef}.
These FFs have been calculated in lattice QCD (LQCD) for high values of $q^2$ (small recoil). 
Since the low-$q^2$ region is not accessible by LQCD calculations as of yet, we use the light-cone sum rules (LCSRs) results in this region.
The inputs used in our analysis are listed in \Sec{sec:th-pred:local}.
We combine LQCD and LCSR results by fitting them to the parametrization used in Ref.~\cite{Bharucha:2015bzk}:
\begin{align}
    \FM{(T),\lambda} (q^2)
    =
    \frac{1}{1 - \frac{q^2}{m_{J^P}^2}}
    \sum_{k=0}^\infty
    \alpha_k^\F
    \left[
        z(q^2)
        -
        z(0)
    \right]^k
    \,,
    \label{eq:zexpOPE}
\end{align}
where the conformal variable $z$ is defined as
\begin{align}
    \label{eq:zdef}
    z(q^2) \equiv \frac{\sqrt{s_+-q^2} - \sqrt{s_+ - s_0^{\phantom{1}}}}{\sqrt{s_+-q^2} + \sqrt{s_+ - s_0^{\phantom{1}}}}
    \,.
\end{align}
The parameter $s_+ \equiv (M_B + M_M)^2$ is fixed and it coincides with the lowest-lying branch point of the corresponding FF. 
The parameter $s_0$ can be chosen arbitrarily in the open interval $(\infty,s_+)$. 
The common choice 
\begin{align}
    s_0 = (M_B + M_M)\left(\sqrt{M_B} - \sqrt{M_M}\right)^2
\end{align} 
minimizes the absolute value of $z$
in the semileptonic region and hence the truncation error.
To ensure interoperability with the results of Ref.~\cite{Bharucha:2015bzk} we use the masses of $s \bar b$ resonances 
as given in Table 3 thereof.

\subsection{Non-Local Form Factors}
\label{sec:nonlocalFFth}

In analogy with the local matrix elements discussed above, the non-local matrix elements $\HM[B\to M]{\mu}$ can be decomposed in terms of the non-local FFs $\HM{\lambda}$: 
\begin{align}
    \HM{p,\mu}(k, q) \propto \sum_\lambda  \HM{p,\lambda}(q^2)\, \SP{\mu}{\lambda}(k, q)\ .
\end{align} 
The explicit form of this decomposition and of the Lorentz structures $\SP{\mu}{\lambda}$ is given in \App{app:FFsdef}.

The non-local FFs $\HM{c,\lambda}$ are more complicated objects than the local FFs $\FM{(T),\lambda}$. 
So far there are no LQCD determinations of these FFs.
Most theory predictions for $\HM{c,\lambda}$ have been obtained in the framework of QCD factorization~\cite{Beneke:2001at,Beneke:2004dp},
which uses a perturbative calculation of the charm loop.
However, this treatment of the charm loop is manifestly invalid close to and above the partonic charm threshold
(i.e., for $q^2 \gtrsim 4 \GeV^2$) and misses potentially relevant power corrections even for $q^2 < 4\,\GeV^2$.
This introduces an uncontrollable systematic uncertainty, sometimes estimated using \emph{ad hoc} models.
Following our previous works~\cite{Bobeth:2017vxj,Gubernari:2020eft}, we use a different strategy to calculate each individual non-local FFs $\HM{c,\lambda}$,
which can be summarized as follows:
\begin{itemize}
    \item \sloppy
    We calculate $\HM{c,\lambda}$ for negative values of the momentum transfer (i.e., at $q^2=\{-7,-5,-3,-1\} \GeV^2$) using a LCOPE. 
    \item We extract the residue of $\HM{c,\lambda}$ at $q^2=M_{J/\psi}^2$ from the measurement of the branching ratios and angular observables in $B\to M J/\psi$ decays.
    \item We obtain data-driven theory predictions by interpolating the two types of inputs for $\HM{c,\lambda}$ over the time-like region
    $0 < q^2 < M_{J/\psi}^2$.
\end{itemize}
In the remainder of this subsection we discuss these steps in further detail.
\\

For $4 m_c^2 - q^2 \gg m_b\,\Lambda_{\rm had}$ the  non-local FFs $\HM{c,\lambda}$ can be expanded in a LCOPE~\cite{Khodjamirian:2010vf,Khodjamirian:2012rm,Gubernari:2020eft}:
\begin{equation}
\begin{aligned}
    \label{eq:masterNLFF}
    \HM{c,\lambda}
    &=
    -\frac{1}{16\pi^2}
    \left(
        \frac{q^2}{2M_B^2}
        \Delta C_9\, \FM{\lambda}
        + \frac{m_b}{M_B} \Delta C_7\,
        \FM{T,\lambda}
    \right)
    +2 \,Q_c\, 
    \left(
        C_2 - \frac{C_1}{2 N_c}
    \right)
    \tV{\lambda}
    \\&
    + \text{ higher-power corrections}
    \,.
\end{aligned}
\end{equation}
The first term on the r.h.s. is the leading-power contribution to the LCOPE, which consists of the local FFs $\FM{(T),\lambda}$ multiplied by the corresponding matching coefficients $\Delta C_{7,9}$. 
These coefficients have been computed to next-to-leading order in QCD~\cite{Asatryan:2001zw,Greub:2008cy,Ghinculov:2003qd,Bell:2014zya,deBoer:2017way,Asatrian:2019kbk}.
The next-to-leading power corrections $\tV{\lambda}$ cannot be expressed in terms of local FFs and hence require the calculation of specific non-local matrix elements.
The quantities $\tV{\lambda}$ have been computed for the first time in~Ref.\cite{Khodjamirian:2010vf}, where the authors have found them to have a considerable impact on the numerical values of the functions $\HM{c,\lambda}$.
However, this calculation is superseded by the one in Ref.~\cite{Gubernari:2020eft}, where it has been shown that the $\tV{\lambda}$ contribution is mostly negligible.
The difference between the two results is well understood and results from (\textit{i}) the inclusion of missing three-particle distributions amplitudes that were not considered in~Ref.\cite{Khodjamirian:2010vf} and (\textit{ii}) an update of the inputs necessary to evaluate these distribution amplitudes~\cite{Braun:2017liq,Nishikawa:2011qk}.
Higher-power corrections in \Eq{eq:masterNLFF} have not been studied yet.
\\

The $B\to M J/\psi$ decay amplitudes are proportional to the residues of the non-local FFs $\HM{c,\lambda}$ at $q^2=M_{J/\psi}^2$~\cite{Khodjamirian:2010vf,Bobeth:2017vxj}:
\begin{align}
    \A_\lambda^{M J/\psi} \propto \underset{q^2\to M_{J/\psi}^2}{\text{Res}} \HM{c,\lambda}(q^2)
    \,.
\end{align}
Therefore, the r.h.s. can be inferred from the measurement of branching ratios and angular observables in $B\to M J/\psi$ decays.
The exact relations between $\A_\lambda^{M J/\psi}$ and $\HM{c,\lambda}$ can be found in \App{app:ampl}.
Although this would yield larger uncertainties in the final results, our approach equally applies even if one does not use this experimental information at the $J/\psi$ pole.
\\

We use the parametrization given in Ref.~\cite{Gubernari:2020eft} to interpolate $\HM{c,\lambda}$.
This parametrization reads
\begin{align}
    \label{eq:H_expansion}
      \HM{c,\lambda}(q^2)
      =
      \frac{1}{\outerF[B\to M]{\lambda}(\hat{z}) \, \P(\hat{z})}
      \sum_{n=0}^\infty  \beta_{\lambda,n}^{B \to M} p_{n}(\hat{z})
      \,,
\end{align}
where the conformal variable $\hat{z}\equiv\hat{z}(q^2)$ is defined as
\begin{align}
    \label{eq:hzdef}
    \hat{z}(q^2) \equiv \frac{\sqrt{\hat{s}_+-q^2} - \sqrt{\hat{s}_+ - \hat{s}_0}}{\sqrt{\hat{s}_+-q^2} + \sqrt{\hat{s}_+ - \hat{s}_0}}
    \,.
\end{align}
The only difference with the definition (\ref{eq:zdef}) is that here $\hat{s}_+ = 4M_D^2$, reflecting a different left-most
branch point for the non-local FFs compared to the local ones.
This implies the optimal value of $\hat{s}_0$ is also different, due to the fact that the parametrization of~Eq.~(\ref{eq:H_expansion}) is valid for $q^2< 4M_D^2$ and not in the whole semileptonic region.
In our analysis we use~Eq.~(\ref{eq:H_expansion}) in the region $-7 \GeV^2 \leq q^2 \leq M_{J/\psi}^2$.
We choose
\eq{\hat{s}_0 = 4\GeV^2\,,}
which implies $|\hat{z}(-7\GeV^2)| \sim |\hat{z}(M_{J/\psi}^2)|$, in order to maximize the convergence of the series in \Eq{eq:H_expansion}.
This value of $\hat{s}_0$ yields $\hat{z}(m_{\psi(2S)}^2) \sim 0.7$, while $|\hat{z}(-7\GeV^2)| \sim |\hat{z}(M_{J/\psi}^2)| \sim 0.2$.
Hence, although our parametrization should in principle also account for the $\psi(2S)$ pole, we cannot expect the same level of accuracy for predictions of the non-local FFs at the $\psi(2S)$ as at negative $q^2$ or at the $J/\psi$ pole.
This observation and the already very large number of parameters convinced us to keep the value of $\HM{c,\lambda}$ at this resonance unfixed for the present work.
Nevertheless, we check the  values of the $\HM{c,\lambda}$ residue at the $\psi(2S)$ pole a-posteriori.
We checked that a change in the value of $\hat{s}_0$ in the range $\sim 4 - 8 \GeV^2$ does not have a noticeable impact on our results for the non-local FFs once the dispersive bound is enforced.

The analytical expression of the outer functions $\outerF[B\to M]{\lambda}$  and Blaschke factor $\P$ are given in Ref.~\cite{Gubernari:2020eft} (see also \App{app:bound} here). 
The polynomials $p_{n}$ are orthonormal with respect to the integration measure
$d\mu = \theta(|\arg z| - \alpha_{H_bH_s}) dz$~\cite{Gubernari:2020eft}. They can conveniently be written using
the Szeg\H{o} recurrence relation~\cite{Simon2004OrthogonalPO}, which is used for their implementation in the public source code of the \EOS software~\cite{EOS:paper,EOS:repo}. Details are given in \App{app:NLFFFit}.

Compared to other approaches proposed in the literature --- for instance the $q^2$ expansion of~Ref.~\cite{Jager:2012uw},
the isobar model of Refs.~\cite{LHCb:2016due,Blake:2017fyh,Egede:2015kha}, the dispersive approach of Ref.~\cite{Khodjamirian:2010vf},
and the ``naive'' $z$ expansion of Ref.~\cite{Bobeth:2017vxj} --- the parametrization
proposed in Ref.~\cite{Gubernari:2020eft}  exhibits a crucial advantage.
The coefficients of this parametrization obey the bound
\begin{equation}
    \label{eq:boundcoeff}
     \sum_{n=0}^\infty 
    \left\{
        2\Big| \beta_{0,n}^{B\to K} \Big|^2
        +
        \sum_{\lambda=\perp,\para,0}
        \left[
            2\Big| \beta_{\lambda,n}^{B\to K^*} \Big|^2
            +
            \Big| \beta_{\lambda,n}^{B_s\to \phi} \Big|^2
    \right]
    \right\}
     < 1\,.
\end{equation}
This \emph{dispersive} (or unitarity) bound has been derived for the first time in Ref.~\cite{Gubernari:2020eft}.
We summarize its derivation in \App{app:bound}, where we also extend it to include penguin operators and one-particle contributions.
The left-hand side of \Eq{eq:boundcoeff} is referred to as the \emph{saturation} of the bound in the following.

\section{Theory Predictions}
\label{sec:th-pred}
\setcounter{equation}{0}

In this section we provide our theory predictions together with technical details and the inputs needed to obtain them.
As a first step, in \Sec{sec:th-pred:local}, we fit the parametrization of the local FFs as given in \Eq{eq:zexpOPE}
to available theory constraints.
As a second step, in \Sec{sec:th-pred:nonlocal}, we perform a Bayesian analysis to provide data-driven predictions
for the non-local FFs, which relies
on the strategy outlined in \Sec{sec:nonlocalFFth} and uses the parametrization \eqref{eq:H_expansion}.
The theory predictions for the non-local FFs critically depend on the results for the local FFs.
Once all the FFs are known we are able to provide predictions for $B\to M\ell^+\ell^-$ observables.
Our SM predictions are then presented and compared to those of previous approaches in \Sec{sec:th-pred:SM}.

\subsection{Local Form Factors}
\label{sec:th-pred:local}

\subsubsection*{Analysis Setup}
The various local $B\to M$ FFs $\FM{(T),\lambda}$ have been calculated using LQCD in Refs.~\cite{Bouchard:2013eph,Bailey:2015dka,Horgan:2013hoa,Horgan:2015vla}. 
These calculations have been performed at small recoil (i.e., large $q^2$).
In principle, one could extrapolate the LQCD results at high-$q^2$ to the low-$q^2$ region.
However, this would introduce a systematic error that cannot be controlled, since dispersive bounds for the local FFs in $b\to s$
transitions have so far not been very constraining~\cite{Bharucha:2010im}.
For this reason we use LCSR results, which are valid only at low $q^2$, to anchor the FFs at both ends of the
phase space.
There are two types of LCSRs that can be used for this purpose: LCSRs with light-meson distribution amplitudes (DAs),
and LCSRs with $B$-meson DAs.
The light-meson LCSRs have currently smaller uncertainties than $B$-meson LCSRs due to smaller parametric uncertainties of the corresponding
light-meson DAs.

Fitting both LQCD and LCSR results to the parametrization (\ref{eq:zexpOPE}), we obtain  all the local $B\to M$ FFs in the whole semileptonic region, that is for $0 \leq q^2 \leq \qsmax\equiv(M_B-M_M)^2$. Such combinations have previously been undertaken in Ref.~\cite{Bharucha:2015bzk,Gubernari:2018wyi}.
\\

For the local $B \to K$ FFs, we use the  light-meson LCSR calculation of Ref.~\cite{Khodjamirian:2017fxg}.
We do not use the $B$-meson LCSR calculation of the $B\to K$ FFs provided in Ref.~\cite{Gubernari:2018wyi} as recommended by the authors,
due to issues with the determination of the sum rule thresholds that are not yet understood.
Our fit uses four synthetic points at $q^2 = \{0,6\}\GeV^2$ for $f_+^{B\to K}$ and $f_T^{B\to K}$, which are generated
using the coefficients and the parametrization provided in Ref.~\cite{Khodjamirian:2017fxg}.

Similarly, we generate eight synthetic points at $q^2 = \{\qsmax-6\GeV^2,\qsmax-3\GeV^2,\qsmax\}$ for all the $B\to K$ FFs using the BGL coefficients of Ref.~\cite{Aoki:2021kgd}, which averages the LQCD results of Refs.\cite{Bouchard:2013eph,Bailey:2015dka}.
The synthetic points are eight instead of nine because we drop the point $q^2 = \{\qsmax-6\GeV^2\}$ for $f_0^{B\to K}$ due to the constraint in \Eq{eq:f+0f00}.
\\

For the local $B\to K^*$ and $B_s\to\phi$ FFs, we use the $B_{(s)}$-meson LCSR calculations of Refs.~\cite{Gubernari:2018wyi,Gubernari:2020eft}. 
Although the results for these FFs from light-meson LCSR calculations are more precise \cite{Bharucha:2015bzk},
we avoid using them due to a conceptional concern about using distribution amplitudes for non-asymptotic states.
The results of Refs.~\cite{Gubernari:2018wyi,Gubernari:2020eft} have similar central values,
but larger uncertainties compared to results of Ref.~\cite{Bharucha:2015bzk}.
Thus, our choice leads to more conservative estimates of the final uncertainties and aims to have a smaller dependence on the LCSR approach.
In addition, it is not entirely clear how to include the finite $K^*$-width effect in the results of Ref.~\cite{Bharucha:2015bzk}, while this can be done consistently in the $B$-meson LCSRs framework, as argued in Ref.~\cite{Descotes-Genon:2019bud}.
Following this reference, we multiply by 1.1 the central values of the synthetic points of the $B\to K^*$ FFs given in ancillary files attached to the arXiv version of Ref.~\cite{Gubernari:2018wyi}.
We only apply the $P$-wave correction factor of 1.1 to the $B\to K^*$ FFs in the low-$q^2$ region, where the analysis
of Ref.~\cite{Descotes-Genon:2019bud} is valid.
The $B_s\to \phi$ FF results attached to the arXiv version of Ref.~\cite{Gubernari:2018wyi} are used as-is.

We generate synthetic points at $q^2=\{16\GeV^2,\qsmax\}$ for each of the local $B\to K^*$ and $B_s\to\phi$ FFs using the coefficients and the parameters of the LQCD calculations of Refs.~\cite{Horgan:2013hoa,Horgan:2015vla}.\footnote{%
    \label{fnote:LQCD}
    Our LQCD inputs stem from private communications with the authors of both references, which provide more
    data points than what is publicly available. No such updates are available for the local $B_s\to \phi$ FFs.
}
For both the LCSRs and LQCD results of $B\to K^*$ and $B_s\to\phi$ FFs, the correlation between the (axial-)vector $\FV{\lambda}$ and tensor $\FV{T,\lambda}$ FFs are not known. Hence, we are forced to treat these two sets of FFs as uncorrelated.
\\

\begin{table}[t!]
    \centering
    \renewcommand{\arraystretch}{0.90}
    \begin{tabular}{lcccc}
        \toprule
            Transition & Constraints \# & Priors \# &  LQCD Ref. & LCSR Ref.\\
        \midrule
             $B\to K$ & 12 & 8 &
             \cite{Aoki:2021kgd} &
             \cite{Khodjamirian:2017fxg} \\[3pt]
             $B\to K^*$ & 117 & 19 &
             \cite{Horgan:2013hoa,Horgan:2015vla} &
             \cite{Gubernari:2018wyi} \\[3pt]
             $B_s\to \phi$ & 47 & 19 &
             \cite{Horgan:2015vla,Horgan:2013hoa} &
             \cite{Gubernari:2020eft} \\
        \bottomrule
    \end{tabular}
    \caption{
    \label{tab:fitLFFsSM}
        Details of the fit of the local $B\to M$ FFs.
        The number of constraints is the total number of synthetic data points from LQCD and LCSR analyses.
        The number of (uniform) priors is the number of independent $\alpha_k^\F$ coefficients.
    }
\end{table}

We combine the LQCD and LCSR synthetic data points discussed above using the parame\-trization \eqref{eq:zexpOPE}.
We truncate the series at $k=2$, taking into account that the number of independent $\alpha_k^\F$ coefficients is reduced by the constraints (\ref{eq:f+0f00}) and (\ref{eq:A00A10})-(\ref{eq:T10T20}).
We checked that truncating the series at $k=3$ does not sensibly impact the results. 
We perform one fit per process. The details of these fits are summarized in \Tab{tab:fitLFFsSM}. 
Posterior samples are drawn with the \EOS software using a Markov chain Monte-Carlo (MCMC) method based on the Metropolis-Hastings algorithm.
A Gaussian mixture density is then iteratively adapted using the Population Monte-Carlo (PMC) algorithm~\cite{pypmc}
to the posterior samples that are obtained with the MCMC method.
Although this step is not strictly necessary, it allows to reuse the posterior for the local FF coefficients in
subsequent analyses and to sample from the posterior in a computationally efficient way.
We then use our local FF results as priors for our theory predictions for
non-local FFs  and within the WET fits.

\subsubsection*{Results}
\begin{figure}[t!]
    \centering
    \includegraphics[width=.32\textwidth]{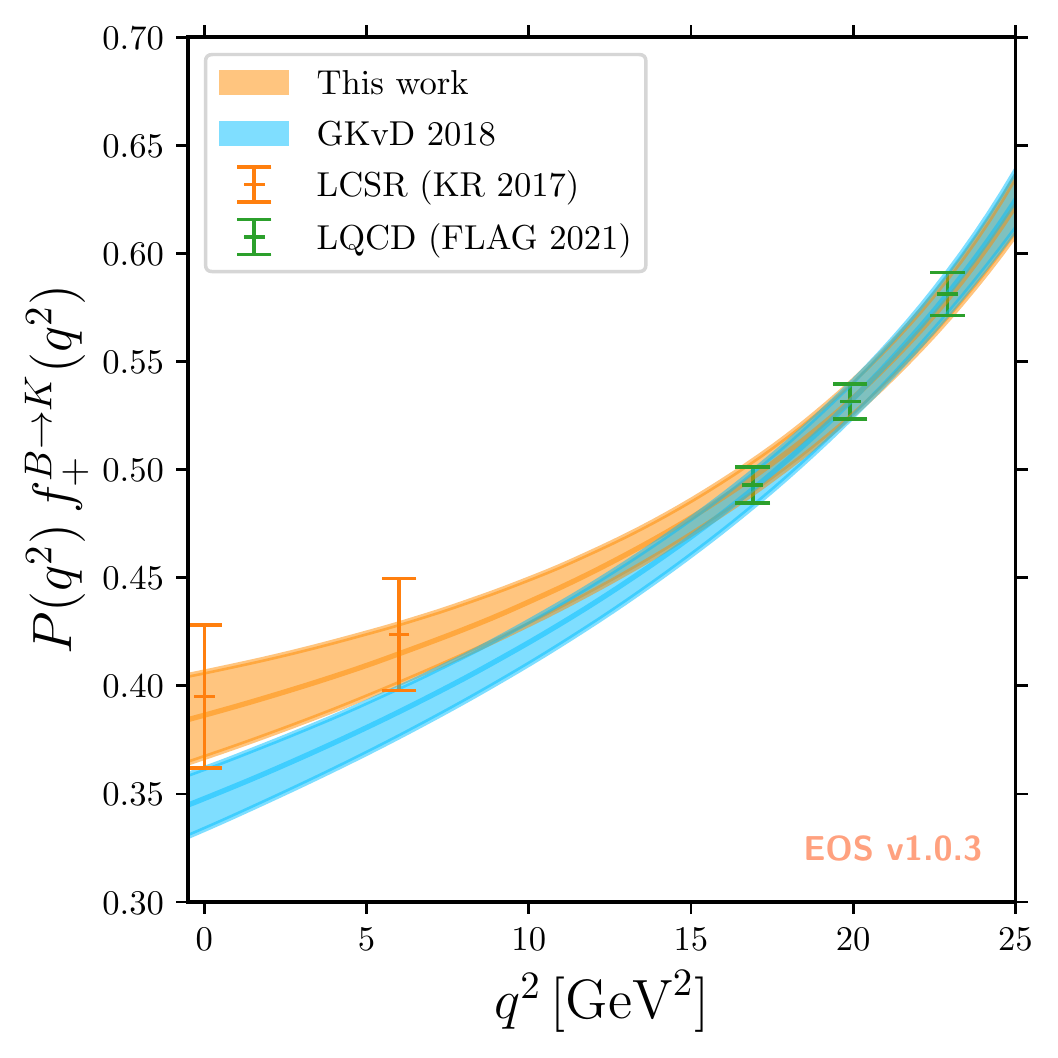}
    \includegraphics[width=.32\textwidth]{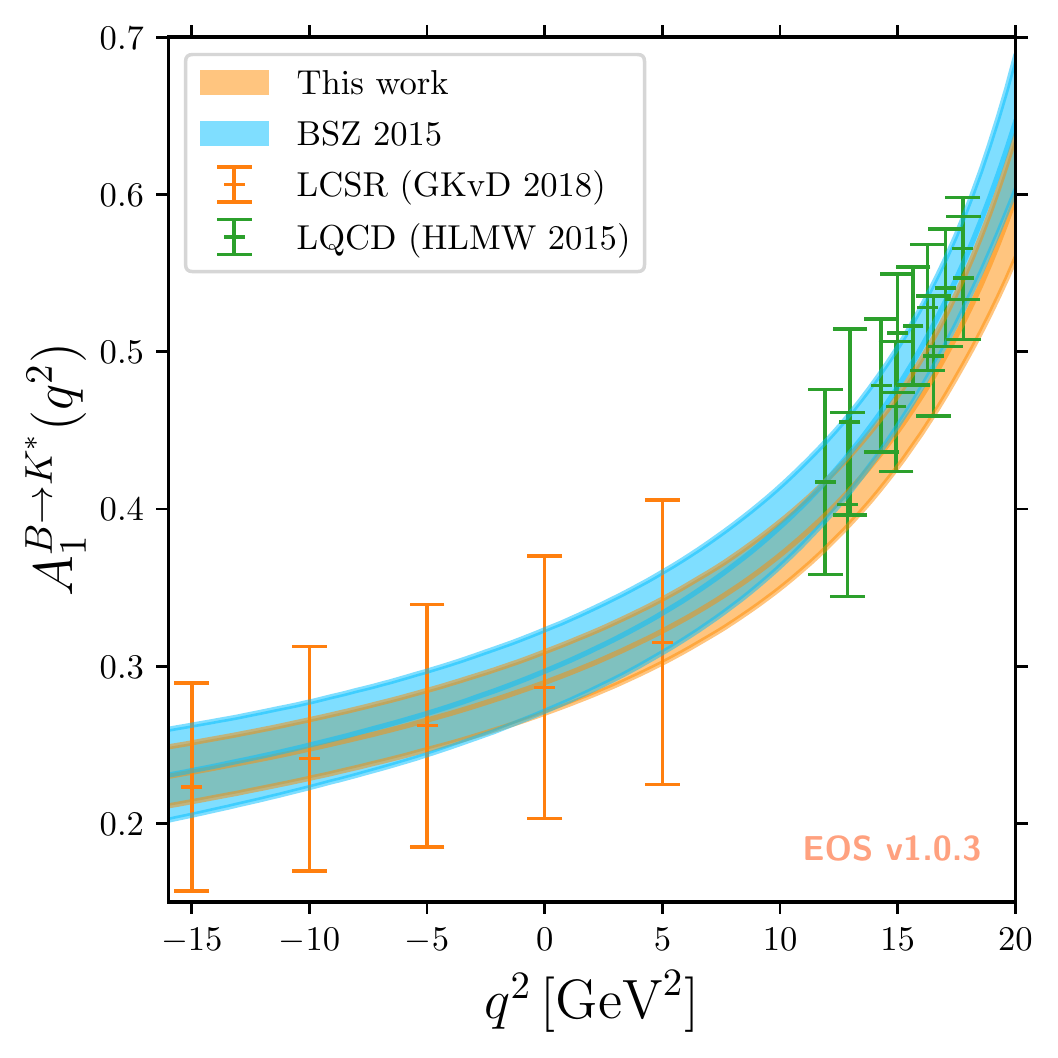}
    \includegraphics[width=.32\textwidth]{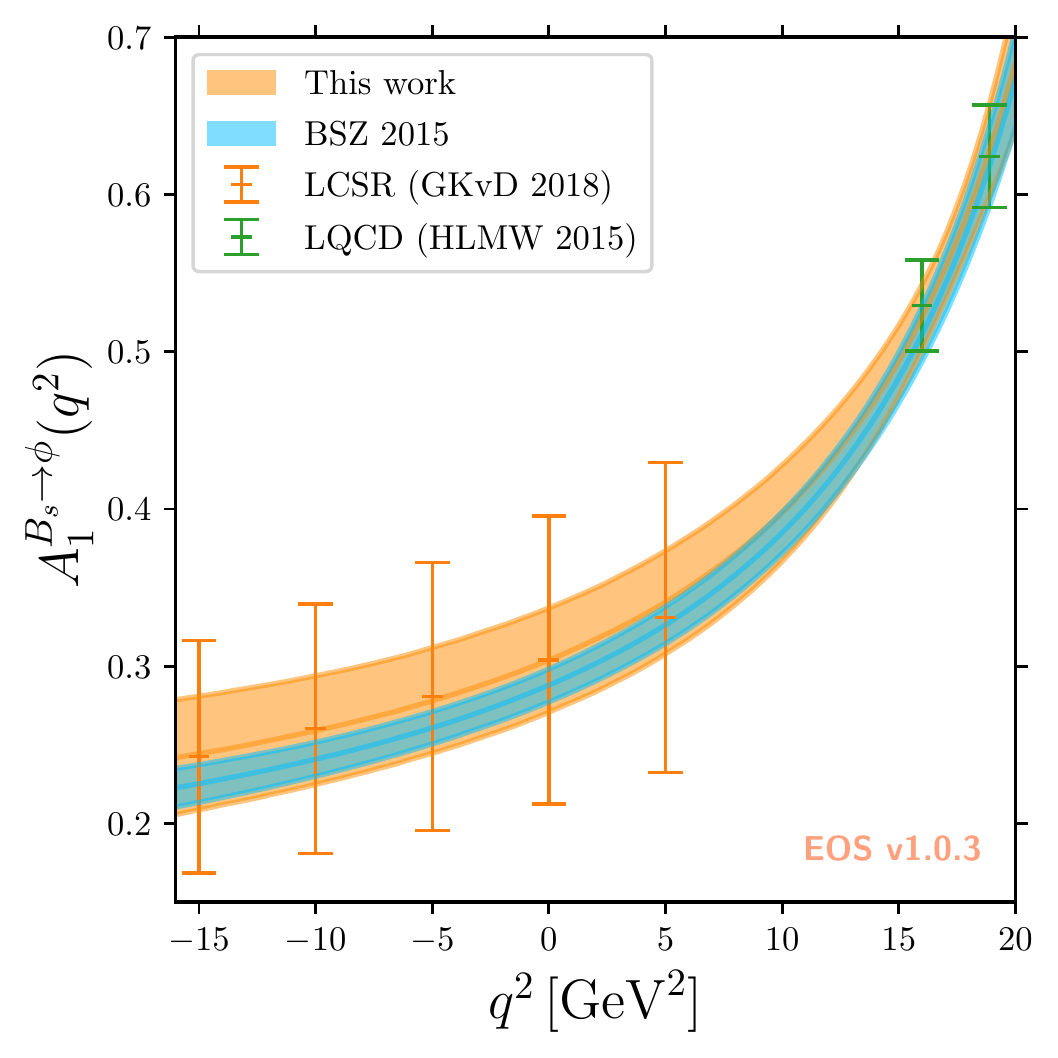}
    \caption{%
        Plots of a selection of the local $B\to M$ FFs discussed in this work.
        Our results (in orange) are juxtaposed with those of  \Reff{Gubernari:2018wyi} for $B\to K$ and \Reff{Bharucha:2015bzk} for $B\to K^*$ and $B_s\to \phi$  (both in blue). We show the medians
        as solid lines and the central $68\%$ uncertainty interval as shaded areas.
        For the $B\to K$ process, we show the product of the FF $f_+^{B\to K}$ and $P(q^2) = 1 - \frac{q^2}{m_{J^P}^2}$
        for readability, see Eq. (\ref{eq:zexpOPE}).
        The conversion between the form-factor basis used for these plots and in the literature and the basis used in this work is presented in \App{app:FFsdef}.
    }
    \label{fig:local_ffs}
\end{figure}

The three fits to the local FF coefficients are excellent: the $p$ values of the best-fit points are $72\%$, $>99\%$, and $>99\%$ for $B\to K$, $B\to K^*$, and $B_s\to \phi$, respectively.
These large $p$ values are due to the sizable uncertainties of the LCSR inputs.
The posterior distributions of the $\alpha_k^\F$ coefficients in each fit are perfectly described by a single multivariate Gaussian distribution:
when adapting a Gaussian mixture density to these posteriors using the PMC algorithm, all but one components are pruned out, and the remaining component
yields perplexities larger than 99\%.
The parameters of these distributions are given in the ancillary files \texttt{BToK-local.yaml}, \texttt{BToKstar-local.yaml} and \texttt{BsToPhi-local.yaml}.
The mean values and uncertainties of the $\alpha_k^\F$ coefficients are also given for convenience in \App{app:plots-tables}.
As an illustration, three of the local FFs are plotted in \Fig{fig:local_ffs}, where we juxtapose our results with those
of Refs.~\cite{Bharucha:2015bzk,Gubernari:2018wyi}.\\

Due to our use of the extended LQCD data set (see footnote \ref{fnote:LQCD}), our results for $B\to K^*$
present comparable uncertainties to those of \Reff{Bharucha:2015bzk} but a slightly different shape.
In contrast, our results for $B_s\to \phi$ are less precise than those of \Reff{Bharucha:2015bzk}
due to our use of the $B$-LCSR results over the light-meson LCSR results therein.

\subsection{Non-Local Form Factors}
\label{sec:th-pred:nonlocal}

\subsubsection*{Analysis Setup}
We produce synthetic correlated data points for the non-local $B\to M$ FFs $\HM{c,\lambda}$ from the LCOPE expression \eqref{eq:masterNLFF},
which holds for $4 m_c^2 - q^2 \gg m_b \Lambda_\text{had}$. 
To ensure a rapid convergence of the LCOPE, we use it only at four negative values of $q^2$: $\{-7,-5,-3,-1\} \GeV^2$.
This approach follows the recommendation in Refs.~\cite{Bobeth:2017vxj,Gubernari:2020eft}.
At leading power, the computation of the non-local FFs requires two inputs (cf. \Eq{eq:masterNLFF}):
the local FFs and the matching coefficients $\Delta C_{7,9}$.

We evaluate the local FFs at each point in $q^2$, using the posterior distributions of the fit described in the previous paragraph. We also determine their covariance matrix across local FFs in the same channel and across different $q^2$ values.
The correlations between (axial-)vector local FFs and the tensor ones are unknown, except for the $B\to K$ LQCD results. 
We encourage the authors of local FF calculations to provide these correlations, which play an important role in rare $B$ decays.

We evaluate the matching coefficients $\Delta C_{7,9}$ using the analytic NLO results provided in Ref.~\cite{Asatrian:2019kbk}.
The numerical evaluation is carried out using the code attached to the arXiv version.
The parametric uncertainty of the matching coefficients is estimated by varying the $c$- and $b$-quark masses
in the $\overline{\rm MS}$ scheme, following the central values and
uncertainties of the world averages~\cite{ParticleDataGroup:2020ssz}.
As explained in \Sec{sec:eft}, we keep the scale $\mu_b = 4.2\GeV$ fixed in our analysis.
We determine the parametric covariance matrix across the two different matching coefficients and across different $q^2$ values.
The imaginary part of the matching coefficient arises first at NLO in $\alpha_s$ and hence
by simply varying its input parameter one is likely to underestimate its total uncertainty.
To compensate for this effect, we add an additional $\alpha_s/\pi \sim 5\%$ systematic uncertainty to the imaginary parts of the matching coefficients. This systematic uncertainty is assumed to be uncorrelated across matching coefficient and $q^2$ points, 
and is therefore added in quadrature to the parametric covariance matrix.\\

At next-to-leading power, the computation of the non-local FFs requires knowledge the non-local
soft-gluon matrix elements $\tV{\lambda}$ (cf. \Eq{eq:masterNLFF}).
It has been shown in Ref.~\cite{Gubernari:2020eft} that  $\tV{\lambda}$ is negligibly small
at negative $q^2$. We choose to only account for these contributions by increasing the uncertainty of the non-local FFs,
by adding the square of their central values to the diagonal of the total covariance matrix.

To account for potentially large contributions beyond next-to-leading power (BNLP), we add an additional
relative uncertainty to the diagonal of total covariance matrix. This additional systematic uncertainty is chosen to be
\begin{equation}
    \Delta \HM{c,\lambda}\big|_\text{BNLP} = \frac{m_b \Lambda_\text{had}}{4m_c^2 - q^2} \times \HM{c,\lambda}\big|_\text{LP}\,,
\end{equation}
where $\Lambda_\text{had} = 0.1\,\GeV$. We hold that our approach yields very conservative estimates for the non-local FFs.\\

We emphasize that there are both (\emph{i}) sources of cross correlations between the local and non-local FFs for the same process
and (\emph{ii}) cross correlations between FFs for different processes.
\begin{itemize}
    \item[(\emph{i})] The first type of correlations arise from the fact that the non-local FFs depend on the local FFs (cf. \Eq{eq:masterNLFF}).
    Most of these correlations can be accounted for by working with the (mostly uncorrelated) set $\{\mathcal{F},\mathcal{F}_T/\mathcal{F},\mathcal{H}/\mathcal{F}\}$
    instead of $\{\mathcal{F},\mathcal{F}_T,\mathcal{H}\}$.
    Accounting for the remaining correlations throughout the different steps in the analysis is computationally expensive.
    To ensure that they are negligible, we perform an iterative procedure.
    The theory points at negative $q^2$ are first computed using the posteriors of the fit to the local FFs.
    We then perform a simultaneous fit of the local and non-local FF parameters and use the best-fit point to update the theory points (the covariance matrix is not updated).
    This procedure is iterated until convergence, that is until the posterior predicted FF values at negative $q^2$ match the input values.
    We finally perform a $\chi^2$ test to quantify the compatibility of the final theory points at negative $q^2$ and the points obtained without this procedure.
    For the three channels under consideration, we find $p$ values exceeding $99\%$ and conclude that the correlations between local and non-local FFs can be safely neglected.
    \item[(\emph{ii})] The second type of correlations arise from the small uncertainties of the matching coefficients $\Delta C_{7,9}$.
    Since these uncertainties are very small, so are the total correlations.
    In principle, also the LCSRs calculations of Refs.~\cite{Gubernari:2018wyi,Gubernari:2020eft} for the $B\to K^*$ and $B_s \to \phi$ local FFs are correlated, since they use a similar set of inputs and distribution amplitudes. However, these correlations are unknown.
    Therefore, we do not account for these correlations in this work either.
\end{itemize}

Our setup yields independent likelihoods for the processes $B\to K\ell^+\ell^-$, $B\to K^*\ell^+\ell^-$, and $B_s\to \phi\ell^+\ell^-$
that feature $8$, $24$, and $24$ degrees of freedom, respectively.
They account for two real-valued degrees of freedom per data point and
four data points per non-local FF.
The multivariate Gaussian distributions describing the values of $\HM{\lambda}/\FM{\lambda}$ at negative $q^2$ are provided as ancillary files
\texttt{BToK-nonlocal-data.yaml}, \texttt{BToKstar-nonlocal-data.yaml}, and \texttt{BsToPhi-nonlocal-data.yaml}.
The normalization to $\FM{\lambda}$ reduces the dependence of these likelihoods on local FFs.\\

In addition to these likelihoods, we also use the residues of the non-local $B\to M$ FFs at $q^2=M_{J/\psi}^2$.
The likelihoods for these residues are obtained from the experimental measurements of the branching ratios and angular observables
in $B\to M J/\psi$ decays, as discussed in \Sec{sec:theory} and detailed in \App{app:ampl}.
This extraction involves elements of the CKM matrix.
The latter are computed in the Wolfenstein parametrization where the parameters are chosen to be Gaussian
and based on the ``Spring 2021'' results by the CKMFitter collaboration~\cite{CKMFitter}:
\begin{align}
    \label{eq:CKM}
    A &= 0.816 \pm 0.009 \,, & \lambda &= 0.22500 \pm 0.00023 \,, \nonumber\\
    \bar{\rho} &= 0.1584 \pm 0.0066 \,, & \bar{\eta} &= 0.3507 \pm 0.0086 \,.
\end{align}
The $B\to M J/\psi$ branching ratios are taken from the most recent PDG world average~\cite{ParticleDataGroup:2020ssz}.
The angular observables are taken from LHCb measurements~\cite{LHCb:2013vga,LHCb:2019nin,LHCb:2021wte},
which are the most precise measurements currently available. Both types of observables suffice to constrain the residues
of the non-local FFs on the $J/\psi$ pole up to a global phase.\\

Within our analysis, we simultaneously fit $\hat{z}$-polynomials to the joint likelihood comprised
of the synthetic data points at negative $q^2$ and the experimental information at $q^2=M_{J/\psi}^2$.
For convenience, we carry out these fits using Lagrange basis polynomials.
Subsequently, the result are converted to the basis of orthogonal polynomials in \Eq{eq:H_expansion}.
This step is documented in detail in \App{app:NLFFFit}.
Since we consider Lagrange basis polynomials of degree five (i.e., $\order{\hat{z}^5}$), there are 12 real-valued fit parameters per non-local FF.
\begin{itemize}
    \item For $B\to K$ we constrain eight real-valued parameters from the synthetic data points at negative $q^2$
    and one real-valued parameter from the $B\to J/\psi K$ branching ratio.
    We are therefore left with three unconstrained parameters, one modulus and two phases.

    \item For $B\to V$ (i.e., $B\to K^*$ and $B_s\to \phi$) we constrain 24 real-valued parameters from the synthetic data points at negative $q^2$
    and five real-valued parameters from the $B\to J/\psi V$ branching ratio and angular observables.
    We are therefore left with seven unconstrained parameters, three moduli and four phases.
\end{itemize}
Therefore, the joint likelihood from the synthetic data points and the nonleptonic decays is \emph{not} sufficient to constrain all fit parameters by construction: our analysis features a total of 17 blind directions.
We parametrize these blind directions as follows:
\begin{itemize}
    \item the moduli and phases of the residues of the non-local FFs at the $\psi(2S)$ pole; and
    \item the phases of the residue of the longitudinal ($\lambda = 0$) non-local FFs at the $J/\psi$ pole.
\end{itemize}
As discussed in Ref.~\cite{Gubernari:2020eft} and revisited in \Sec{sec:nonlocalFFth}, the dispersive bound for the
non-local FFs provides some control over the truncation error of the expansion (\ref{eq:H_expansion}), which
has now become a parametric uncertainty.
Application of this bound is explained in \Sec{sec:th-pred:SM}, where SM predictions are derived.
In our analysis, the use of this bound is essential for phenomenological applications, since it provides
means to constrain the seventeen otherwise ``blind'' parameters, which would otherwise yields unbounded
uncertainties.

We can now motivate the choice to expand the non-local FFs up to order $\hat{z}^5$:
it is the lowest order that allows a nontrivial use of the dispersive bound. 
Truncating the expansion at lower order would not yield sufficiently many unconstrained fit parameters
and therefore fall short of a maximally conservative estimate of the uncertainties due to the
non-local FFs.
We have explicitly checked that expanding the non-local FFs up to $\hat{z}^6$ leads to
virtually the same size of parametric uncertainties; albeit with minor distortions of the shape
that are compatible with the $\hat{z}^5$ results within their respective uncertainties.

\subsubsection*{Results}

\begin{table}[t!]
    \centering
    \renewcommand{\arraystretch}{0.90}
    \begin{tabular}{lcccccc}
        \toprule
            Transition  & Constraints \#    & Priors \# & Comment       & Experimental Ref.
            & $\chi^2$  & $p$ value [\%] \\
        \midrule
             $B\to K$   & 10                & 9         & $B^0$\&$B^+$  & \cite{ParticleDataGroup:2020ssz}  & 2.6 & 11 \\[3pt]
             $B\to K^*$ & 30                & 29        & $B^0$\&$B^+$  & \cite{ParticleDataGroup:2020ssz,LHCb:2013vga} & 0.34 & 56 \\[3pt]
             $B_s\to \phi$
                        & 33                & 29        &               & \cite{ParticleDataGroup:2020ssz,LHCb:2019nin,LHCb:2021wte}& 0.48 & 98\\
        \bottomrule
    \end{tabular}
    \caption{
    \label{tab:fitNLFFsSM}
        Summary of the goodness of fit for the posteriors of the non-local $B\to M$ FFs.
        The 17 unconstrained parameters discussed in the text do not enter the likelihood, which
        makes a goodness-of-fit discussion possible.
        We use the neutral and charged modes for the branching ratios of $B\to K^{(*)}J/\psi$,
        as well as the two angular analyses of $B_s\to \phi J/\psi$ decays,
        which differ in the decay of the $J/\psi$ to either muon or electron pairs.
    }
\end{table}

The quality of the fits to non-local FF for the three processes $B\to K$, $B\to K^*$ and $B_s\to \phi$ is very satisfactory,
with $p$ values in excess of $11\%$.
A summary of the inputs and the goodness of these fits is given in \Tab{tab:fitNLFFsSM}.
For further discussion of our results, including determinations of uncertainties and discussions of the saturation of the dispersive bound,
we draw posterior samples for all non-local FF parameters.
The parameters included in the likelihood of each fit follow a multivariate Gaussian distribution.
The remaining parameters with blind directions, i.e., the longitudinal phases of the residues on the $J/\psi$ pole
and all phases and moduli describing the residues on the $\psi(2S)$ pole, follow their respective priors.
For these phases, we draw from independent uniform priors on the interval $[0, 2\pi)$.
For these moduli, we draw from independent uniform priors whose intervals are conservatively
chosen to contain at least $99\%$ of the samples that are allowed by the bound.

\begin{figure}
    \centering
    \includegraphics[width=.4\textwidth]{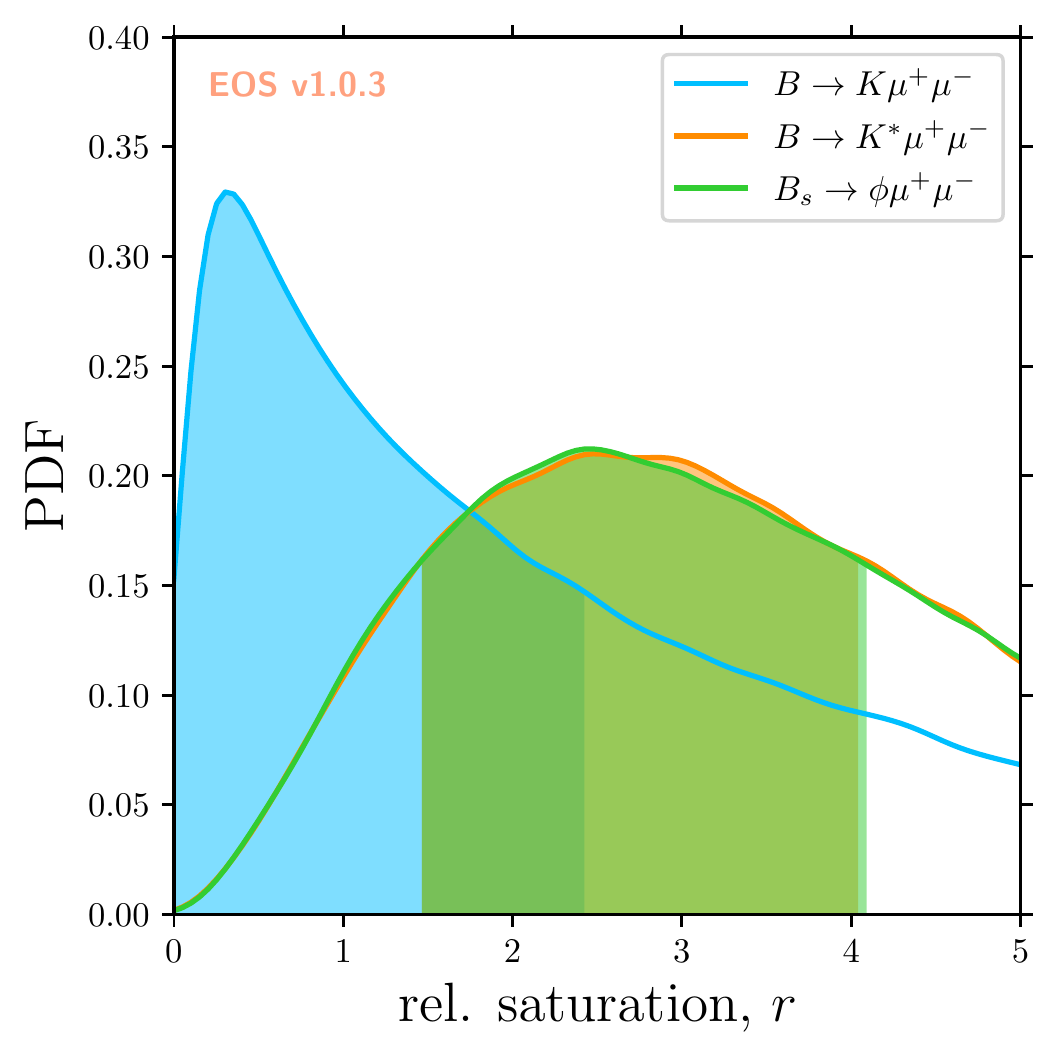}
    \hspace{7mm}
    \includegraphics[width=.4\textwidth]{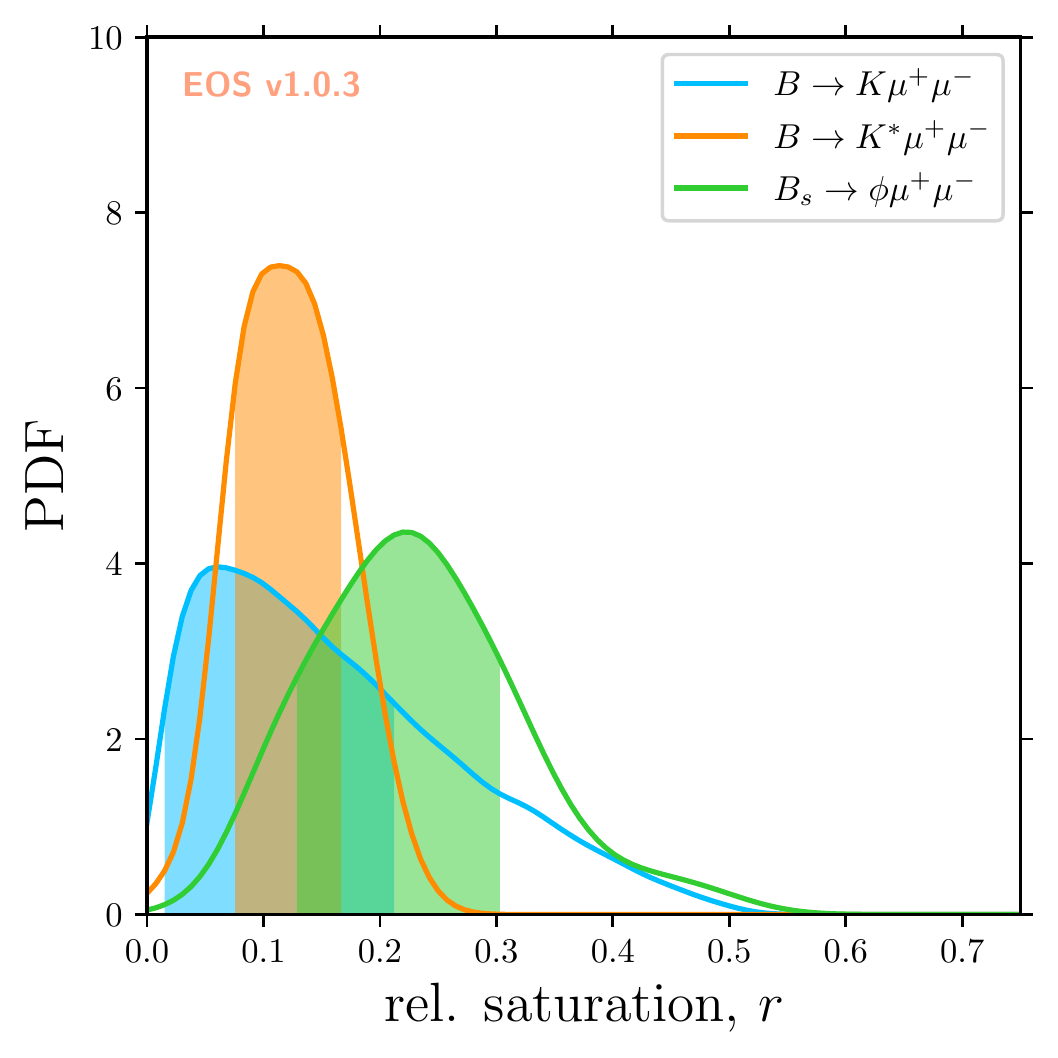}
    \caption{%
        (left) Posterior-predictive PDFs~$p_\mathcal{B}^{B\to M}(r)$ for the relative saturation
        $r$ of the dispersive bound by a single process $B\to M$ \emph{in the absence of applying the bound}.
        We only display the interval $r \in [0, 5]$.
        (right) Reweighted PDFs~$p_\mathcal{B}^{B\to M}(r)$ as defined in \Eq{eq:reweighting},
        that is after applying the bound.
        The changes in the shape of $B\to K^*$ and $B_s\to \phi$ in the right-hand plot are due to different factors in the penalty function in \Eq{eq:reweighting}.
    }
    \label{fig:saturation}
\end{figure}

We present the central value and standard deviation of the analytic functions $\hHM{c,\lambda}$ --- defined in \Eqs{eq:HhatBK}{eq:HhatBV} --- in \Tabs{tab:NLFFposteriors:BToK}{tab:NLFFposteriors:BsToPhi}.
The central values and covariances are additionally provided in the three ancillary files
\texttt{\justify BToK-hatH.yaml}, \texttt{\justify BToKstar-hatH.yaml}, and \texttt{\justify BsToPhi-hatH.yaml}.
These distributions do not yet respect the dispersive bound. To illustrate this, we
produce posterior-predictive distributions for the saturation of the bound by each process.
These distributions are illustrated in the left-hand plot of \Fig{fig:saturation}.
For $B\to K$ we find a clear peak at $\sim 40\%$ saturation. For $B\to K^*$ and $B_s\to \phi$
the bound is clearly violated, and we find clear peaks at $\sim 250\%$ saturation.
This is not surprising, since the bound has not been taken into consideration for the sampling
of the parameters.

\subsection[{Standard Model Predictions of $B_q\to M\mu^+\mu^-$ Observables}]{Standard Model Predictions of $\boldsymbol{B_q\to M\mu^+\mu^-}$ Observables}
\label{sec:th-pred:SM}

\subsubsection*{Analysis Setup}
Using the preceding results within this section, we compute data-driven SM predictions for a variety of observables.
In addition to the parametrized approach to non-local FFs $\HM{c,\lambda}$,
we account for non-local FFs $\HM{sb,\lambda}$ that arise from four-quark operators with and without charm fields
and from the chromomagnetic operator, as discussed in \Sec{sec:eft} and detailed in \App{app:Hsb}.
Our predictions arise from re-weighted importance samples of the posterior-predictive distributions.\\

As discussed earlier in this section, the application of the dispersive bound for the non-local FFs
is central to our approach. We apply it to the samples for each of the three processes, and we use
the process $B\to K$ to illustrate how we apply it.
Let $r^{B\to K}$ be the saturation of the bound due to the longitudinal and only non-local FF in $B\to K\ell^+\ell^-$:
\begin{equation}
    r^{B\to K} \equiv \sum_n |\beta_{0,n}^{B\to K}|^2\,.
\end{equation}
We compute this saturation for each of the importance samples for this process. Since we assumed the processes to be uncorrelated, we 
can re-weight each sample with a relative weight $w(r^{B\to K})$, which is computed as
\begin{multline}
    \label{eq:reweighting}
    w(r^{B\to K})
        = \int dr^{B\to K^*} \int dr^{B_s\to\phi} \, p_\mathcal{B}^{B\to K^*}(r^{B\to K^*}) \, p_\mathcal{B}^{B_s\to \phi}(r^{B_s\to \phi}) \,
            \\\times P\left(2 \, r^{B\to K} + 2 \, r^{B\to K^*} +  r^{B_s\to\phi} \right) \,.    
\end{multline}
Above, $p_\mathcal{B}^{B\to K^*}$ and $p_\mathcal{B}^{B_s\to \phi}$ are the PDF of the raw saturation of the bound
due the other processes under consideration, as shown in left-hand plot of \Fig{fig:saturation},
and $P$ is a penalty function that enforces the strong dispersive bound.
The application to weights for the other processes is straight forward.
For a strict handling of the bound, the penalty function would read $P(r) \equiv \theta(1 - r)$.
However, to account for the perturbative uncertainty in the calculation of $\chi^\text{OPE}$ ---
which sets the scale for the bound,  see \App{app:bound} for the definition --- a different penalty function can be used.
Here, we use the approach of \Reff{Bordone:2019vic}, which applies the penalty function
\begin{equation}
     \label{eq:penalty_likelihood}
    -2\ln P(r) 
        = \begin{cases}
            0    &   \text{if}\, r < 1, \\
            \dfrac{\left(r - 1\right)^2}{\sigma^2}
                 & \text{otherwise.}
         \end{cases}
\end{equation}
We use $\sigma = 5\%$, which comfortably accounts for the perturbative uncertainty of \Reff{Gubernari:2020eft}.
We show the PDF for the process-specific saturation of the re-weighted importance samples in the
right-hand plot of \Fig{fig:saturation}. As shown, the samples with saturation in excess of $100\%$ have been
effectively removed by re-weighting.\\

\subsubsection*{Results}
\begin{figure}[t!]
    \centering
    \includegraphics[width=.32\textwidth]{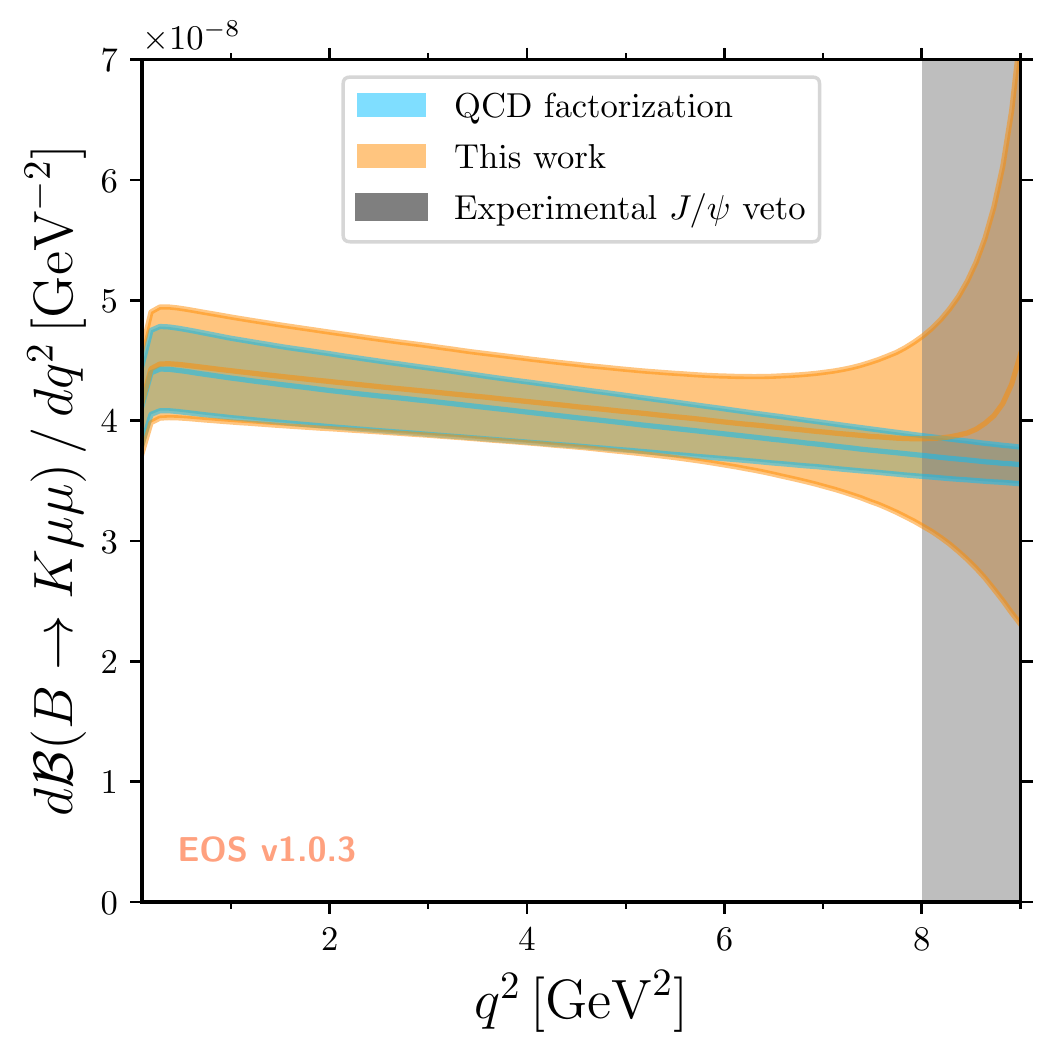}
    \includegraphics[width=.32\textwidth]{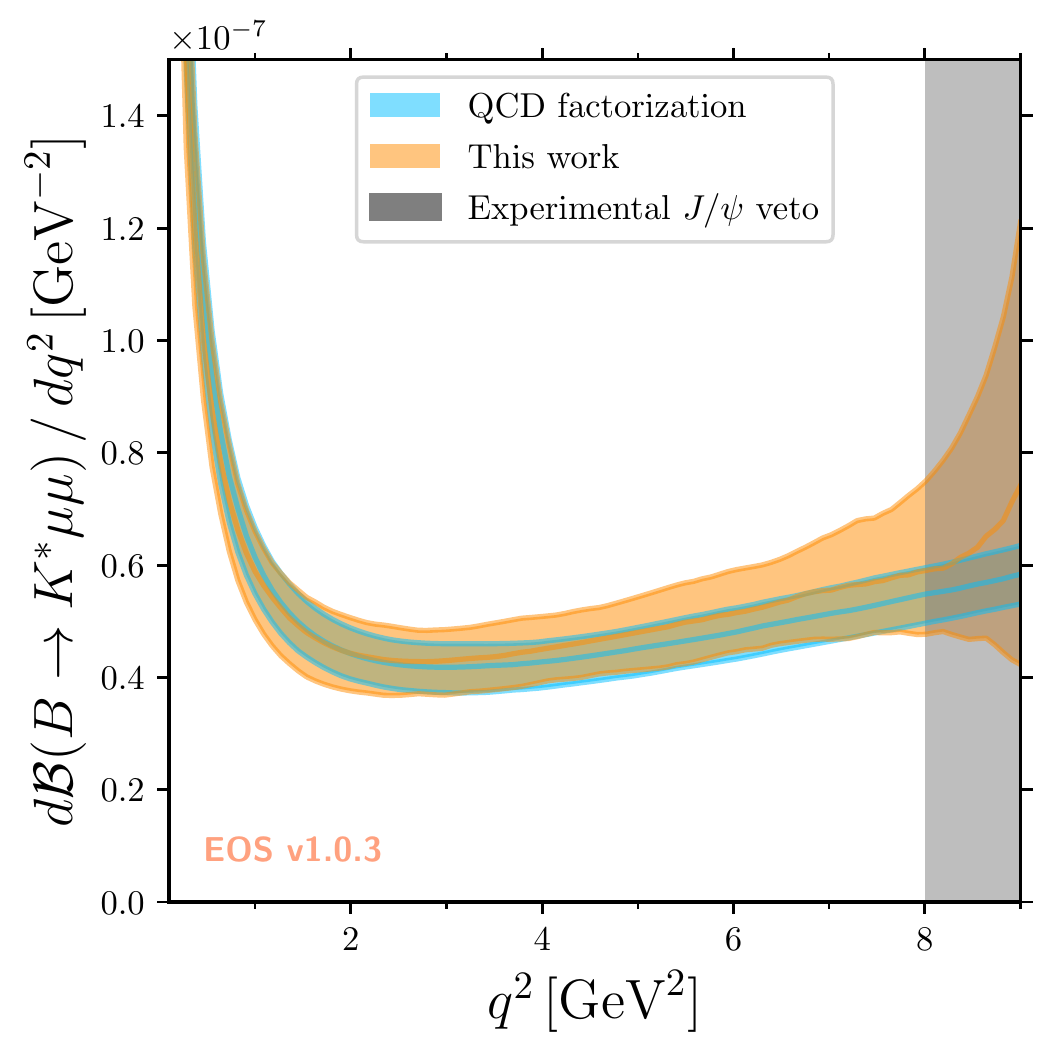}
    \includegraphics[width=.32\textwidth]{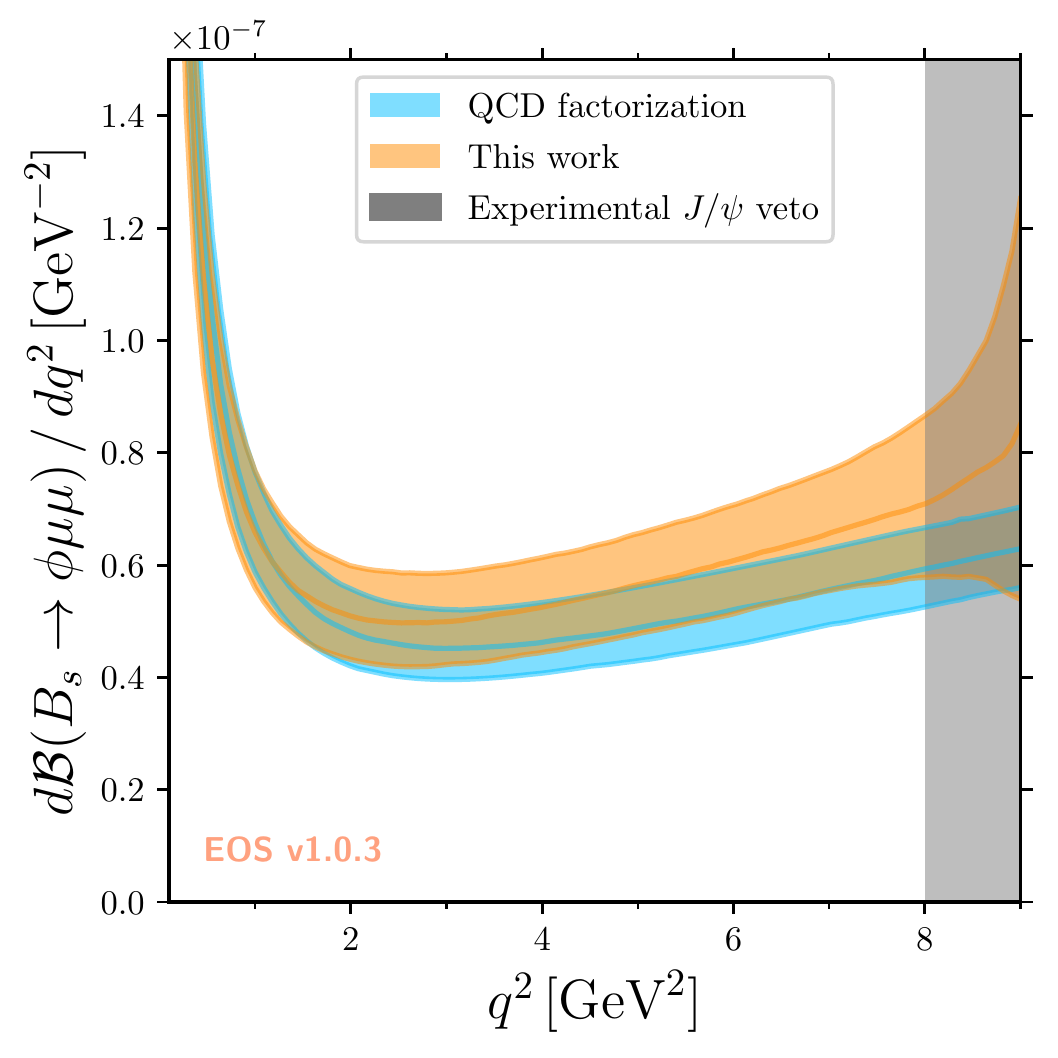}
    \caption{
    Comparison between our SM predictions for the differential branching ratio in the three channels and the result obtained in QCD factorization~\cite{Beneke:2001at, Beneke:2004dp}, including the form-factor relations in the large-energy limit.
    }
    \label{fig:SM_vsBFS}
\end{figure}

We begin with the predictions for the differential branching ratios for the three processes under consideration,
which we illustrate in \Fig{fig:SM_vsBFS}. Our predictions are juxtaposed with the predictions obtained
in the QCD factorization framework of Refs.~\cite{Beneke:2001at, Beneke:2004dp}.
In these references, the large-energy symmetry relations to the local FFs (the ``factorizable contributions'') are used.
The likelihoods used to constrain the remaining local FFs are described in \Sec{sec:th-pred:local}.
Predictions for further observables, including the angular observables, are obtained and provided in \App{app:plots-tables}.

Several observations regarding \Fig{fig:SM_vsBFS} are in order:
\begin{itemize}
    \item The central values of the two different predictions are in excellent agreement within uncertainties.
    However, our predictions show a larger uncertainty than the QCDF predictions.
    \item Our predictions exhibit increasing uncertainties when approaching the $J/\psi$ pole, as expected.
    No such increase is present in the QCDF approach, since it does not account for this resonant effect.
    \item We find very similar shapes across both approaches for $B\to K^{(*)}\mu^+\mu^-$.
    However, the shapes of the
    two predictions for $B_s\to \phi\mu^+\mu^-$ differ significantly, with a larger slope visible in our predictions.
\end{itemize}

We verify that our predictions for the $B\to M\psi(2S)$ branching ratios are consistent with experimental data.
Although we obtain central values an order of magnitude larger, the sizable uncertainties of $\order{100\%}$ make our results trivially compatible with data.

As a last remark, we emphasize the theory predictions in our approach can be systematically improved with more precise local FF calculations --- which are the largest contribution to the uncertainties in our SM predictions --- and/or saturating the dispersive bound adding more channels, for instance $\Lambda_b\to\Lambda\mu^+\mu^-$.

\section{Confrontation with Data}
\label{sec:confrontation}

Having correlated predictions within the SM at hand, we confront them with the available
experimental measurements in \Sec{sec:confrontation:SM}.
We then continue with a simple BSM analysis of the data in \Sec{sec:confrontation:BSM}.

\subsection{Compatibility of the SM Results}
\label{sec:confrontation:SM}

\begin{table}[p]
    \centering
    \renewcommand{\arraystretch}{0.95}
    \resizebox{\textwidth}{!}{
    \begin{tabular}{l c c c c c}
        \toprule
        Process         & Description           & \# of obs.         & Collaboration & Comment       & Ref.\\
        \midrule
        \multirow{3}{*}{$B\to K\mu^+\mu^-$}
                        & binned $\mathcal{B}$  & 2                  & BaBar         &               & \cite{BaBar:2012mrf}\\
                        & binned $\mathcal{B}$  & 1                  & Belle         &               & \cite{Belle:2019xld}\\
                        & binned $\mathcal{B}$  & 10                 & LHCb          & $B^0$\&$B^+$  & \cite{Aaij:2014pli}\\
        \midrule
        \multirow{7}{*}{$B\to K^*\mu^+\mu^-$}
                        & binned ang.~obs.      & 12                 & ATLAS         & $6$ obs.      & \cite{ATLAS:2018gqc}\\
                        & binned ang.~obs.      & 16                 & CMS           & $B^0$\&$B^+$; $A_{\rm FB}\&F_{\rm L}$
                                                                                                    & \cite{CMS:2013mkz, CMS:2015bcy, CMS:2020oqb}\\
                        & binned ang.~obs.      & 64                 & LHCb          & $B^0$\&$B^+$; $8$ obs.
                                                                                                    & \cite{LHCb:2020gog, LHCb:2020lmf}\\
        \cmidrule{2-6}
                        & binned $\mathcal{B}$  & 2                  & BaBar         &               & \cite{BaBar:2012mrf}\\
                        & binned $\mathcal{B}$  & 2                  & Belle         & $B^0$\&$B^+$  & \cite{Belle:2019oag}\\
                        & binned $\mathcal{B}$  & 7                  & CMS           &               & \cite{CMS:2013mkz, CMS:2015bcy}\\
                        & binned $\mathcal{B}$  & 7                  & LHCb          & $B^0$\&$B^+$  & \cite{Aaij:2014pli, LHCb:2016ykl}\\
        \midrule
        \multirow{2}{*}{$B_s\to \phi\mu^+\mu^-$}
                        & binned ang.~obs.      & 12                 & LHCb          & $4$ obs.      & \cite{LHCb:2021xxq}\\
        \cmidrule{2-6}
                        & binned $\mathcal{B}$  & 4                  & LHCb          &               & \cite{LHCb:2021zwz}\\
        \midrule
        $B_s\to \mu^+\mu^-$
                        & $\mathcal{B}$         & 1                  & ATLAS+CMS+LHCb
                                                                                     & world average & \cite{LHCb:2020zud}\\
        \bottomrule
    \end{tabular}}
    \caption{%
        \label{tab:likelihoods}
        Summary of the experimental likelihoods used in our analysis.
        Binned branching ratios (\BR) and angular observables (ang.~obs.) are binned in the squared dimuon invariant mass $q^2$.
        The measurements cover the range from $q^2\sim 1\,\GeV^2$ to $q^2\simeq 8\,\GeV$.
        Present measurements of the angular observables below $\sim 1\,\GeV^2$ suffer from inconsistent
        treatment of muon mass effects, and are not used here.
        Bins above $\simeq 8\,\GeV^2$ and below the $J/\psi$ pole are presently not available.
        For the $B_s\to\phi\mu^+\mu^-$ angular analysis of Ref.~\cite{LHCb:2021xxq},
        we marginalized the likelihood w.r.t. the CP-violating observables $\mathcal{A}_i$,
        which are not relevant to our analysis.
    }
    \vspace*{2em}
    \newcommand{\mc}[3]{\multicolumn{#1}{#2}{#3}}
    \resizebox{\textwidth}{!}{
    \begin{tabular}{l c c c c c c c c c c c}
        \toprule
        \multirow{2}{*}{Analysis} & \mc{5}{c}{SM}                                                & & \mc{5}{c}{BSM$_{9,10}$} \\
                                  & \mc{2}{c}{$\chi^2$} & d.o.f.    & \mc{2}{c}{$p$ value [\%]}  & & \mc{2}{c}{$\chi^2$} & d.o.f.    & \mc{2}{c}{$p$ value [\%]} \\
        \midrule
        $B\to K\mu^+\mu^-$        & $54$  & $(23)$      & $10$      & $< 10^{-5}$  & $(1.1)$     & & \mc{5}{c}{---}\\
        $B\to K^*\mu^+\mu^-$      & $106$ & $(103)$     & $103$     & $41$         & $(48)$      & & $99$ & $(99)$       & $101$     & $53$  & $(54)$\\  
        $B_s\to \phi\mu^+\mu^-$   & $19$  & $(13)$      & $9$       & $2.7$        & $(16)$      & & $13$ & $(13)$       & $7$       & $6.9$ & $(7.8)$\\
        \midrule
        $B_s\to \mu^+\mu^-$ \& $B\to K\mu^+\mu^-$
                                  & $59$  & $(28)$      & $11$      & $< 10^{-5}$  & $(0.33)$    & & $20$ & $(20)$       & $9$       & $1.7$ & $(1.8)$\\
        \bottomrule
    \end{tabular}}
    \let\mc\undefined
    \caption{%
        \label{tab:gof}
        Summary of the goodness of fit at the respective best fit points for the individual analyses.
        We provide two sets of values, the main numbers include both the theory priors and the experimental likelihoods,
        while the values in parenthesis only account for the latter.
    }
\end{table}

\subsubsection*{Analysis Setup}
We confront our SM predictions obtained in \Sec{sec:th-pred:SM} with the presently available experimental measurements from the
ATLAS, BaBar, Belle, CMS, and LHCb collaborations. For this purpose, we restrict our analysis to the
processes $B\to K\mu^+\mu^-$, $B\to K^*\mu^+\mu^-$, and $B_s\to \phi\mu^+\mu^-$. A summary of the three likelihoods that describe the data
is presented in \Tab{tab:likelihoods}.

The experimental constraints on the branching ratios are driven by LHCb measurements.
For each $B\to M$ transition, these measurements are performed relative to the branching ratio of the normalization channel $B\to M J/\psi$.
Since our setup also predicts these non-leptonic branching fractions, we use LHCb measurements of the ratios $\mathcal{B}(B\to M\mu^+\mu^-) / \mathcal{B}(B\to MJ/\psi)$.\footnote{
    These ratios are extracted from LHCb publications using:
    \begin{equation*}
        \mu_\mathrm{ratio} = \frac{\mu_{\ell\ell}}{\mu_{J/\psi}} \,, \qquad
        \sigma^\mathrm{stat}_\mathrm{ratio} = \frac{\sigma^\mathrm{stat}_{\ell\ell}}{\mu_{J/\psi}} \,, \qquad
        \sigma^\mathrm{syst}_\mathrm{ratio} = \frac{1}{\mu_{J/\psi}} \left(
            \sigma^\mathrm{syst}_{\ell\ell}
            - \frac{\mu_{\ell\ell}}{\mu_{J/\psi}} \sqrt{ (\sigma^\mathrm{stat}_{J/\psi})^2 + (\sigma^\mathrm{syst}_{J/\psi})^2 }
        \right) \,,
    \end{equation*}
where $\mu, \sigma^\mathrm{stat}$ and $\sigma^\mathrm{syst}$ are the mean, the statistical uncertainties and the systematic uncertainties, while $\ell\ell$ and $J/\psi$ denote $\mathcal{B}(B\to M\mu^+\mu^-)$ and $\mathcal{B}(B\to MJ/\psi)$, respectively.
}
This avoids double-counting of the experimental uncertainty of the normalization mode in our setup.

\begin{figure}[t!]
    \centering
    \includegraphics[width=.4\textwidth]{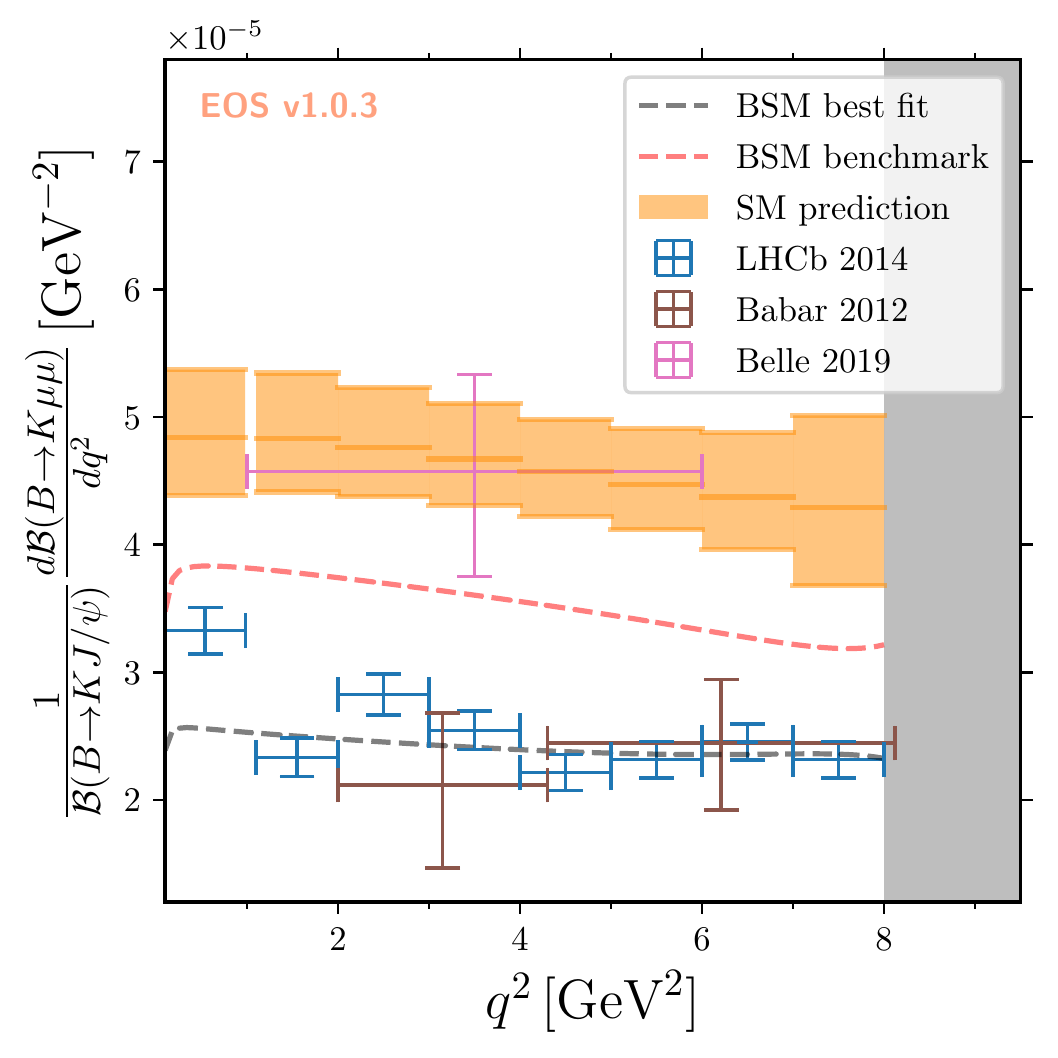}
    \includegraphics[width=.4\textwidth, height=.398\textwidth]{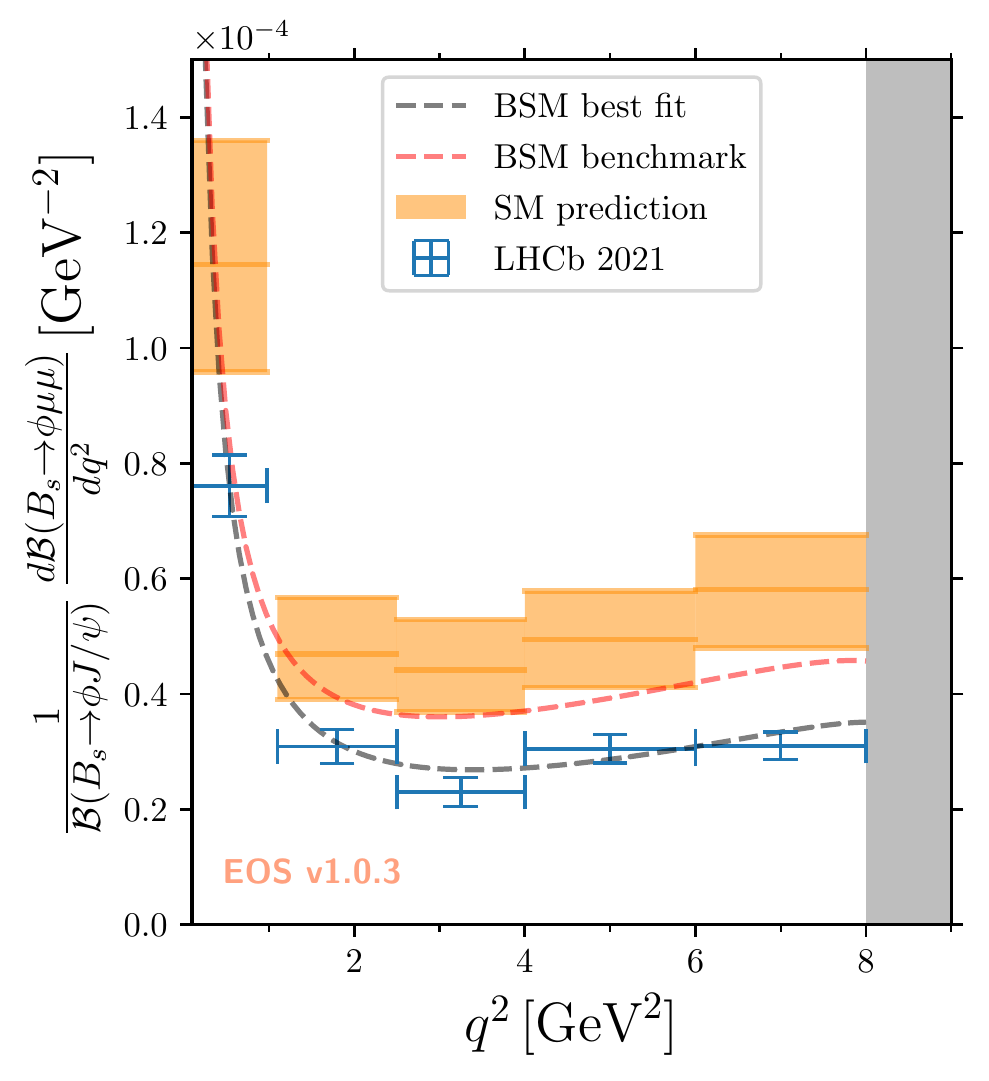} \\
    \includegraphics[width=.4\textwidth]{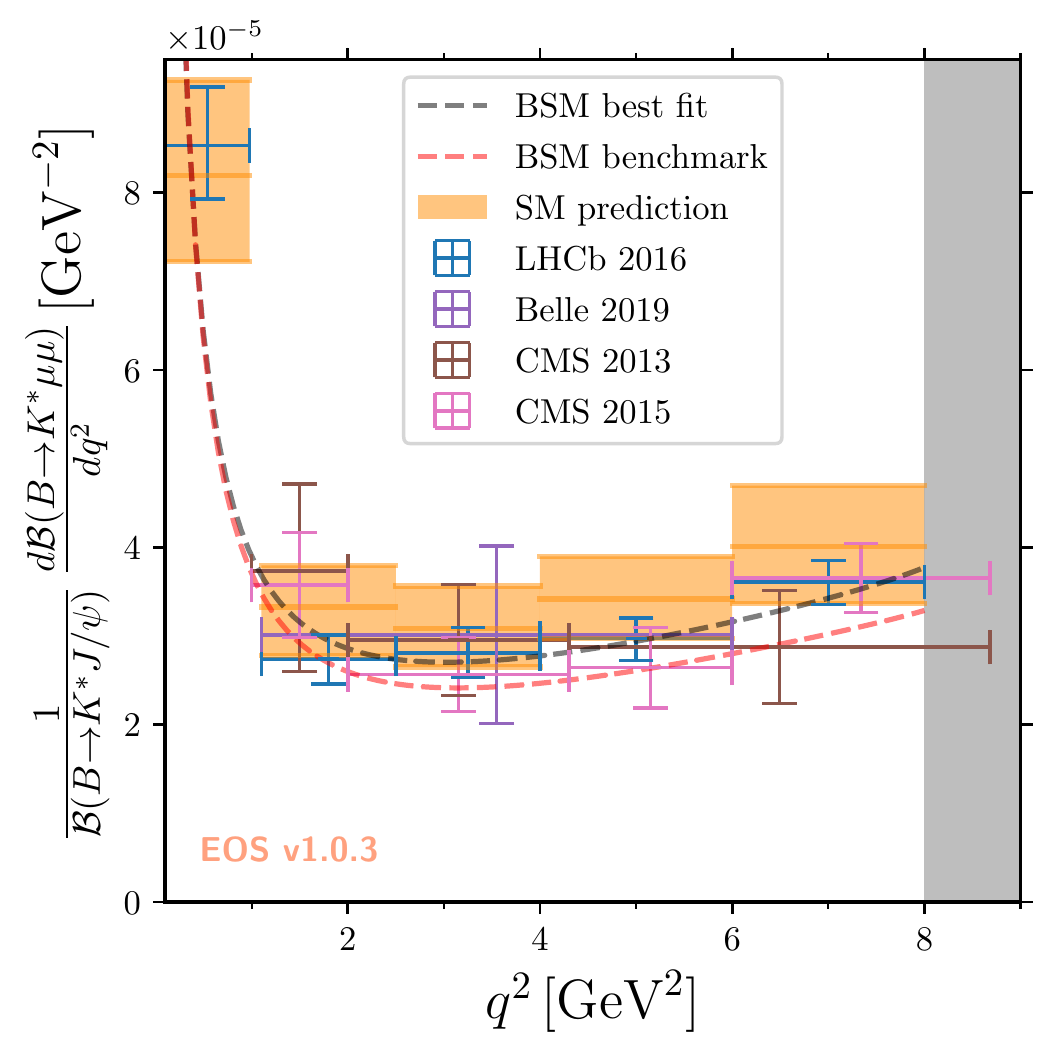}
    \includegraphics[width=.4\textwidth, height=.393\textwidth]{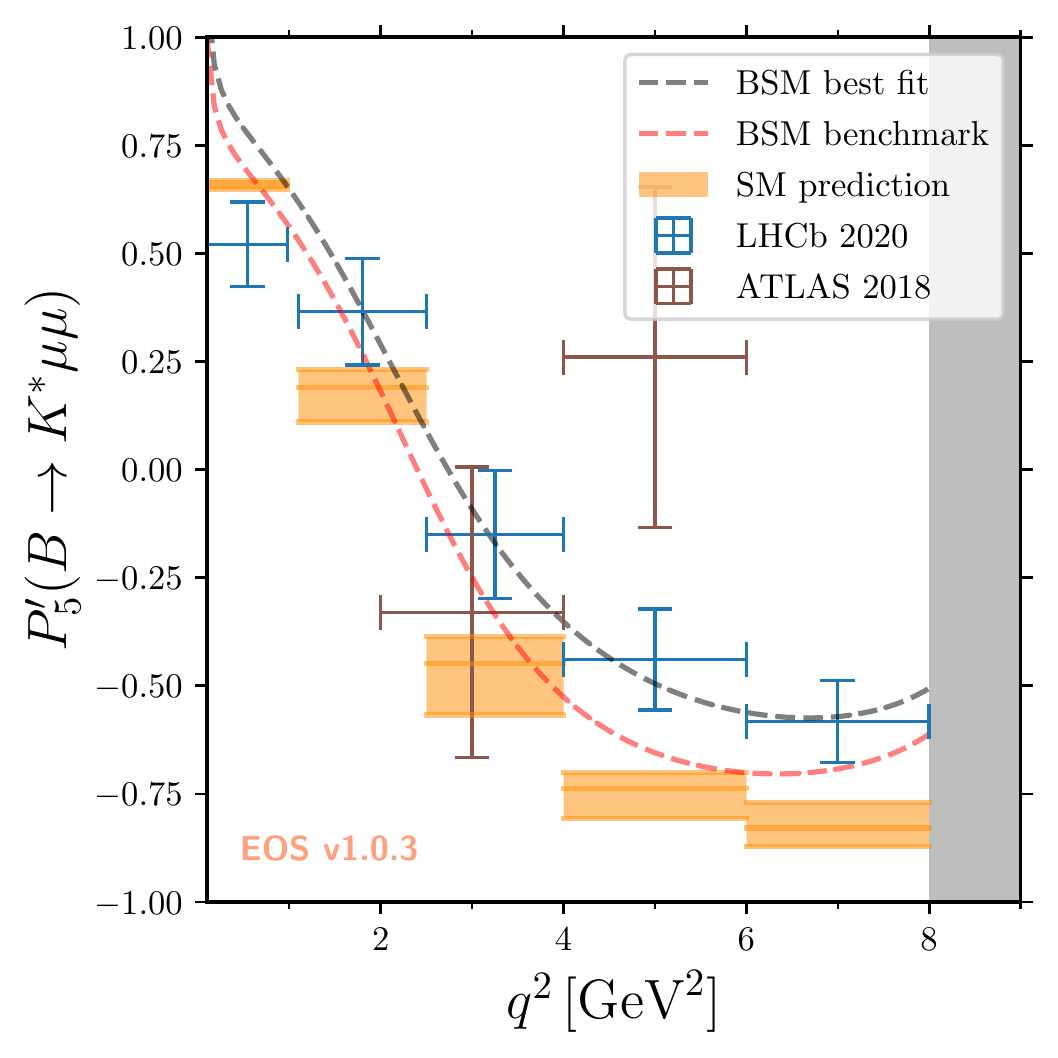}
    \caption{%
        Updated SM predictions for the normalized differential branching ratios and the optimized angular observable $P'_5$,
        which we overlay with two BSM scenarios. The scenario labeled ``BSM best fit'' corresponds to the \emph{process-specific} BSM best-fit point
        of the likelihoods of Fig.~\ref{fig:BSM_likelihoods}.
        ``BSM benchmark'' is obtained by setting $C_9^\mathrm{BSM} = - C_{10}^\mathrm{BSM} = -0.5$ and adapting
        all hadronic parameters.
        The small uncertainty in the first bin of $P'_5$ compared to the literature is due to a smaller
        soft gluon contribution~\cite{Gubernari:2020eft}.
    }
    \label{fig:SM_Plots}
\end{figure}

\subsubsection*{Results}
In \Fig{fig:SM_Plots} we compare our predictions with the available experimental data of the branching ratios and the $P'_5$ observable for $B \to K^*\mu^+\mu^-$ in bins of $q^2$.
Further plots confronting our SM predictions of the remaining angular observables with the data are provided in Appendix \ref{app:plots-tables}.
The bins are chosen to align with those of the LHCb measurements for ease of comparison.
We find a clear discrepancy between the central values of the predictions and the measurements of certain observables.
The compatibility of the data with the SM predictions is determined with a goodness-of-fit test at
the best-fit point of the hadronic parameters for the local and non-local FFs. We emphasize
that we fit \emph{all} the hadronic parameters, i.e., we permit the fit to move away
from the prior prediction of the FF coefficients as obtained in \Sec{sec:th-pred:local} and \Sec{sec:th-pred:nonlocal}.
A summary of our fit results is shown in the ``SM'' columns of \Tab{tab:gof}.
In this table we give two values per each channel.
The first value is obtained by evaluating the $\chi^2$ of the experimental likelihoods
and the multivariate priors for the local and non-local FF parameters.
The second value, given in parenthesis, only accounts for the experimental likelihoods.
Our findings can be summarized as follows:

\begin{itemize}
    \item In our approach, the SM predictions of each individual $B\to K^*\mu^+\mu^-$ observable is in good agreement with the data.
    The $p$ value of the SM best-fit point exceed $40\%$ and doesn't require a large departure from the local and non-local FF priors.
    This observation differs from what can be found in the literature (see \Reff{Descotes-Genon:2015uva} for a channel specific analysis).
    The reason for this is partially explained by the input used for the local FF fit.
    Performing the same analysis using the FF parameters of \Reff{Bharucha:2015bzk} results in an increase of the tension.
    The tension between the SM prediction and the data is dominated by the full angular distribution,
    which yields $\sim 90\%$ of the overall $\chi^2$.
    \item A similar conclusion also apply, although to a lesser extend, to $B_s\to\phi\mu^+\mu^-$.
    The SM best-fit point can account for the experimental data with a $p$ value of $16\%$.
    However, this agreement requires a sizable distortion of the hadronic parameters, and especially of the local FFs.
    Here again, the tension is smaller than in the literature due to local FF inputs (see \Reff{Descotes-Genon:2015uva} for a channel specific analysis).
    As can be seen in \Fig{fig:local_ffs}, the inputs used in the present work imply larger uncertainty on these FFs,
    hereby reducing the tension between the SM predictions and the measurements.
    \item For $B\to K\mu^+\mu^-$, the tension between the SM and the data is sizable.
    The SM best-fit point can barely account for the semi-leptonic branching ratios, and this low agreement is only obtained through a large distortion of the local FFs.
    Note that contrary to $B\to K^*\mu^+\mu^-$ and $B_s\to \phi\mu^+\mu^-$, for the local FFs in $B\to K\mu^+\mu^-$ we use light-meson LCSR results that have smaller uncertainties than the corresponding $B$-meson LCSR results (cf. \Sec{sec:th-pred:local}).
    \item We finally observe that for the three processes under consideration, the SM best-fit points show an upper deviation with respect to the prior
    describing the moduli on the $\psi(2S)$. This produces a significant tension with respect to the measurements of the $B\to M\psi(2S)$ branching ratios.
    This behavior is expected as it allows to decrease the predicted semi-leptonic branching ratios in the physical region, as required by the data.
\end{itemize}

\subsection{BSM Analysis}
\label{sec:confrontation:BSM}

\subsubsection*{Analysis Setup}

As an outgrowth of our SM predictions, we conduct a simple BSM analysis, accounting model-independently for effects in the operators $\cO_{9}^\mu$ and $\cO_{10}^\mu$ only. Clearly, this simple analysis should be followed up with an analysis
that takes into account the full basis of $sb\mu\mu$ operators at mass dimension six. However, it would require a magnitude more in computing resources and time than what has been used in the present analysis.\\

Our analysis involves the theoretical likelihoods for the local and non-local FFs as discussed in the previous section, the experimental likelihoods for the various measurements listed in \Tab{tab:likelihoods},
and non-informative (i.e., flat) priors for all fit parameters.
Since we only fit for Wilson coefficients in the $sb\mu\mu$ sector of the WET
and we wish to emphasize the effects of the non-local FFs,
we choose not to include the lepton universality testing ratios $R_{K^{(*)}}$.\\

Given the large number of parameters, we choose to conduct our BSM analysis as three individual analyses,
labeled $B\to K\mu^+\mu^-$ + $B_s\to\mu^+\mu^-$, $B\to K^*\mu^+\mu^-$, $B_s\to \phi\mu^+\mu^-$.
The combination of $B\to K\mu^+\mu^-$ and $B_s\to\mu^+\mu^-$ data is required to simultaneously constraint
$\mathcal{C}_9$ and $\mathcal{C}_{10}$.
All three analyses share independent but identical uniform priors for the Wilson coefficients $\mathcal{C}_9$ and $\mathcal{C}_{10}$.

Since we choose to conduct three individual analyses, our application of the dispersive bound is limited to the process under study, thereby weakening the bound somewhat. This leads to posterior regions that are slightly more conservative than they could be.
The three analyses differ in their respective likelihoods and in the sets of hadronic parameters.
Due to the use of the absolute branching ratio of $B_s\to\mu^+\mu^-$ decay, the
simultaneous analysis of $B\to K\mu^+\mu^-$ + $B_s\to\mu^+\mu^-$
requires additionally priors for the Wolfenstein CKM parameters.
The latter are chosen to be Gaussian as given in \Eq{eq:CKM}.
Except for the $B_s$ decay constant $f_{B_s}$, all hadronic parameters follow from our results in the previous section.
For this decay constant we use a univariate Gaussian prior of $f_{B_s} = 0.2303 \pm 0.0013 \GeV$ \cite{Aoki:2021kgd}.\\

\subsubsection*{Results}
\begin{figure}
    \centering
    \includegraphics[width=.5\textwidth]{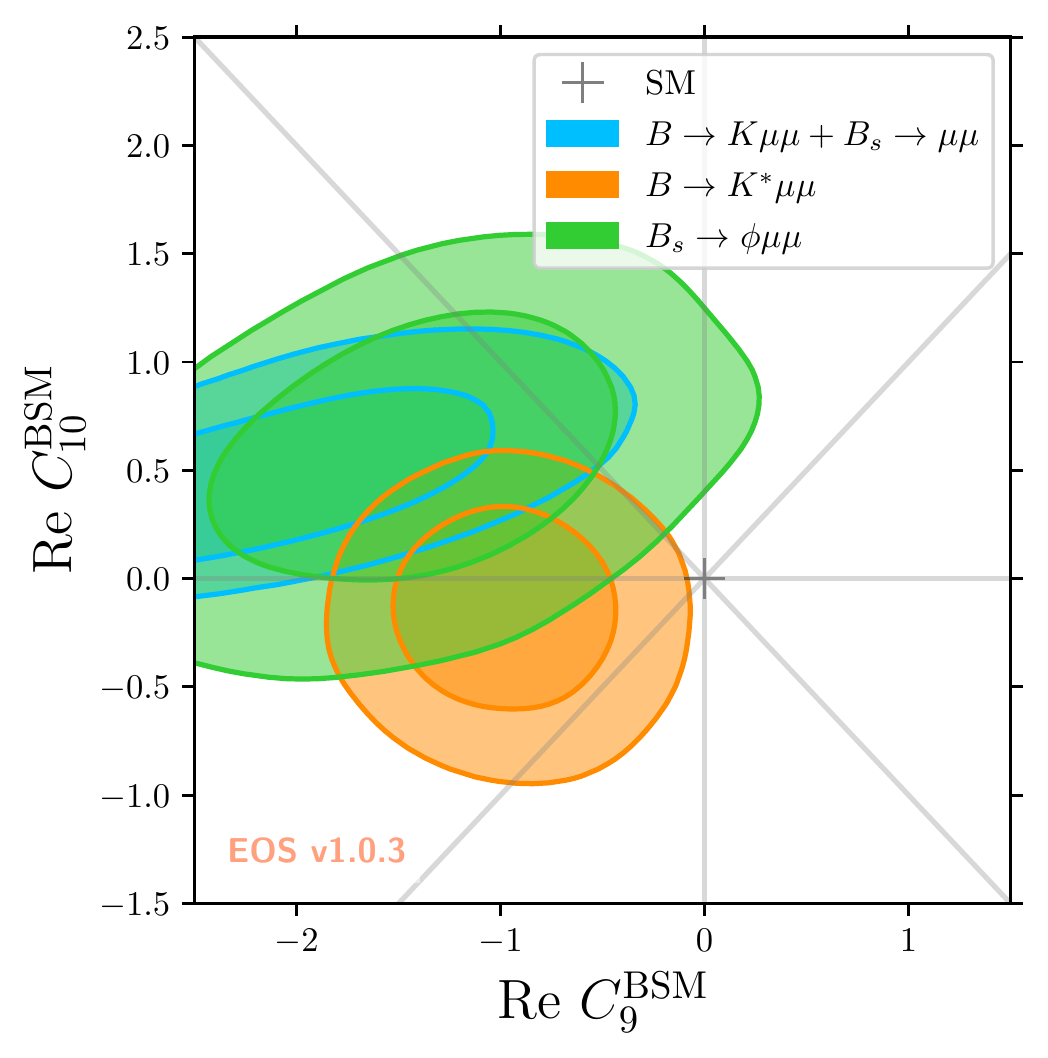}
    \caption{$1$ and $2\sigma$ contours of the posterior samples of the $C_9^\mathrm{BSM}, C_{10}^\mathrm{BSM}$ fit.
    All other Wilson coefficients are assumed SM-like.
    The strong dispersive bound is applied to all samples.
    The pulls are $5.7\sigma$, $2.7\sigma$ and $2.6\sigma$ for $B\to K\mu^+\mu^-$ + $B_s\to\mu^+\mu^-$, $B\to K^*\mu^+\mu^-$, and $B_s\to \phi\mu^+\mu^-$, respectively.}
    \label{fig:BSM_likelihoods}
\end{figure}

We show the three marginalized 2D posteriors for the individual likelihoods in \Fig{fig:BSM_likelihoods}.
We observe that the channel-specific results are in good agreement with each other: the point
$(\Re{C_9}^\mathrm{BSM}, \Re{C_{10}}^\mathrm{BSM}) = (-1.0, +0.4)$ is compatible with all channels at the $\sim 1\sigma$
level. This shift away from the SM point is compatible with albeit somewhat larger than what has been discussed
previously in the literature~\cite{Alguero:2021anc,Altmannshofer:2021qrr,Ciuchini:2020gvn,Hurth:2021nsi,Du:2015tda}.
To compute the SM-pull in the marginalized posterior plane, we approximate the posterior distributions with
Gaussian mixture densities and compute the isobar of the distribution corresponding to the SM point.
We find pulls of $5.7\sigma$, $2.7\sigma$ and $2.6\sigma$ for $B\to K\mu^+\mu^-$ + $B_s\to\mu^+\mu^-$, $B\to K^*\mu^+\mu^-$, and $B_s\to \phi\mu^+\mu^-$, respectively.

A summary of our fit results is shown in the ``$\mathrm{BSM}_{9,10}$'' columns of \Tab{tab:gof}.
We observe a small improvement of the goodness-of-fit in $B\to K^*\mu^+\mu^-$ with respect to the SM fit, as expected from our previous comments.
For $B_s\to \phi\mu^+\mu^-$, the global $\chi^2$ also improved, resulting in larger $p$ value, but the one associated to the experimental likelihood only changed marginally.
As can be inferred from the number in parenthesis, the best-fit point can now be obtained without distortion of the hadronic parameters.
The $B\to K\mu^+\mu^-$ fit is also improved in the presence of BSM physics, but a tension remains.
We find that the large $\chi^2$ value is driven by Belle 2019 measurement of the semi-leptonic branching ratio.
Being in agreement with SM predictions, this measurement is \emph{de facto} in tension with the measurements of the other collaborations.

From our results we conclude that the non-local FFs are not the source of the tension between SM predictions and data:
floating these FFs is insufficient to bring the three processes in agreement with the SM. We also find that the local
FFs are driving the uncertainties. For the process $B_s\to \phi\mu^+\mu^-$ in particular, the tension with the
SM increases substantially when we use light-meson LCSR results~\cite{Bharucha:2015bzk} instead of the the
$B$-LCSR results~\cite{Gubernari:2020eft} for the local FFs; see the discussion in \Sec{sec:confrontation:SM}.

\section{Conclusions and Outlook}
\label{sec:conclusions}
\setcounter{equation}{0}

We have carried out the first global analysis of exclusive $b\to s\mu^+\mu^-$ data
within the framework of parametrizable non-local form factors (FFs)~\cite{Bobeth:2017vxj,Gubernari:2020eft}.
Our results include correlated theory predictions for the local (\Tabs{tab:LFFposteriors:BToK}{tab:LFFposteriors:BsToPhi})
and non-local ( \Tabs{tab:NLFFposteriors:BToK}{tab:NLFFposteriors:BsToPhi}) FFs as well
as SM predictions of the branching ratios (\Tab{tab:BR_predictions}) and the angular observables (\Tab{tab:AO_predictions})
for the three processes $B\to K\mu^+\mu^-$, $B\to K^*\mu^+\mu^-$, and $B_s\to \phi\mu^+\mu^-$.
Numerical results for these quantities are attached to this preprint as ancillary files.
All of our numerical results have been obtained using the \EOS software in version 1.0.3,
which has been modified for this purpose. Through \EOS, we make both our analytical
and numerical results available to the community at large, for immediate use or for comparison with
other software.\\

We compare our SM predictions to those obtained within the QCD factorization framework, which is ubiquitously used in
the literature. We find overall good agreement between them, albeit with substantially larger uncertainties
within our approach.
We confront our SM predictions with the large body of data available from various experiments.
For $B\to K\mu^+\mu^-$, we find substantial tensions between the SM predictions and the available
experimental data. For $B\to K^*\mu^+\mu^-$ and $B_s\to\phi\mu^+\mu^-$, the tension is less pronounced,
dominantly due to the substantial uncertainties of their respective local FFs.
\\

Our results further include simple, process-specific model-independent BSM analyses of three available data sets.
Our analyses are in mutual agreement with a BSM-induced shift to the $C_9$ and $C_{10}$ Wilson coefficients
in the $sb\mu\mu$ sector of the Weak Effective Theory:
\begin{equation*}
    (\Re{C_9}^\mathrm{BSM}, \Re{C_{10}}^\mathrm{BSM}) \simeq (-1.0, +0.4)\,.
\end{equation*}
We have not yet achieved a \emph{simultaneous} global analysis of the data, which poses serious
computational obstacles. Chief among them ranks the fact that sampling from a posterior with 
$\sim 140$ parameters requires magnitudes more computation power and time than the analyses presented here.

Our approach shows that a parametrization of the non-local
FFs is viable large-scale statistical analyses of exclusive $b\to s\ell^+\ell^-$ processes,
although nontrivial and computationally expensive.
This holds for producing SM predictions and (simple) BSM analyses of the available data, which
we illustrate at the hand of $b\to s\mu^+\mu^-$.
Our results clearly illustrate that systematic uncertainties due to the non-local FFs have now been
transformed into parametric uncertainties. Our results also show that the onus has been shifted
toward the local FFs, due to their dual role: they enter the observables directly at
timelike $q^2$, and they produce the presently largest uncertainty in the prediction of the
non-local FFs at spacelike $q^2$.
\\

Our work is only the first step toward controlling the hadronic uncertainties due to non-local FFs
in a statistical analyses of the available data. It should and will be followed up in future analyses,
e.g., by
\begin{itemize}
    \item performing a \emph{simultaneous} analysis of the three decay modes;
    \item using LFU-probes such as $R_K$, $R_{K^*}$ and $R_\phi$ in the BSM analyses;
    \item using $\Lambda_b\to\Lambda\mu^+\mu^-$ data to obtain complementary constraints on the BSM Wilson coefficients.
\end{itemize}

We wish to end on the message that our analysis is systematically improvable. Reducing the uncertainties
of the local FFs will reduce the overall uncertainties of the theory predictions both
within and beyond the SM more than linearly. We look forward to impending updated and new lattice QCD analyses
of the local $B\to K^{(*)}$ and $B_s\to \phi$ FFs.

\section*{Acknowledgments}

We a very grateful to Wolfgang Altmannshofer, Andrzej Buras, Ulrik Egede, Joaquim Matias, and Peter Stangl for helpful discussions and insightful comments on the manuscript.
\\

The work of NG is supported by the Deutsche Forschungsgemeinschaft (DFG, German Research Foundation)
under grant 396021762 -- TRR 257.
The work of MR and DvD is supported by the DFG within the Emmy Noether Programme under grant DY-130/1-1
and the Sino-German Collaborative Research Center TRR110 ``Symmetries and the Emergence of Structure in QCD''
(DFG Project-ID 196253076, NSFC Grant No. 12070131001, TRR 110).
JV acknowledges funding from the Spanish MINECO through the ``Ram\'on y Cajal'' program RYC-2017-21870, the “Unit of Excellence Mar\'ia de Maeztu 2020-2023” award to the Institute of Cosmos Sciences (CEX2019-000918-M) and from the grants PID2019-105614GB-C21 and 2017-SGR-929.
NG, MR, and DvD were supported for parts of this work by the Munich Institute for Astro- and Particle Physics (MIAPP),
which is funded by the DFG under Germany’s Excellence Strategy -- EXC-2094 -- 390783311.

\newpage


\appendix
\addtocontents{toc}{\protect\setcounter{tocdepth}{1}}

\renewcommand{\theequation}{\thesection.\arabic{equation}}

\allowdisplaybreaks


\section{Definitions of Local and Non-Local Form Factors}
\label{app:FFsdef}
\setcounter{equation}{0}

Form factors (FFs) are scalar-valued functions, which parametrize hadronic matrix elements
of exclusive semileptonic decay processes.
In this appendix, we collect our definitions of local and non-local FFs that are relevant
to $B\to K^{(*)}\ell^+\ell^-$ and $B_s\to \phi \ell^+\ell^-$ decays. 
Our definitions coincide with the ones in \Reff{Gubernari:2020eft}. 
Throughout this work, the FFs are understood to be functions of the squared dimuon mass
$m_{\mu^+\mu^-}^2 \equiv q^2$, i.e., $\FM{(T),\lambda}\equiv \FM{(T),\lambda}(q^2)$ and $\HM{\lambda}\equiv \HM{\lambda}(q^2)$.
Even though we focus on the $B\to K$ and $B_{(s)}\to \{K^*,\phi\}$ transitions,
the same definitions can be used for any $B_{(s)} \to P$(seudoscalar) and $B_{(s)} \to V$(ector) transitions, respectively.
Hence, in the following we do not specify the exclusive final states but only their spin and parity.

\subsection{$\boldsymbol{B\to P}$ Transitions}

We define the $B\to P$ FFs as follows:
\begin{align}
    \label{eq:def:hel-ff-v-BK}
    \FP{\mu}(k, q)
        & \equiv 
    \bra{P(k)} \bar{s}\gamma_\mu P_L\,b \ket{\bar{B}(q+k)} = \frac{1}{2}\left[ \SP{\mu}{0} \, \FP{0} + \SP{\mu}{t} \, \FP{t} \right] \,,
    \\
    \label{eq:def:hel-ff-t-BK}
    \FP{T,\mu}(k, q)
        & \equiv 
    \bra{P(k)} \bar{s}\sigma_{\mu\nu} q^\nu P_R\,b\ket{\bar{B}(q+k)}
        = \frac{i}{2}\, M_B \,  \SP{\mu}{0} \, \FP{T,0}
        \,,\\
    \HP{\mu}(k, q)
        & \equiv i\,\int d^4x\, e^{iq\cdot x}
        \nonumber\\*
        & \times
                 \bra{P(k)} \T\left\{ j^\text{em}_\mu(x), 
                 \left(
                    C_1 \cO_1^c +
                    C_2 \cO_2^c
                    + 
                    \sum_{i=3}^{6} C_i \cO_i 
                    +
                    C_8 \cO_8
                    \right)\!\!(0)
                 \right\} \ket{\bar{B}(q+k)} 
        \hspace{1cm}\nonumber\\*
    \label{eq:def:nonlocBK}
        & = M_B^2\, \SP{\mu}{0}\, \HP{0} \,,
\end{align}
where $j^\text{em}_\mu= \sum_p Q_p \bar p \gamma_\mu p$.
In these definitions we have used the two independent $B \to P$ Lorentz structures $\SP{\mu}{\lambda}$:
\begin{align}
\label{eq:def:SP}
    \SP{\mu}{0}(k, q) & \equiv 2 k_\mu - \frac{2 (q\cdot k)}{q^2} q_\mu \,, &
    \SP{\mu}{t}(k, q) & \equiv \frac{M_B^2 - M_P^2}{q^2} q_\mu\,,
\end{align}
where $\lambda = 0,t$ denotes either longitudinal or
timelike polarization of the underlying current.

Our FF basis is related to traditional basis of local FFs (defined in, e.g., Refs.~\cite{Gubernari:2018wyi,Khodjamirian:2006st})
and the non-local FF \HP{ } of~\Reff{Khodjamirian:2010vf} by
\begin{align}
    \label{eq:rel:BK-ff-to-us}
    \FP{0}        & = f_+^{B\to P},
    \\
    \FP{t}        & = f_0^{B\to P}, 
    \\
    \FP{T,0}      & = \frac{q^2}{M_B(M_B+M_K)} f_T^{B\to P}, 
    \\
    \HP{0} &= - Q_c \frac{q^2}{2 M_B^2} \HP{ }
    \,.
\end{align}
The following identity holds at maximum recoil:
\begin{align}
    f_+^{B\to P}(q^2=0) = f_0^{B\to P}(q^2=0)
    \,.
    \label{eq:f+0f00}
\end{align}

\subsection{$\boldsymbol{B\to V}$ Transitions}

We define the $B\to V$ FFs as follows:
\begin{align}
    \label{eq:deflocBtoV}
    \FV{\mu}(k, q)
    & \equiv 
     \bra{V(k, \eta)} \bar{s}\gamma_\mu P_L \,b \ket{\bar{B}(q+k)}
     \hspace{300cm}
    \nonumber\\
        & =\frac{1}{2}\, \eta^{*\alpha} \left[ \SP{\alpha\mu}{\perp} \FV{\perp}
        -  \SP{\alpha\mu}{\para} \FV{\para} - \SP{\alpha\mu}{0} \FV{0} - \SP{\alpha\mu}{t} \FV{t}\right]
    \,,\\
    \FV{T,\mu}(k, q)
    & \equiv
     \bra{V(k, \eta)} \bar{s}\sigma_{\mu\nu} q^\nu P_R\, b\ket{\bar{B}(q+k)}
   \nonumber \\ 
    &= \frac{i}{2}\, M_B\, \eta^{*\alpha}\, \left[\SP{\alpha\mu}{\perp} \FV{T,\perp}
    -\SP{\alpha\mu}{\para} \FV{T,\para} - \SP{\alpha\mu}{0} \FV{T,0}\right]\,,\\
    \HV{\mu}(k, q)
        & \equiv i\,\int d^4x\, e^{iq\cdot x}
        \nonumber\\*
        & \times
                 \bra{V(k, \eta)} \T\left\{ j^\text{em}_\mu(x), \left(
                    C_1 \cO_1^c +
                    C_2 \cO_2^c
                    + 
                    \sum_{i=3}^{6} C_i \cO_i 
                    +
                    C_8 \cO_8
                    \right)\!\!(0)
                 \right\} \ket{\bar{B}(q+k)}
        \nonumber\\*
        & = M_B^2\, \eta^{*\alpha}\,\left[
        \SP{\alpha\mu}{\perp} \HV{\perp} - \SP{\alpha\mu}{\para} \HV{\para} - 
        \SP{\alpha\mu}{0} \HV{0}\right]\,.
    \label{eq:defnonlocBtoV}
\end{align}
In the above, $\eta$ is the polarization vector of the vector meson.
In these definitions we have used the four independent $B \to V$ Lorentz structures $\SV{\alpha\mu}{\lambda}$:
\begin{equation}
\begin{aligned}
    \label{eq:def:S}
    \SV{\alpha\mu}{\perp}(k, q) & = \frac{\sqrt{2}\, M_B}{\sqrt{\lamkin}} \epsilon_{\alpha\mu k q}\,, &
    \SV{\alpha\mu}{\para}(k, q) & = \frac{i\, M_B}{\sqrt{2}}\left[g_{\alpha\mu} - \frac{4(q\cdot k)}{\lamkin} q_\alpha k_\mu
                          + \frac{4 M_V^2}{\lamkin} q_\alpha q_\mu\right]\,, \\
    \SV{\alpha\mu}{t}(k, q)     & = \frac{2i\, M_V}{q^2} q_\alpha q_\mu\,, &
    \SV{\alpha\mu}{0}(k, q)     & = \frac{4i\, M_V M_B^2}{q^2 \lamkin}
                            \left[q^2 q_\alpha k_\mu - (q\cdot k)\,q_\alpha q_\mu\right]\,,
\end{aligned}    
\end{equation}
where $\lambda = \perp,\para,0,t$ denotes the different polarisations of the underlying current, and $\lamkin\equiv\lambda(M_B^2,M_V^2,q^2)$ is the K\"all\'en function.

Our FF basis is related to the traditional basis of local FFs (defined in, e.g., Refs.~\cite{Gubernari:2018wyi,Khodjamirian:2006st}) and the non-local FFs \HV{i} of~\Reff{Khodjamirian:2010vf} by
\begin{align}
    \FV{\perp}        & = \frac{\sqrt{2\,\lamkin}}{M_B (M_B + M_V)} V^{B\to V}\,, 
    \\
    \FV{\para}        & = \frac{\sqrt{2}\,(M_B + M_V)}{M_B} A_1^{B\to V}\,,       
    \\
    \FV{0}            & = \frac{(M_B^2 - M_V^2 - q^2)(M_B + M_V)^2 A_1^{B\to V} - \lamkin A_2^{B\to V}}{2 M_V M_B^2 (M_B + M_V) }\,, 
    \\
    \FV{t}            & = A_0^{B\to V}\,,
    \\
    \FV{T,\perp}  & = \frac{\sqrt{2\, \lamkin}}{M_B^2} T_1^{B\to V}\,, 
    \\
    \FV{T,\para}  & = \frac{\sqrt{2} (M_B^2 - M_V^2)}{M_B^2} T_2^{B\to V}\,, 
    \\
    \FV{T,0}      & = \frac{q^2 (M_B^2 + 3 M_V^2 - q^2)}{2 M_B^3 M_V} T_2^{B\to V}
                                - \frac{q^2\lamkin}{2 M_B^3 M_V (M_B^2 - M_V^2)} T_3^{B\to V}\,,
    \\
    \HV{\perp}  & = Q_c\, \frac{\sqrt{\lamkin}}{\sqrt{2} M_B^3} \, \HV{1}\,, 
    \\
    \HV{\para} & = -\sqrt{2}\,Q_c\,\frac{M_B^2 - M_V^2}{M_B^3} \,  \HV{2}\,, 
    \\
    \HV{0}     & =
    -Q_c\,\frac{q^2}{2 M_B^4 M_V}\, 
                                \left[(M_B^2 + 3 M_V^2 - q^2)\HV{2} - \frac{\lamkin}{M_B^2 - M_V^2} \HV{3} \right]\,.
\end{align}
The following identities hold at maximum recoil:
\begin{align}
    A_0^{B\to V}(q^2=0) & =
    \frac{M_B+M_V}{2M_V} A_1^{B\to V}(q^2=0) 
    -
    \frac{M_B-M_V}{2M_V} A_2^{B\to V}(q^2=0) 
    \,,
    \label{eq:A00A10}
    \\
    T_1^{B\to V}(q^2=0) & = T_2^{B\to V}(q^2=0)
    \,.
    \label{eq:T10T20}
\end{align}

\section{Decay and Transversity Amplitudes}
\label{app:ampl}
\setcounter{equation}{0}

In this appendix we list our definitions of the decay amplitudes and the transversity amplitudes.
We give our formulas explicitly for  $B\to K$ and $B \to K^*$ transitions, but analogous formulas apply to, for example, $B_s \to \phi$.
We use the notation introduced in \App{app:FFsdef}.

\subsection{$\boldsymbol{B\to K\ell^+\ell^-}$}

We decompose the amplitude of $\bar{B}(p)\to \bar{K}(k)\ell^+(q_1) \ell^-(q_2)$ as
\begin{align}
    \A^{K\ell\ell}
        &\equiv 
        \frac{G_F\, \alpha_e\, V_{tb}^{} V_{ts}^*}{\sqrt{2} \pi}
        \bigg\{ (C_9 \,L^\mu_{V} + C_{10} \,L^\mu_{A})\  \FM[B\to K]{\mu} 
        -  \frac{L^\mu_{V}}{q^2} \Big[  2 i m_b C_7\,\FM[B\to K]{T,\mu}  + 16\pi^2 \HM[B\to K]{\mu} \Big]   \bigg\}
        \nonumber\\
        &\equiv 
        \frac{G_F}{\sqrt{2}}  \frac{\alpha_e}{4\pi}\,
            \frac{V_{tb}^{} V_{ts}^*}{\N^{K\ell\ell}}
            \left\lbrace
                  (L_V^\mu - L_A^\mu) \SP{\mu}{0} \A_{0,L}^{K\ell\ell}
                + (L_V^\mu + L_A^\mu) \SP{\mu}{0} \A_{0,R}^{K\ell\ell}
                - L_A^\mu \SP{\mu}{t} \A_t^{K\ell\ell}
            \right\rbrace \,,
    \label{eq:ampBP}
\end{align}
where
\begin{align}
    &
    L_{V(A)}^\mu \equiv \bar u_\ell(q_1) \gamma^\mu(\gamma_5) v_\ell(q_2)
    \,,
    &&
    \N^{K\ell\ell}
    = G_F \alpha_e V_{tb}^{} V_{ts}^* \sqrt{\frac{\beta_\ell \lamkin^{3/2}}{3\cdot 2^{10} \pi^5 M_B^3}}
    \,,
    &&
    \beta_\ell\equiv \sqrt{1 -\frac{4m_\ell^2}{q^2}}
    \,.
    &
\end{align}
The transversity amplitudes in \Eq{eq:ampBP} read
\begin{align}
    \A^{K\ell\ell}_{\lambda,L(R)}
        & = \N^{K\ell\ell} \left\lbrace(C_9 \mp C_{10}) \FP[K]{\lambda}
          + \frac{2 m_b M_B}{q^2} \left[C_7 \FP[K]{T,\lambda} - 16 \pi^2 \frac{M_B}{m_b} \HP[K]{\lambda} \right]\right\rbrace\,,
\end{align}
with $\lambda=0,\,t$ and $\FP[K]{T,t} =\HP[K]{t} =0 $.
Since we restrict ourselves to the case where the leptons in the final state have the same mass, the equality $\A^{K\ell\ell}_{t} = \A^{K\ell\ell}_{t,L} - \A^{K\ell\ell}_{t,R}$ holds.
The definitions above --- neglecting lepton masses --- imply
\begin{equation}
    \frac{d\Gamma(\bar{B}\to \bar{K}\ell^+\ell^-)}{dq^2} 
    = \left| \A^{K\ell\ell}_{0,L} \right|^2
    + \left| \A^{K\ell\ell}_{0,R} \right|^2.
\end{equation}
In our numerical evaluation we account for nonzero lepton masses.

\subsection{$\boldsymbol{B\to K^*\ell^+\ell^-}$ and $\boldsymbol{B_s\to \phi \ell^+\ell^-}$}

We decompose the amplitude of $\bar{B}(p)\to \bar{K}^*(k, \eta)\, \ell^+(q_1) \ell^-(q_2)$ as
\begin{align}
    \A^{K^*\ell\ell}
    &\equiv 
        \frac{G_F\, \alpha_e\, V_{tb}^{} V_{ts}^*}{\sqrt{2} \pi}
        \bigg\{ (C_9 \,L^\mu_{V} + C_{10} \,L^\mu_{A})\  \FM[B\to K^*]{\mu} 
        -  \frac{L^\mu_{V}}{q^2} \Big[  2 i m_b C_7\,\FM[B\to K^*]{T,\mu}  + 16\pi^2 \HM[B\to K^*]{\mu} \Big]   \bigg\}
        \nonumber\\*
        & \equiv 
        \frac{G_F}{\sqrt{2}} \frac{\alpha_e}{4\pi}\,
            \frac{V_{tb}^{} V_{ts}^*}{\N^{K^*\ell\ell}}\,\eta^{*\alpha}
        \left\lbrace
                \left(L_V^\mu - L_A^\mu\right)
                \left[\SV{\alpha\mu}{\perp} \A_{\perp,L}^{K^*\ell\ell}
                    - \SV{\alpha\mu}{\para} \A_{\para,L}^{K^*\ell\ell}
                    - 
                    \SV{\alpha\mu}{0} \A_{0,L}^{K^*\ell\ell}
                \right]
            \right.\nonumber\\*
        & + \left.
                \left(L_V^\mu + L_A^\mu\right)
                \left[\SV{\alpha\mu}{\perp} \A_{\perp,R}^{K^*\ell\ell}
                    - \SV{\alpha\mu}{\para} \A_{\para,R}^{K^*\ell\ell}
                    -  \SV{\alpha\mu}{0} \A_{0,R}^{K^*\ell\ell}
                \right]
                + L_A^\mu \SV{\alpha\mu}{t} \A_t^{K^*\ell\ell}
            \right\rbrace \,.
    \label{eq:ampBV}
\end{align}
where
\begin{align}
    &
    L_{V(A)}^\mu \equiv \bar u_\ell(q_1) \gamma^\mu(\gamma_5) v_\ell(q_2)
    \,,
    &&
    \N^{K^*\ell\ell}
        = G_F \alpha_e V_{tb}^{} V_{ts}^* \sqrt{\frac{q^2 \beta_\ell \sqrt{\lamkin}}{3\cdot 2^{10} \pi^5 M_B}}
    \,.
    &
\end{align}
The transversity amplitudes in \Eq{eq:ampBV} read
\begin{align}
    \A^{K^*\ell\ell}_{\lambda,L(R)}
        & = \N^{K^*\ell\ell} \left\lbrace(C_9 \mp C_{10}) \FV[B\to K^*]{\lambda} + \frac{2 m_b M_B}{q^2} \left[C_7 \FV[B\to K^*]{T,\lambda} - 16 \pi^2 \frac{M_B}{m_b} \HV[B\to K^*]{\lambda} \right]\right\rbrace
        \,,
    \label{eq:tampBKstar}
\end{align}
with $\lambda=\perp,\,\para,\,0,\,t$  and $\FV[B\to K^*]{T,t} =\HV[B\to K^*]{t} =0 $.
Since we restrict ourselves to the case where the leptons in the final state have the same mass, the equality $\A^{K^*\ell\ell}_{t} = \A^{K^*\ell\ell}_{t,L} - \A^{K^*\ell\ell}_{t,R}$ holds.
The definitions above --- neglecting lepton masses --- imply
\begin{equation}
    \frac{d\Gamma(\bar{B}\to \bar{K}^*\ell^+\ell^-)}{d q^2} = \sum_{\chi = L,R} \left[ 
    \left| \A^{K^*\ell\ell}_{\lambda,\chi}\right|^2 +
    \left| \A^{K^*\ell\ell}_{\lambda,\chi}\right|^2 + \frac{M_B^2}{q^2}
    \left| \A^{K^*\ell\ell}_{\lambda,\chi}\right|^2
    \right].
\end{equation}
In our numerical evaluation we account for nonzero lepton masses.

The transversity amplitudes $\A^{K^*\ell\ell}_{\lambda,L(R)}$ defined in \Eq{eq:tampBKstar} are related to the transversity amplitudes $A^{K^*\ell\ell}_{\lambda,L(R)}$ defined in, for instance, Refs.~\cite{Altmannshofer:2008dz,Bobeth:2012vn} via
\begin{equation}
\begin{aligned}
    A^{K^*\ell\ell}_{\perp,L(R)} & = \A^{K^*\ell\ell}_{\perp,L(R)} \,,
    &
    A^{K^*\ell\ell}_{\para,L(R)} & = - \A^{K^*\ell\ell}_{\para,L(R)} \,,
    &
    \\
    A^{K^*\ell\ell}_{0,L(R)} & = - \frac{M_B}{\sqrt{q^2}}\A^{K^*\ell\ell}_{0,L(R)} \,,
    &
    A^{K^*\ell\ell}_{t} & = - \frac{1}{M_B}\sqrt{\frac{\lamkin}{q^2}}\A^{K^*\ell\ell}_{t} \,.
    &
\end{aligned}
\end{equation}

\subsection{$\boldsymbol{B\to K\psi}$}

We decompose the $\bar{B}(p)\to \bar{K}(k) \psi(q, \eps)$ amplitude, with $\psi =J/\psi,\psi(2S)$, as
\begin{align}
    \A^{K\psi}
        & \equiv \frac{4 G_F}{\sqrt{2}} V_{tb}^{} V_{ts}^* \bra{\bar{K}(k) \, \psi(q, \eps)} C_1 \cO_1^c + C_2 \cO_2^c 
                    + 
                    \sum_{i=3}^{6} C_i \cO_i+
    C_8 \cO_8 \ket{\bar{B}(p)} + O(V_{ub} V_{us}^*)\nonumber\\
        & \equiv \frac{4 G_F}{\sqrt{2}} \frac{V_{tb}^{} V_{ts}^*}{\N^{K\psi}}\, \eps^{*\mu} \SP{\mu}{0} \A_0^{K\psi} + O(V_{ub} V_{us}^*)\,.
    \label{eq:ampBPpsi}
\end{align}
where $\eps$ is the polarization vector of the vector charmonium $\psi$, and
\begin{equation}
    \N^{K\psi} \equiv G_F V_{tb}^{} V_{ts}^* \sqrt{\frac{\lamkin^{3/2}}{2\pi M_B^3 M_\psi^2}}
    \,.
\end{equation}
The transversity amplitude in \Eq{eq:ampBPpsi} reads 
\begin{align}
    \A_0^{K\psi} = \N^{K\psi} \frac{M_B^2}{M_\psi f_\psi}\, \underset{q^2\to M_\psi^2}{\text{Res}} \HP[K]{0}(q^2)
    \,,
\end{align}
where $f_\psi$ is the decay constant of the charmonium state considered.
Note that in this case the transversity amplitude only depends on the non-local FF $\HP[K]{0}$ and that
$$
\underset{q^2\to M_{J/\psi}^2}{\text{Res}} \HM{\lambda}(q^2)
=
\underset{q^2\to M_{J/\psi}^2}{\text{Res}} \HM{c,\lambda}(q^2)
\,.
$$
This formula holds for  $B\to K^*\psi$ and $B_s\to \phi\psi$ decays as well.
The definitions above imply
\begin{equation}
    \Gamma(\bar{B}\to \bar{K}\psi) = 
    \left| \A^{K\psi}_0\right|^2.
\end{equation}

\subsection{$\boldsymbol{B\to K^*\psi}$ and $\boldsymbol{B_s\to \phi\psi}$}

We decompose the $\bar{B}(p)\to \bar{K}^*(k, \eta) \psi(q, \eps)$ amplitude, with $\psi =J/\psi,\psi(2S)$, as
\begin{align}
    \A^{K^*\psi}
        & \equiv \frac{4 G_F}{\sqrt{2}} V_{tb}^{} V_{ts}^* \bra{\bar{K}^*(k, \eta) \, \psi(q, \eps)} C_1 \cO_1^c + C_2 \cO_2^c 
                    + 
                    \sum_{i=3}^{6} C_i \cO_i +
    C_8 \cO_8 \ket{\bar{B}(p)} + O(V_{ub} V_{us}^*)\nonumber\\
        & \equiv \frac{4 G_F}{\sqrt{2}} \frac{V_{tb}^{} V_{ts}^*}{\N^{K^*\psi}}\, \eta^{*\alpha} \eps^{*\mu} \left[\SV{\alpha\mu}{\perp} \A_\perp^{K^*\psi}
                - \SV{\alpha\mu}{\para} \A_\para^{K^*\psi} - \SV{\alpha\mu}{0} \A_0^{K^*\psi}\right] + O(V_{ub} V_{us}^*)\,,
        \label{eq:ampBVpsi}
\end{align}
where $\eps$ is the polarization vector of the vector charmonium $\psi$, and
\begin{equation}
    \N^{K^*\psi} \equiv G_F V_{tb}^{} V_{ts}^* \sqrt{\frac{\sqrt{\lamkin}}{2\pi M_B}}
    \,.
\end{equation}
The transversity amplitudes in \Eq{eq:ampBVpsi} read 
\begin{align}
    \A_\lambda^{K^*\psi} = \N^{K^*\psi} \frac{M_B^2}{M_\psi f_\psi}\, \underset{q^2\to M_\psi^2}{\text{Res}} \HV[B\to K^*]{\lambda}(q^2)
    \,.
\end{align}
The definitions above imply
\begin{equation}
    \Gamma(\bar{B}\to \bar{K}^*\psi) =
    \left|\A^{K^*\psi}_\perp\right|^2
    +
    \left|\A^{K^*\psi}_\para\right|^2
    +
    \frac{M_B^2}{M_\psi^2}
    \left|\A^{K^*\psi}_0\right|^2
    .
\end{equation}
Our notation for transversity amplitudes $\A^{K^*\psi}_\lambda$ defined in \Eq{eq:ampBVpsi} is related to notation for the transversity amplitudes $A^{K^*\psi}_\lambda$ adopted by the LHCb collaboration (see, e.g., Ref.~\cite{LHCb:2013vga}) through the equalities
\begin{equation}
\begin{aligned}
    &
    \left|A^{K^*\psi}_\perp\right| = 
    \frac{\left|\A^{K^*\psi}_\perp\right|}{\sqrt{\Gamma(\bar{B}\to \bar{K}^*\psi)}}\,,
    &&
    \left|A^{K^*\psi}_\para\right| = 
    \frac{\left|\A^{K^*\psi}_\para\right|}{\sqrt{\Gamma(\bar{B}\to \bar{K}^*\psi)}}\,,
    &&
    \left|A^{K^*\psi}_0\right| = \frac{M_B}{M_\psi}
    \frac{\left|\A^{K^*\psi}_0\right|}{\sqrt{\Gamma(\bar{B}\to \bar{K}^*\psi)}}\,.
    &
\end{aligned}
\end{equation}

\subsection{$\boldsymbol{B\to K^*\gamma}$ and $\boldsymbol{B_s\to \phi\gamma}$}

We decompose the amplitude of $\bar{B}(p)\to \bar{K}^*(k, \eta)\,\gamma(q, \eps)$ as
\begin{align}
    \A_{K^*\gamma}
        & \equiv -i\frac{4 G_F}{\sqrt{2}} \frac{e}{16\pi^2} V_{tb}^{} V_{ts}^*\, \eps^{*\mu}\left\lbrace
            2 m_b C_7 
            \FV[B\to K^*]{T,\mu}
            -i 16 \pi^2 
            \HV[B\to K^*]{\mu}
        \right\rbrace\nonumber\\
        & \equiv \frac{4 G_F}{\sqrt{2}} \frac{e}{16\pi^2} \frac{V_{tb}^{} V_{ts}^*}{\N^{K^*\gamma}} \,\eta^{*\alpha} \eps^{*\mu}\,
            \left[\SV{\alpha\mu}{\perp} \A_\perp^{K^*\gamma} - \SV{\alpha\mu}{\para} \A_\para^{K^*\gamma} \right]\,,
    \label{eq:ampBVgamma}
\end{align}
where $\eps$ is the polarization vector of the photon, and
\begin{equation}
    \N^{K^*\gamma} = G_F V_{tb}^{} V_{ts}^* \sqrt{\frac{\alpha_e M_B (M_B^2 - M_{K^*}^2)}{2^7 \pi^4}}
    \,.
\end{equation}
The transversity amplitudes in \Eq{eq:ampBVgamma} read 
\begin{align}
    \A^{K^*\gamma}_\lambda
        & \equiv \N^{K^*\gamma} \left(m_b C_7 \FV[B\to K^*]{T,\lambda} - 16 \pi^2 M_B \HV[B\to K^*]{\lambda}\right)\,,
\end{align}
with $\lambda=\perp,\,\para$.
The definitions above imply
\begin{equation}
    \Gamma(\bar{B} \to \bar{K}^* \gamma) 
    = \left| \A_\perp^{K^*\gamma} \right|^2 
    + \left| \A_\para^{K^*\gamma} \right|^2.
\end{equation}

\section{Contribution of $\boldsymbol{\HM{sb,\mu}}$ in QCD Factorization}
\label{app:Hsb}
\setcounter{equation}{0}

Our approach allows to parametrize exclusively the non-local matrix element $\HM{c,\mu}$ defined in \Eq{eq:Hpmu}.
The matrix element $\HM{sb,\mu} \equiv \HM{s,\mu}+\HM{b,\mu}$ is added to our results for $\HM{c,\mu}$.
The latter is computed using formulas given in Refs.~\cite{Beneke:2001at,Beneke:2004dp}, which are obtained in the framework of QCD factorization. 
In this appendix, we give explicitly the formulas that we use to evaluate  $\HM{sb,\mu}$.
We neglect $\HM{u,\mu}$ and $\HM{d,\mu}$, because their contributions are suppressed by small Wilson coefficients.

Starting from Eq.~(15) of Ref.~\cite{Beneke:2001at}, we compute $\HM{sb,\mu}$ with the following adaptations:
\begin{itemize}
    \item We remove all contributions proportional to $Q_c$, since they are already accounted for by $\HM{c,\mu}$.
    \item We remove all factorizable contributions, since they are already accounted for by the local FFs $\FM{\mu}$ and $\FM{T,\mu}$.
\end{itemize}

\subsection{$\boldsymbol{B\to K\ell^+\ell^-}$}

Using the results of Refs.~\cite{Beneke:2001at,Bobeth:2007dw}, one obtains
\begin{equation}
    \HP[K]{sb,0} =
        - \frac{1}{16 \pi^2} t\,\FP[K]{0} 
        - \frac{1}{16 \pi^2} t_T\,\FP[K]{0,T} 
        - t_{wa}
        \,,
\end{equation}
where
\begin{align}
    t ~=~ & \frac{s}{2 M_B^2} \left(
    Y_{sb}^{(t)}
         - \frac{\alpha_s}{4 \pi} \left(
            C_2 F_{2,sb}^{(9)} + C_1 F_{1,sb}^{(9)}
            + C_8^\textrm{eff} F_8^{(9)}
        \right) \right)
        \,,\\
    t_T ~=~ & - \frac{\alpha_s}{4 \pi} \frac{m_b}{M_B} \left(
            \left(C_2 - \frac{C_1}{6}\right) F_{2,sb}^{(7)}
            + C_8^\textrm{eff} F_8^{(7)}
        \right)
    \,,\\
    t_{wa} ~=~ & \frac{m_b\,s\,f_B f_P}{16 M_B^4 N_c}
            \int \frac{d\omega}{\omega} \Phi_{B,-}(\omega)
            \int_0^1 du \, \Phi_K(u) T_{0,-}(u, \omega)
    \,.
\end{align}
The $F_{i,sb}^{(j)}$ functions are given in Ref.~\cite{Asatrian:2019kbk}, while the functions $F_8^{(9)}$ and $T_{0,-}(u, \omega) = -T_{\parallel,-}(u, \omega)$ are given
in Ref.~\cite{Beneke:2001at}.
We have also introduce the notation
\begin{align}
    Y_{sb}^{(t)} =~ & -\frac{1}{2} \left(7\,C_3 + \frac{4}{3}\,C_4 + 76\,C_5 + \frac{64}{3}\,C_6\right) h(q^2, m_b) \nonumber\\
    \label{eq:quarkloop}
    & -\frac{1}{2} \left(C_3 + \frac{4}{3}\,C_4 + 16\,C_5 + \frac{64}{3}\,C_6\right) h(q^2, 0) \\
    & +\frac{2}{9} \left(6\,C_3 + 32\,C_5 + \frac{32}{3}\,C_6 \right) \,, \nonumber
\end{align}
where $h(q^2, m)$ is defined in Eq.~(11) of Ref.~\cite{Beneke:2001at}.

\subsection{$\boldsymbol{B\to K^*\ell^+\ell^-}$ and $\boldsymbol{B_s\to\phi\ell^+\ell^-}$}

Using the results of Ref.~\cite{Beneke:2001at}, one obtains
\begin{equation}
    \HM[B\to K^*]{sb,\lambda} = 
        - \frac{1}{16 \pi^2} t \, \FM[B\to K^*]{\lambda} 
        - \frac{1}{16 \pi^2} t_T \,\FM[B\to K^*]{T,\lambda} 
        - \delta_{\lambda0} t_{wa}
        \,,
    \label{eq:BFSH}
\end{equation}
where
\begin{align}
    t ~=~ & \frac{s}{2 M_B^2} 
    \left(Y_{sb}^{(t)}
         - \frac{\alpha_s}{4 \pi} \left(
            C_2 F_{2,sb}^{(9)} + C_1 F_{1,sb}^{(9)}
            + C_8^\textrm{eff} F_8^{(9)}
        \right) \right)
    \,,\\
    t_T ~=~ & - \frac{\alpha_s}{4 \pi} \frac{m_b}{M_B} \left(
            \left(C_2 - \frac{C_1}{6}\right) F_{2,sb}^{(7)}
            + C_8^\textrm{eff} F_8^{(7)}
        \right)
    \,,\\
    t_{wa} ~=~ & -\frac{m_b s \lamkin}{32 M_B^5 N_c (M_B^2 - m_V^2)} \frac{f_B f_V}{E}
            \int \frac{d\omega}{\omega} \Phi_{B,-}(\omega)
            \int_0^1 du\, \Phi_{V,\parallel}(u) T_{\parallel,-}(u, \omega)
    \,.
    \label{eq:Kstwa}
\end{align}
In the equations above, we have used the same notation as in the $B\to K \ell^+\ell^-$ case.

%

For the evaluation the non-local FFs $\HM[B_s\to \phi]{sb,\lambda}$, we use \Eqs{eq:BFSH}{eq:Kstwa} replacing the $B\to K^*$ local FFs with the  $B_s\to \phi$ ones and taking into account that the function $T_{\parallel,-}(u, \omega)$ in the weak annihilation term is slightly different~\cite{Beneke:2004dp}.

\section{Dispersive Bound}
\label{app:bound}
\setcounter{equation}{0}

Dispersive bounds (or unitarity bounds) are constraints on the FFs.
They have been originally applied to kaon decays more than half a century ago~\cite{Meiman:1963,Okubo:1971jf,Okubo:1972ih,Bourrely:1980gp}, and then in the nineties to $B$-meson decays (see, e.g., Refs.~\cite{deRafael:1992tu,deRafael:1993ib,Boyd:1994tt,Boyd:1995cf}).
They have been and still are successfully used to control the truncation error of the various parametrizations needed to extrapolate (or interpolate) the local FFs in the whole physically-allowed region.

In Ref.~\cite{Gubernari:2020eft}, three of us extended this method to the non-local FFs in $B \to M\ell^+\ell^-$ decays for the first time.
The formulation of a dispersive bound for non-local FFs is much more involved than the one for local FFs. 
This is due to the complicated analytic structure of the non-local FFs, where the energy of the first branch point at $4 M_D^2$ is smaller than the threshold $(M_B + M_M)^2$.
Conversely, the lowest-energy branch point of the local FFs coincides with the threshold $(M_B + M_M)^2$.
\\

In this appendix we summarize the derivation of the bound of Ref.~\cite{Gubernari:2020eft}, to which we refer for further details.
We also complement the derivation Ref.~\cite{Gubernari:2020eft} by including the one-particle contribution and the penguin operators $\cO_{3,\dots,6}$; the operator $\cO_8$ does not contribute at the precision we are working at.

To obtain the dispersive bound for non-local FFs $\HM{c,\lambda}$ we define the correlator
\begin{equation}
\begin{aligned}
\label{eq:Pi-general}
    \Pi^{\mu\nu}(q)
    \equiv
    i \int  d^4 x\, e^{i q\cdot x}\bra{0} \T\left\lbrace O^\mu(q;x), O^{\nu,\dagger}(q;0) \right\rbrace  \ket{0} 
    =
    \left(\frac{q^\mu q^\nu}{q^2} - g^{\mu\nu}\right) \Pi(q^2)\,,
\end{aligned}
\end{equation}
where
\begin{equation}
\begin{aligned}
    \label{eq:non-loc-oper}
    O^\mu(q;x)
        & = \left(\frac{-16\pi^2 i}{q^2}\right)\int d^4y\, e^{+i q \cdot y}\ 
            \T\left\{ 
            \bar{c}\gamma^\mu c(x+y), 
            \left(
            C_1 \cO_1^c +
            C_2 \cO_2^c
            + 
            \sum_{i=3}^{6} C_i \cO_i 
            \right)\!\!(x)
            \right\}
            \,,\\
    O^{\nu,\dagger}(q;0)
        & = \left(\frac{+16\pi^2 i}{q^2}\right)\int d^4z\, e^{-i q\cdot z}\ 
            \T\left\{ 
            \bar{c}\gamma^\mu c(z), 
            \left(
            C_1 \cO_1^c +
            C_2 \cO_2^c
            + 
            \sum_{i=3}^{6} C_i \cO_i 
            \right)\!\!(0)
            \right\}\,.
\end{aligned}
\end{equation}
We are only interested in the discontinuity of the correlator $\Pi$ due to on-shell intermediate states
with flavor quantum numbers $B = -S = -1$, which we label $\Disc \Pi$.
We use it to define a new function $\chi$ via a doubly-subtracted dispersion relation: 
\begin{align}
    \label{eq:def-chi}
    \chi(-m_b^2)
        \equiv \frac{1}{2 i\pi} \int\limits_{0}^\infty ds\ \frac{\Disc \Pi(s)}{(s+m_b^2)^3}
    \,,
\end{align}
where we have already chosen $-m_b^2$ as subtraction point.
\\

The function $\chi$ can be calculated using the local OPE proposed in Refs.~\cite{Grinstein:2004vb,Beylich:2011aq}. 
This calculation is described in detail in Ref.~\cite{Gubernari:2020eft}.
The inclusion of the penguin operators changes only the prefactor of the function $f^{(9)}_{\text{LO}}$ defined in Ref.~\cite{Asatrian:2019kbk}:
\begin{align}
    C_F C_1 + C_2 \to C_F C_1 + C_2 + 6 C_3 + 60 C_5
    \,.
\end{align}
The $\order{\alpha_s}$ corrections of the penguin operators to the corresponding matching coefficient have not yet been calculated and hence we neglect them.
The Wilson coefficients $C_3$ and $C_5$ are numerically small, thus the value of $\chi$ found in Ref.~\cite{Gubernari:2020eft} remains unchanged:
\begin{equation}
\begin{aligned}
    \chi^\OPE(-m_b^2) & = (1.81\pm0.02)\cdot 10^{-4}\ \text{GeV}^{-2}\,.
    \label{eq:OPEchi}
\end{aligned}
\end{equation}
\\

The discontinuity $\Disc \Pi$ can be expressed in terms of hadronic quantities using unitarity:
\begin{align}
    \label{eq:ImPi-hadronic0}
    \Disc\Pi^\text{had}(s)
        & = 
            i        
    \left(\frac{q^\mu q^\nu}{s} - g^{\mu\nu}\right)\! 
            \sum \!\!\!\!\!\!\!\!\! \int \limits_{\Gamma } d\rho_\Gamma\, (2\pi)^4 \delta^{(4)}(p_\Gamma - q)
            \braket{0| O^\mu (q; 0)| \Gamma }
            \!
            \braket{\Gamma  | O^{\dagger,\nu}(q; 0) | 0}
            ,
\end{align}
where $s=q\cdot q$ and the sum runs over all on-shell hadronic states $\Gamma$.
The $B_s^*$ meson is the only one-particle state below the threshold $(M_B+M_M)^2$ with a non-vanishing
contribution.\footnote{
    Resonances above the threshold $(M_B+M_M)^2$ are considered features of a multi-body branch cut,
    rather than one-particle contributions.
}
Its contribution to $\Disc\Pi$ reads
\begin{align}
\Disc^{\rm 1-pt} \Pi^\text{had}(s)
        & =  \frac{2}{3}i \pi \left(\frac{q^\mu q^\nu}{s} - g^{\mu\nu}\right) 
        \braket{0| O^\mu| B_s^*}
            \braket{B_s^* | O^{\dagger,\nu} | 0}
        \delta(s-M_{B_s^*}^2)\,.
\end{align}
The matrix element in this equation can be estimated using again the local OPE:
\begin{align}
    \braket{0| O^\mu (q; 0)| B_s^*}
    \simeq
    \frac{1}{2} C_{3,1}(M_{B_s^*}^2)
    \left(g^{\mu\alpha} - \frac{q^\mu q^\alpha}{M_{B_s^*}^2}\right)
    \braket{0| \bar{s} \gamma_\alpha b | B_s^*}
    \,,
\end{align}
where $C_{3,1}$ is defined as in Eq.~(3.6) of Ref.~\cite{Gubernari:2020eft}.
Using the definition of the $B_s^*$ decay constant
\begin{align}
    \braket{0| \bar{s} \gamma_\alpha b | B_s^*} = i \eta_\alpha M_{B_s^*} f_{B_s^*}\,,
\end{align}
where $\eta_\alpha$ is the $B_s^*$ polarization vector, the function $\Disc^{\rm 1-pt} \Pi$ can be written as
\begin{align}
\Disc^{\rm 1-pt} \Pi^\text{had}(s)
        & = 
        i\frac{\pi}{2} \left|C_{3,1}(M_{B_s^*}^2) \right|^2   
        M_{B_s^*}^2 f_{B_s^*}^2
        \delta(s-M_{B_s^*}^2)
        \,.
\end{align}
We can now estimate the numerical contribution of the $B_s^*$ to $\chi^\text{had}$.
We find
\begin{align}
    \chi^\text{1-pt}(-m_b^2) 
    =
    1.53\cdot10^{-6}\GeV^{-2}
    \,.
\end{align}
This contribution is very small compared to $\chi^\OPE$ and hence we conclude that the one-particle contribution has no impact on the dispersive bound. 
A large one-particle contribution --- like in the case of the dispersive bounds for the local $B\to D^*$ FFs --- would have made the dispersive bound for the FFs $\HM{c,\lambda}$ more constraining. 
\\

Following Ref.~\cite{Gubernari:2020eft}, we replace the definitions of the $B\to M$ non-local FFs of \App{app:FFsdef} in \Eq{eq:ImPi-hadronic0}.
Thus, the contribution to $\Disc \Pi$ of the $BK,\,BK^*,$ and $B_s\phi$ states can be written as
\eqa{
&& \hspace{-3mm}
    \frac{3}{32 i\pi^3}\, \Disc \Pi^\text{had}(s)
        = \frac{2 M_B^4\,\lambda^{3/2}(M_B^2, M_K^2, s)}{ s^4} \left|\HP[K]{c,0}(s)\right|^2
            \theta(s - s_+^{B\to K^*})
\nonumber\\
&&        +\ \frac{2 M_B^6\,\sqrt{\lambda(M_{B_{}}^2, M_{K^*}^2, s)}}{ s^3} \left(
            \left|\HV[B\to K^*]{c,\perp}(s)\right|^2
            + \left|\HV[B\to K^*]{c,\para}(s)\right|^2
            + \frac{M_B^2}{s}\left|\HV[B\to K^*]{c,0}(s)\right|^2
        \right) \theta(s - s_+^{B\to K^*})
\nonumber\\
&&        +\ \frac{M_B^6\,\sqrt{\lambda(M_{B_s}^2, M_{\phi}^2, s)}}{ s^3} \left(
            \left|\HV[B_s\to\phi]{c,\perp}(s)\right|^2
            + \left|\HV[B_s\to\phi]{c,\para}(s)\right|^2
            + \frac{M_{B_s}^2}{s} \left|\HV[B_s\to\phi]{c,0}(s)\right|^2
        \right) \theta(s - s_+^{B_s\to \phi})
\nonumber\\[2mm]
&&        +\dots\,  .
\label{eq:ImPi-hadronic}
}
Here, $s_+^{B\to M}\equiv \hat{s}_+\equiv (M_B + M_M)^2$ and the ellipsis denote the contribution of further states with the right quantum numbers.
\\

We can now insert the OPE result (\ref{eq:OPEchi}) and the hadronic representation of $\Disc \Pi$ of \Eq{eq:ImPi-hadronic} into the dispersion relation \eqref{eq:def-chi}:
\begin{align}
\chi^\OPE(-m_b^2)
&= \frac{32 \pi^2}{3} \int\limits_{s_+^{B\to K}}^{\infty} ds\, 
                \frac{M_B^4\,\lambda^{3/2}(M_{B}^2, M_K^2, s)}{ s^4 (s+m_b^2)^3}\, \left|\HP[K]{c,0}(s)\right|^2
\nonumber\\
&\hspace{-10mm} + \frac{32 \pi^2}{3} \int\limits_{s_+^{B\to K^*}}^{\infty} ds\,
                \frac{M_B^6\,\sqrt{\lambda(M_{B_{}}^2, M_{K^*}^2, s)}}{ s^3(s+m_b^2)^3} 
                \left(\sum_{\lambda=\perp,\para} \left|\HV[B\to K^*]{c,\lambda}(s)\right|^2
                +\frac{M_B^2}{s}\left|\HV[B\to K^*]{c,0}(s)\right|^2
                \right)
\nonumber\\
&\hspace{-10mm}+ \frac{16 \pi^2}{3} \int\limits_{s_+^{B_s\to \phi}}^{\infty} ds\,
        \frac{M_{B_s}^6\,\sqrt{\lambda(M_{B_s}^2, M_{\phi}^2, s)}}{ s^3(s+m_b^2)^3} 
        \left(\sum_{\lambda=\perp,\para} \left|\HV[B_s\to\phi]{c,\lambda}(s)\right|^2
        +\frac{M_{B_s}^2}{s}\left|\HV[B_s\to \phi]{c,0}(s)\right|^2
        \right)
\nonumber\\*[2mm]
&\hspace{-10mm} + \dots\,.
\label{eq:dispbound}
\end{align}
We call this equation the \emph{dispersive bound}.
Using the change of variable (\ref{eq:hzdef}), the whole complex $s$-plane is mapped into the unit disk of the $\hat{z}$-plane.
In particular, the integration between $\hat{s}_+$ and infinity becomes an integration over an arc of the unit circle. 
We then define the functions
\eqa{
\label{eq:HhatBK}
\hHP[P]{c,0}(\hat{z}) & \equiv & \outerF[B\to P]{0}(\hat{z}) \, \P(\hat{z}) \,\HP[P]{c,0}(\hat{z})\,,
\\*
\label{eq:HhatBV}
\hHV[B\to V]{c,\lambda}(\hat{z}) & \equiv & \outerF[B\to V]{\lambda}(\hat{z}) \, \P(\hat{z}) \, \HV[B\to V]{c,\lambda}(\hat{z}) \,,
}
which allow us to write \Eq{eq:dispbound} as (using $\hat{z}=e^{i\alpha}$)
\begin{multline}
\label{eq:dispbound2}
\!\!\!\!
1 > 2 \!\!\!\int\limits_{-\alpha^{B\to K}}^{+\alpha^{B\to K}} \!\!\!\!d\alpha \left|\hHP[K]{c,0}(e^{i\alpha})\right|^2
+
\sum_{\lambda}
\left[
\,2\!\!\!
\int\limits_{-\alpha^{B\to K^*}}^{+\alpha^{B\to K^*}} \!\!\!\!\!d\alpha
 \left|\hHV[B\to K^*]{c,\lambda}(e^{i\alpha}) \right|^2
    +\!\!\! \int\limits_{-\alpha^{B_s\to\phi}}^{+\alpha^{B_s\to\phi}} \!\!\!\!\! d\alpha
    \left|\hHV[B_s\to \phi]{c,\lambda}(z(\alpha)) \right|^2\, 
    \!\right].
\end{multline}
where $\alpha^{B\to M} \equiv \big|\arg \hat{z}(\hat{s}_+)\big|$.
The outer functions $\outerF[B\to M]{\lambda}$ are defined such that they are free of kinematical singularities and the dispersive bound (\ref{eq:dispbound}) can be written in the simpler form of \Eq{eq:dispbound2}.
Their analytical expression can be found in Appendix C of Ref.~\cite{Gubernari:2020eft}.
The Blaschke factor $\P$ removes the dynamical singularities of $\HM{c,\lambda}$, so that the $\hHM{c,\lambda}$ is analytic in the open unit disk.
They are defined as in Ref.~\cite{Gubernari:2020eft}:
\begin{equation}
\P(\hat{z}) \equiv \prod_{\psi=J/\psi,\psi(2S)} \frac{\hat{z} - \hat{z}_\psi}{1 - \hat{z} \, \hat{z}_\psi^*}\,,
\end{equation}
where $\hat{z}_\psi = \hat{z}(s = M_\psi^2)$.

The functions $\hHV{c,\lambda}$ are expanded in a series of orthonormal polynomials in the arc of the unit circle between $-\alpha^{B\to M}$ and $+\alpha^{B\to M}$:
\begin{equation}
    \label{eq:expansionHhat}
    \hHM{\lambda}(\hat{z})
        = \sum_{n=0}^\infty  \beta_{\lambda,n}^{B\to M} p_{n}(\hat{z})\,.
\end{equation}
The polynomials up to order seven are available in the public \EOS code~\cite{EOS:paper}, see also \App{app:NLFFFit}.
Inserting the expansion (\ref{eq:expansionHhat}) into \Eq{eq:dispbound2} the dispersive bound can be finally written as
\begin{equation*}
     \sum_{n=0}^\infty 
    \left\{
        2\Big| \beta_{0,n}^{B\to K} \Big|^2
        +
        \sum_{\lambda=\perp,\para,0}
        \left[
            2\Big| \beta_{\lambda,n}^{B\to K^*} \Big|^2
            +
            \Big| \beta_{\lambda,n}^{B_s\to \phi} \Big|^2
    \right]
    \right\}
     < 1\,.
\end{equation*}
This inequality is of paramount importance for the study of the non-local FFs $\HM{c,\lambda}$.
It implies that the coefficients of the polynomials $p_{n}$ are bounded by model independent constraints.
Clearly, the bound becomes more constraining when considering $B\to K$, $B\to K^*$, and $B_s\to\phi$ non-local FFs simultaneously.
In principle, one could also add additional channels to saturate the bound, such as $\Lambda_b \to \Lambda\ell^+\ell^-$ or any multibody decay mediated by $b\to s \ell^+\ell^-$ transitions.

\section{Details on the Parametrization of Non-Local Form Factors}
\label{app:NLFFFit}

The polynomials $p_{n}$ of the parametrization \eqref{eq:H_expansion} are --- up to a normalization factor --- Szeg\H{o} polynomials~\cite{Simon2004OrthogonalPO}.
As such, they can be obtained via the following recurrence relation:
\begin{equation}
\begin{aligned}
    \Phi_0(\hat{z})
        & = 1\,, &
    \Phi^*_0(\hat{z})
        & = 1\,, \\
    \Phi_n(\hat{z})
        & = \hat{z} \Phi_{n - 1} - \rho_{n - 1} \Phi^*_{n - 1}\,, &
    \Phi^*_n(\hat{z})
        & = \Phi^*_{n - 1} - \rho_{n - 1} \hat{z} \Phi_{n - 1}\,, &
\end{aligned}
\end{equation}
which holds for real $z$. 
The orthonormal polynomials then follow from
\begin{equation}
\begin{aligned}
    p_n(\hat{z}) & = \frac{\Phi_n(\hat{z})}{N_n}\,, &
    N_n    & = \left[2\alpha^{B\to M} \prod_{i=0}^{n-1} \left(1 - \rho_i^2\right)\right]^{1/2}\,,
\end{aligned}
\end{equation}
where $\alpha^{B\to M} \equiv \big|\arg \hat{z}(\hat{s}_+)\big|$ and the Verblunsky coefficients $\rho_i$
are obtained from an orthogonalization procedure. 
For $\hat{s}_0 = 4\GeV^2$ we find
\begin{align}
    2\,\alpha^{B\to M}, & \left\lbrace \rho_0, \dots \rho_5 \right\rbrace = \\
    & \begin{cases}
     2.482, \left\lbrace 0.7623, -0.7982, 0.8072, -0.8101, 0.8114, -0.8121 \right\rbrace \qquad \text{for } B \to K \,, \nonumber  \\
     2.276, \left\lbrace 0.7977, -0.8298, 0.8372, -0.8396, 0.8406, -0.8412 \right\rbrace \qquad \text{for } B \to K^* \,, \nonumber \\
     2.183, \left\lbrace 0.8129, -0.8432, 0.8500, -0.8522, 0.8531, -0.8536 \right\rbrace \qquad \text{for } B_s \to \phi \,. \nonumber
     \end{cases}
\end{align}
\\

The coefficients $\beta_{\lambda,n}^{B \to M}$ of the parametrization \eqref{eq:H_expansion} are constrained by the dispersive bound \Eq{eq:boundcoeff}.
However, the non-linear experimental constraints at the $J/\psi$ pole as well as the parametric correlations due to the presence of the bound imply that the posterior distributions of these coefficients are not Gaussian.
This makes the estimation of parametric uncertainties very challenging.
Hence, we introduce an intermediate step to perform our fits:
the analytic functions $\hHM{c,\lambda}$ defined in \Eqs{eq:HhatBK}{eq:HhatBV} are first fitted using \emph{ad hoc} Lagrange basis polynomials in $\hat{z}$.
Our fifth order Lagrange basis polynomials are defined such that they interpolate $\hHM{c,\lambda}$ at the following points:
\begin{equation}
    \hat{z}_i \in \{
        \hat{z}(-7\GeV^2), \hat{z}(-5\GeV^2), \hat{z}(-3\GeV^2), \hat{z}(-1\GeV^2), \hat{z}(m_{J/\psi}^2), \hat{z}(m_{\psi(2S)}^2)
    \}\,.
\end{equation}
The resulting parametrization reads
\begin{equation}
    \label{eq:LagrangeHhat}
    \left. \hHM{c,\lambda}(\hat z) \right|_{n \leq 5}
        = \sum_{i=0}^5  h_{\lambda,i}^{B\to M} \ell_{i}(\hat z)\,,
\end{equation}
where $\ell_{i}(\hat{z}_j) \equiv\delta_{ij}$ are the Lagrange basis polynomials of degree five,
and $h_{\lambda,i}^{B\to M} \equiv \hHM{c,\lambda}(\hat{z}_i)$.
For convenience, the complex $\hHM{c,\lambda}$ values at negative $q^2$ are described with Cartesian coordinates, while
residues on the charmonium poles are described with polar coordinates.
Since no phenomenological constraint is used at the $\psi(2S)$ pole, the corresponding values are left unconstrained and allow an estimation of the parametric uncertainty.

The use of Lagrange basis polynomials ensures that the posterior distributions of the parameters $h_{\lambda,i}^{B\to M}$ are Gaussian distributed \emph{before} the application of the bound.
This is confirmed empirically by the excellent description of these distributions by multivariate Gaussian distributions. The perplexity of the posterior samples is in excess of $99\%$.
The ancillary files \texttt{BToK-hatH.yaml}, \texttt{BToKstar-hatH.yaml}, \texttt{BsToPhi-hatH.yaml} contain
the multivariate Gaussian distributions that describe the posteriors for the three channels.
The mean values and standard deviations of the parameters are also given in \Tabs{tab:NLFFposteriors:BToK}{tab:NLFFposteriors:BsToPhi}.
Note that we have \emph{not} yet enforced the dispersive bound to the intermediate results presented in either
the ancillary files or in the tables; users are expected to apply the bounds in their analyses.
We apply the bound to the coefficients $\beta_{\lambda,n}^{B \to M}$ of \Eq{eq:H_expansion} through the following procedure:
\begin{enumerate}
    \item We draw samples of the Lagrange polynomials parameters from the distributions described in the ancillary files.
    The parameters that are not fixed in our fit are varied uniformly in the ranges provide in \Tabs{tab:NLFFposteriors:BToK}{tab:NLFFposteriors:BsToPhi}.
    
    \item We perform a change of basis from the Lagrange polynomials to the orthonormal polynomials $p_{n}$.
    The change of basis is achieved by evaluating the r.h.s. of Eqs. (\ref{eq:expansionHhat}) and (\ref{eq:LagrangeHhat})
    at an interpolated point $\hat{z}_i$. It reads
    \begin{equation}
        \beta_{\lambda,n}^{B \to M} = \left( P^{-1} \right)_{ni} h_i^{B \to M} \qquad \mathrm{with} \; P_{in} \equiv p_{n}(\hat{z}_i) \,.
    \end{equation}
    We emphasize that the $\beta_{\lambda,n}^{B \to M}$ samples are \emph{not} Gaussian distributed.

    \item
    We subsequently apply the dispersive bound to the
    $\beta_{\lambda,n}^{B\to M}$ samples using the procedure discussed in \Sec{sec:th-pred:SM}.

\end{enumerate}
Our theory predictions are then obtained using these weighted samples of the $\beta_{\lambda,n}^{B \to M}$ coefficients.

\begin{table}[t!]
    \centering
    \renewcommand{\arraystretch}{0.90}
    \begin{tabular}{ccc}
        \toprule
            $\hat{z}_i$ & Parameter & Posterior\\
        \midrule
            \multirow{2}{*}{$\hat{z}(-7\GeV^2)$}        & $\mathrm{Re}  \, \hat{\mathcal{H}}^{B\to K}_{c,0}$  & $  ( 4.0 \pm 2.1) \times 10^{-5}$    \\
                                                   & $\mathrm{Im}  \, \hat{\mathcal{H}}^{B\to K}_{c,0}$  & $  (-4.2 \pm 1.0) \times 10^{-6}$    \\[3pt]
            \multirow{2}{*}{$\hat{z}(-5\GeV^2)$}        & $\mathrm{Re}  \, \hat{\mathcal{H}}^{B\to K}_{c,0}$  & $  (10.0 \pm 2.1) \times 10^{-5}$    \\
                                                   & $\mathrm{Im}  \, \hat{\mathcal{H}}^{B\to K}_{c,0}$  & $  (-5.2 \pm 1.2) \times 10^{-6}$    \\[3pt]
            \multirow{2}{*}{$\hat{z}(-3\GeV^2)$}        & $\mathrm{Re}  \, \hat{\mathcal{H}}^{B\to K}_{c,0}$  & $  (16.3 \pm 2.1) \times 10^{-5}$    \\
                                                   & $\mathrm{Im}  \, \hat{\mathcal{H}}^{B\to K}_{c,0}$  & $  (-6.6 \pm 1.4) \times 10^{-6}$    \\[3pt]
            \multirow{2}{*}{$\hat{z}(-1\GeV^2)$}        & $\mathrm{Re}  \, \hat{\mathcal{H}}^{B\to K}_{c,0}$  & $  (23.0 \pm 3.0) \times 10^{-5}$    \\
                                                   & $\mathrm{Im}  \, \hat{\mathcal{H}}^{B\to K}_{c,0}$  & $  (-8.4 \pm 1.7) \times 10^{-6}$    \\[3pt]
            \multirow{2}{*}{$\hat{z}(m_{J/\psi}^2)$}    & $\mathrm{Abs} \, \hat{\mathcal{H}}^{B\to K}_{c,0}$  & $  (1.225 \pm 0.013) \times 10^{-3}$ \\
                                                   & $\mathrm{Arg} \, \hat{\mathcal{H}}^{B\to K}_{c,0}$  & $  [0, 2\pi]^{(\star)}$              \\[3pt]
            \multirow{2}{*}{$\hat{z}(m_{\psi(2S)}^2)$}  & $\mathrm{Abs} \, \hat{\mathcal{H}}^{B\to K}_{c,0}$  & $  [0, 0.5]^{(\star)}$               \\
                                                   & $\mathrm{Arg} \, \hat{\mathcal{H}}^{B\to K}_{c,0}$  & $  [0, 2\pi]^{(\star)}$              \\
        \bottomrule
    \end{tabular}
    \caption{
      \label{tab:NLFFposteriors:BToK}
      Posteriors of the $B\to K$ non-local FFs fit.
      The ranges of the parameters that are not constrained in the fit are marked with a star ${}^{(\star)}$.
      These large ranges contain more than 99\% of the posterior probability.
      They are suggested by the authors to reproduce the results of this work, as described in the text.
      The multinormal distribution describing the constrained parameters is given in the ancillary file \texttt{BToK-hatH.yaml}.
      }
\end{table}

\begin{table}[t!]
    \centering
    \renewcommand{\arraystretch}{0.90}
    \resizebox{\textwidth}{!}{
    \begin{tabular}{cccccc}
        \toprule
            \multirow{2}{*}{$\hat{z}_i$} & \multirow{2}{*}{Parameter} & \multicolumn{3}{c}{Posterior}\\
                                       &                            & $\lambda=\perp$ & $\lambda=\para$ & $\lambda=0$\\
        \midrule
            \multirow{2}{*}{$\hat{z}(-7\GeV^2)$}        & $\mathrm{Re}  \, \hat{\mathcal{H}}^{B\to K^*}_{c,\lambda}$ & $(1.781 \pm 0.092) \times 10^{-4}$ & $(1.801 \pm 0.086) \times 10^{-4}$ & $(-0.6 \pm 2.3) \times 10^{-5}$ \\
                                                   & $\mathrm{Im}  \, \hat{\mathcal{H}}^{B\to K^*}_{c,\lambda}$ & $(5.88 \pm 0.60) \times 10^{-6}$ & $(5.87 \pm 0.58) \times 10^{-6}$ & $(-4.9 \pm 1.1) \times 10^{-6}$ \\[3pt]
            \multirow{2}{*}{$\hat{z}(-5\GeV^2)$}        & $\mathrm{Re}  \, \hat{\mathcal{H}}^{B\to K^*}_{c,\lambda}$ & $(1.861 \pm 0.097) \times 10^{-4}$ & $(1.882 \pm 0.088) \times 10^{-4}$ & $(4.4 \pm 2.2) \times 10^{-5}$ \\
                                                   & $\mathrm{Im}  \, \hat{\mathcal{H}}^{B\to K^*}_{c,\lambda}$ & $(6.90 \pm 0.73) \times 10^{-6}$ & $(6.87 \pm 0.70) \times 10^{-6}$ & $(-6.0 \pm 1.2) \times 10^{-6}$ \\[3pt]
            \multirow{2}{*}{$\hat{z}(-3\GeV^2)$}        & $\mathrm{Re}  \, \hat{\mathcal{H}}^{B\to K^*}_{c,\lambda}$ & $(1.87 \pm 0.10) \times 10^{-4}$ & $(1.887 \pm 0.093) \times 10^{-4}$ & $(9.6 \pm 2.1) \times 10^{-5}$ \\
                                                   & $\mathrm{Im}  \, \hat{\mathcal{H}}^{B\to K^*}_{c,\lambda}$ & $(8.18 \pm 0.91) \times 10^{-6}$ & $(8.14 \pm 0.087) \times 10^{-6}$ & $(-7.5 \pm 1.4) \times 10^{-6}$ \\[3pt]
            \multirow{2}{*}{$\hat{z}(-1\GeV^2)$}        & $\mathrm{Re}  \, \hat{\mathcal{H}}^{B\to K^*}_{c,\lambda}$ & $(1.78 \pm 0.13) \times 10^{-4}$ & $(1.78 \pm 0.12) \times 10^{-4}$ & $(1.55 \pm 0.28) \times 10^{-4}$ \\
                                                   & $\mathrm{Im}  \, \hat{\mathcal{H}}^{B\to K^*}_{c,\lambda}$ & $(9.8 \pm 1.2) \times 10^{-6}$ & $(9.7 \pm 1.2) \times 10^{-6}$ & $(-9.5 \pm 1.8) \times 10^{-6}$ \\[3pt]
            \multirow{2}{*}{$\hat{z}(m_{J/\psi}^2)$}    & $\mathrm{Abs} \, \hat{\mathcal{H}}^{B\to K^*}_{c,\lambda}$ & $(3.491 \pm 0.096) \times 10^{-4}$ & $(3.71 \pm 0.11) \times 10^{-4}$ & $(1.104 \pm 0.020) \times 10^{-3}$ \\
                                                   & $\mathrm{Arg} \, \hat{\mathcal{H}}^{B\to K^*}_{c,\lambda}$ & $2.940 \pm 0.028$ & $3.343 \pm 0.036$ & $[0, 2\pi]^{(\star)}$ \\ [3pt]
            \multirow{2}{*}{$\hat{z}(m_{\psi(2S)}^2)$}  & $\mathrm{Abs} \, \hat{\mathcal{H}}^{B\to K^*}_{c,\lambda}$ & $[0, 0.3]^{(\star)}$ & $[0, 0.3]^{(\star)}$ & $[0, 0.3]^{(\star)}$ \\
                                                   & $\mathrm{Arg} \, \hat{\mathcal{H}}^{B\to K^*}_{c,\lambda}$ & $[0, 2\pi]^{(\star)}$ & $[0, 2\pi]^{(\star)}$ & $[0, 2\pi]^{(\star)}$ \\
        \bottomrule
    \end{tabular}
    }
    \captionsetup{width=0.95\textwidth}
    \caption{
      \label{tab:NLFFposteriors:BToKstar}
      Posteriors of the $B\to K^*$ non-local FFs fit.
      The ranges of the parameters that are not constrained in the fit are marked with a star ${}^{(\star)}$.
      These large ranges contain more than 99\% of the posterior probability.
      They are suggested by the authors to reproduce the results of this work, as described in the text.
      The multinormal distribution describing the constrained parameters is given in the ancillary file \texttt{BToKstar-hatH.yaml}.
    }
\end{table}

\begin{table}[t!]
    \centering
    \renewcommand{\arraystretch}{0.90}
    \resizebox{\textwidth}{!}{
    \begin{tabular}{cccccc}
        \toprule
            \multirow{2}{*}{$\hat{z}_i$} & \multirow{2}{*}{Parameter} & \multicolumn{3}{c}{Posterior}\\
                                       &                            & $\lambda=\perp$ & $\lambda=\para$ & $\lambda=0$\\
        \midrule
            \multirow{2}{*}{$\hat{z}(-7\GeV^2)$}        & $\mathrm{Re}  \, \hat{\mathcal{H}}^{B_s\to\phi}_{c,\lambda}$ & $(2.03 \pm 0.18) \times 10^{-4}$ & $(2.02 \pm 0.15) \times 10^{-4}$ & $(-1.8 \pm 4.4) \times 10^{-5}$ \\
                                                   & $\mathrm{Im}  \, \hat{\mathcal{H}}^{B_s\to\phi}_{c,\lambda}$ & $(6.74 \pm 0.88) \times 10^{-6}$ & $(6.58 \pm 0.80) \times 10^{-6}$ & $(-6.1 \pm 1.9) \times 10^{-6}$ \\[3pt]
            \multirow{2}{*}{$\hat{z}(-5\GeV^2)$}        & $\mathrm{Re}  \, \hat{\mathcal{H}}^{B_s\to\phi}_{c,\lambda}$ & $(2.11 \pm 0.19) \times 10^{-4}$ & $(2.11 \pm 0.16) \times 10^{-4}$ & $(4.2 \pm 4.0) \times 10^{-5}$ \\
                                                   & $\mathrm{Im}  \, \hat{\mathcal{H}}^{B_s\to\phi}_{c,\lambda}$ & $(7.9 \pm 1.0) \times 10^{-6}$ & $(7.69 \pm 0.92) \times 10^{-6}$ & $(-7.3 \pm 1.9) \times 10^{-6}$ \\[3pt]
            \multirow{2}{*}{$\hat{z}(-3\GeV^2)$}        & $\mathrm{Re}  \, \hat{\mathcal{H}}^{B_s\to\phi}_{c,\lambda}$ & $(2.1 \pm 0.21) \times 10^{-4}$ & $(2.11 \pm 0.16) \times 10^{-4}$ & $(1.10 \pm 0.38) \times 10^{-4}$ \\
                                                   & $\mathrm{Im}  \, \hat{\mathcal{H}}^{B_s\to\phi}_{c,\lambda}$ & $(9.3 \pm 1.3) \times 10^{-6}$ & $(9.1 \pm 1.1) \times 10^{-6}$ & $(-9.0 \pm 2.2) \times 10^{-6}$ \\[3pt]
            \multirow{2}{*}{$\hat{z}(-1\GeV^2)$}        & $\mathrm{Re}  \, \hat{\mathcal{H}}^{B_s\to\phi}_{c,\lambda}$ & $(1.99 \pm 0.23) \times 10^{-4}$ & $(1.99 \pm 0.18) \times 10^{-4}$ & $(1.84 \pm 0.42) \times 10^{-4}$ \\
                                                   & $\mathrm{Im}  \, \hat{\mathcal{H}}^{B_s\to\phi}_{c,\lambda}$ & $(1.10 \pm 0.17) \times 10^{-5}$ & $(1.10 \pm 0.15) \times 10^{-5}$ & $(-1.14 \pm 0.25) \times 10^{-5}$ \\[3pt]
            \multirow{2}{*}{$\hat{z}(m_{J/\psi}^2)$}    & $\mathrm{Abs} \, \hat{\mathcal{H}}^{B_s\to\phi}_{c,\lambda}$ & $(3.62 \pm 0.14) \times 10^{-4}$ & $(3.56 \pm 0.13) \times 10^{-4}$ & $(9.90 \pm 0.37) \times 10^{-4}$ \\
                                                   & $\mathrm{Arg} \, \hat{\mathcal{H}}^{B_s\to\phi}_{c,\lambda}$ & $2.62 \pm 0.15$ & $3.078 \pm 0.066$ & $[0, 2\pi]^{(\star)}$ \\ [3pt]
            \multirow{2}{*}{$\hat{z}(m_{\psi(2S)}^2)$}  & $\mathrm{Abs} \, \hat{\mathcal{H}}^{B_s\to\phi}_{c,\lambda}$ & $[0, 0.3]^{(\star)}$ & $[0, 0.3]^{(\star)}$ & $[0, 0.3]^{(\star)}$ \\
                                                   & $\mathrm{Arg} \, \hat{\mathcal{H}}^{B_s\to\phi}_{c,\lambda}$ & $[0, 2\pi]^{(\star)}$ & $[0, 2\pi]^{(\star)}$ & $[0, 2\pi]^{(\star)}$ \\
        \bottomrule
    \end{tabular}
    }
    \captionsetup{width=0.95\textwidth}
    \caption{
      \label{tab:NLFFposteriors:BsToPhi}
      Posteriors of the $B_s\to\phi$ non-local FFs fit.
      The ranges of the parameters that are not constrained in the fit are marked with a star ${}^{(\star)}$.
      These large ranges contain more than 99\% of the posterior probability.
      They are suggested by the authors to reproduce the results of this work, as described in the text.
      The multinormal distribution describing the constrained parameters is given in the ancillary file \texttt{BsToPhi-hatH.yaml}.
    }
\end{table}

\section{Our Results in a Nutshell}
\label{app:plots-tables}
\setcounter{equation}{0}

For the reader's convenience, in this appendix we organize our results schematically and refer to the table, figure or ancillary file where they can be found.
We also provide supplemental tables and figures.
Additional material can be obtained upon request.

\subsubsection*{Local form factors}

\Tabs{tab:LFFposteriors:BToK}{tab:LFFposteriors:BsToPhi} contain the central value and the standard deviation of the coefficients of the parametrization~\eqref{eq:zexpOPE}.
The correlations between these coefficients are given in the ancillary files \texttt{BToKstar-local.yaml}, \texttt{BsToPhi-local.yaml} and \texttt{BToK-local.yaml}.
We compare our results with those of Refs.~\cite{Gubernari:2018wyi,Bharucha:2015bzk} for the FFs $f_+^{B\to K}$, $A_1^{B\to K^*}$, and $A_1^{B_s\to\phi}$ in \Fig{fig:local_ffs}.

\begin{table}[t!]
    \centering
    \renewcommand{\arraystretch}{0.90}
    \begin{tabular}{cccccc}
        \toprule
            Coefficient & This work & Ref.~\cite{Gubernari:2018wyi} & Coefficient & This work & Ref.~\cite{Gubernari:2018wyi}\\
        \midrule
            $\alpha^+_0$ & $0.39 \pm 0.02$ & $0.33 \pm 0.02$   &  $\alpha^T_0$ & $0.36 \pm 0.02$ & $0.30 \pm 0.03$ \\
            $\alpha^+_1$ & $-0.56 \pm 0.16$ & $-0.87 \pm 0.14$ &  $\alpha^T_1$ & $-0.71 \pm 0.19$ & $-0.77 \pm 0.15$ \\
            $\alpha^+_2$ & $0.36 \pm 0.37$ & $0.01 \pm 0.75$   &  $\alpha^T_2$ & $0.07 \pm 0.42$ & $0.01 \pm 0.87$ \\[3pt]
            $\alpha^0_1$ & $0.72 \pm 0.14$ & $0.20 \pm 0.17$   &&&\\
            $\alpha^0_2$ & $0.66 \pm 0.25$ & $-0.45 \pm 0.41$ &&&\\
        \bottomrule
    \end{tabular}
    \caption{
      \label{tab:LFFposteriors:BToK}
      Posteriors of the $B\to K$ local FFs fit.
      We have used the exact relation \eqref{eq:f+0f00} to reduce the number of parameters to 8.
      The posterior distribution is accurately described with a single multivariate Gaussian distribution, whose parameters are given in the ancillary file \texttt{BToK-local.yaml}.
      We compare our results to those of Ref.~\cite{Gubernari:2018wyi}.
      }
\end{table}

\begin{table}[t!]
    \centering
    \renewcommand{\arraystretch}{0.90}
    \begin{tabular}{crrcrr}
        \toprule
            Coefficient & This work & Ref.~\cite{Bharucha:2015bzk}   & Coefficient & This work & Ref.~\cite{Bharucha:2015bzk} \\
        \midrule
            $\alpha^{A_0}_0$ & $0.34 \pm 0.03$ & $0.37 \pm 0.03$    &  $\alpha^{T_1}_0$ & $0.32 \pm 0.02$ & $0.31 \pm 0.03$    \\
            $\alpha^{A_0}_1$ & $-1.12 \pm 0.20$ & $-1.37 \pm 0.26$  &  $\alpha^{T_1}_1$ & $-0.95 \pm 0.14$ & $-1.01 \pm 0.19$  \\
            $\alpha^{A_0}_2$ & $2.18 \pm 1.76$ & $0.13 \pm 1.63$    &  $\alpha^{T_1}_2$ & $2.11 \pm 1.28$ & $1.53 \pm 1.64$    \\[3pt]
            $\alpha^{A_1}_0$ & $0.29 \pm 0.02$ & $0.30 \pm 0.03$    &  $\alpha^{T_{23}}_0$ & $0.62 \pm 0.03$ & $0.67 \pm 0.06$ \\
            $\alpha^{A_1}_1$ & $0.46 \pm 0.13$ & $0.39 \pm 0.19$    &  $\alpha^{T_{23}}_1$ & $0.97 \pm 0.32$ & $1.32 \pm 0.22$ \\
            $\alpha^{A_1}_2$ & $1.22 \pm 0.73$ & $1.19 \pm 1.03$    &  $\alpha^{T_{23}}_2$ & $1.81 \pm 2.45$ & $3.82 \pm 2.20$ \\[3pt]
            $\alpha^{A_{12}}_1$ & $0.55 \pm 0.34$ & $0.53 \pm 0.13$ &  $\alpha^{T_2}_1$ & $0.60 \pm 0.18$ & $0.50 \pm 0.17$    \\
            $\alpha^{A_{12}}_2$ & $0.58 \pm 2.08$ & $0.48 \pm 0.66$ &  $\alpha^{T_2}_2$ & $1.70 \pm 0.99$ & $1.61 \pm 0.80$    \\[3pt]
            $\alpha^V_0$ & $0.36 \pm 0.03$ & $0.38 \pm 0.03$        &&&\\
            $\alpha^V_1$ & $-1.09 \pm 0.17$ & $-1.17 \pm 0.26$      &&&\\
            $\alpha^V_2$ & $2.73 \pm 1.99$ & $2.42 \pm 1.53$        &&&\\
        \bottomrule
    \end{tabular}
    \caption{
      \label{tab:LFFposteriors:BToKstar}
      Posteriors of the $B\to K^*$ local FFs fit.
      We used the exact relations \eqref{eq:A00A10}-\eqref{eq:T10T20} to reduce the number of parameters to 19.
      The posterior distribution is accurately described with a single multivariate Gaussian distribution, whose parameters are given in the ancillary file \texttt{BToKstar-local.yaml}.
      We compare our results to those of Ref.~\cite{Bharucha:2015bzk}.
      }
\end{table}

\begin{table}[t!]
    \centering
    \renewcommand{\arraystretch}{1.1}
    \begin{tabular}{@{}crrcrr@{}}
        \toprule
            Coefficient & This work & Ref.~\cite{Bharucha:2015bzk}   & Coefficient & This work & Ref.~\cite{Bharucha:2015bzk} \\
        \midrule
            $\alpha^{A_0}_0$ & $0.38 \pm 0.04$ & $0.42 \pm 0.02$    &  $\alpha^{T_1}_0$ & $0.34 \pm 0.04$ & $0.30 \pm 0.01$     \\
            $\alpha^{A_0}_1$ & $-1.26 \pm 0.40$ & $-0.98 \pm 0.24$  &  $\alpha^{T_1}_1$ & $-0.77 \pm 0.32$ & $-1.10 \pm 0.08$   \\
            $\alpha^{A_0}_2$ & $2.83 \pm 3.83$ & $3.27 \pm 1.36$    &  $\alpha^{T_1}_2$ & $0.93 \pm 2.76$ & $0.58 \pm 1.00$     \\[3pt]
            $\alpha^{A_1}_0$ & $0.30 \pm 0.03$ & $0.29 \pm 0.01$    &  $\alpha^{T_{23}}_0$ & $0.64 \pm 0.06$ & $0.65 \pm 0.04$  \\
            $\alpha^{A_1}_1$ & $0.48 \pm 0.30$ & $0.35 \pm 0.10$    &  $\alpha^{T_{23}}_1$ & $1.18 \pm 0.69$ & $2.10 \pm 0.33$  \\
            $\alpha^{A_1}_2$ & $1.99 \pm 1.84$ & $1.70 \pm 0.79$    &  $\alpha^{T_{23}}_2$ & $1.78 \pm 5.14$ & $6.74 \pm 1.80$  \\[3pt]
            $\alpha^{A_{12}}_1$ & $0.33 \pm 0.54$ & $0.95 \pm 0.13$ &  $\alpha^{T_2}_1$ & $0.74 \pm 0.43$ & $0.40 \pm 0.08$     \\
            $\alpha^{A_{12}}_2$ & $-0.60 \pm 3.19$ & $2.15 \pm 0.48$ &  $\alpha^{T_2}_2$ & $1.93 \pm 2.59$ & $1.04 \pm 0.61$     \\[3pt]
            $\alpha^V_0$ & $0.38 \pm 0.05$ & $0.36 \pm 0.01$      &&& \\
            $\alpha^V_1$ & $-0.91 \pm 0.39$ & $-1.22 \pm 0.16$    &&& \\
            $\alpha^V_2$ & $3.80 \pm 4.09$ & $3.74 \pm 1.73$      &&& \\
        \bottomrule
    \end{tabular}
    \caption{
      \label{tab:LFFposteriors:BsToPhi}
      Posteriors of the $B_s\to\phi$ local FFs fit.
      We used the exact relations \eqref{eq:A00A10}-\eqref{eq:T10T20} to reduce the number of parameters to 19.
      The posterior distribution is accurately described with a single multivariate Gaussian distribution, whose parameters are given in the ancillary file \texttt{BsToPhi-local.yaml}.
      We compare our results to those of Ref.~\cite{Bharucha:2015bzk}.
      }
\end{table}

\subsubsection*{Non-Local form factors}

\Tabs{tab:NLFFposteriors:BToK}{tab:NLFFposteriors:BsToPhi} contain the central value and the standard deviation of the posteriors of the analytic functions $\hHM{c,\lambda}$ defined in \Eqs{eq:HhatBK}{eq:HhatBV} at $q^2=\{-7\GeV^2,-5\GeV^2, -3\GeV^2,\allowbreak -1\GeV^2,\ m_{J/\psi}^2,m_{\psi(2S)}^2\}$ .
The correlations of each of these functions across the different $q^2$ points are given in the ancillary files \texttt{\justify BToK-hatH.yaml}, \texttt{\justify BToKstar-hatH.yaml}, and \texttt{\justify BsToPhi-hatH.yaml}.
These results are prior to the dispersive bound application.

\subsubsection*{Standard Model Predictions}

\begin{table}[t!]
    \centering
    \renewcommand{\arraystretch}{1.30}
    \resizebox{\textwidth}{!}{
    \begin{tabular}{@{}cc rrrr rrrr@{}}
        \toprule
            Process
                & $q^2$ bin     & $F_L [10^{-1}]$        & $S_3 [10^{-2}]$          & $S_4 [10^{-1}]$         & $S_5 [10^{-1}]$
                & $A_\text{FB} [10^{-1}]$ & $S_7 [10^{-2}]$         & $S_8 [10^{-3}]$         & $S_9 [10^{-4}]$ \\
        \midrule
            \multirow{5}{*}{\rotatebox{90}{$B^0\to K^{*0}\mu^+\mu^-$}}
                & $[0.1, 0.98]$ & $3.00^{+0.45}_{-0.32}$ &  $1.17^{+0.28}_{-0.24}$  &  $0.92^{+0.02}_{-0.04}$ &  $2.39^{+0.08}_{-0.07}$
                & $-0.97^{+0.05}_{-0.05}$ & $-0.99^{+0.09}_{-0.10}$ & $-6.62^{+0.39}_{-0.69}$ & $-3.77^{+1.06}_{-1.78}$ \\[1pt]
                & $[1.1, 2.5]$  & $7.68^{+0.30}_{-0.41}$ &  $0.40^{+0.21}_{-0.16}$  & $-0.09^{+0.12}_{-0.08}$ &  $0.66^{+0.20}_{-0.25}$
                & $-1.53^{+0.21}_{-0.27}$ & $-1.17^{+0.25}_{-0.25}$ & $-6.22^{+1.25}_{-1.70}$ & $-4.04^{+2.55}_{-3.00}$ \\[1pt]
                & $[2.5, 4.0]$  & $8.08^{+0.19}_{-0.28}$ & $-1.02^{+0.43}_{-0.34}$  & $-1.42^{+0.13}_{-0.09}$ & $-1.87^{+0.27}_{-0.35}$
                & $-0.29^{+0.16}_{-0.27}$ & $-0.89^{+0.35}_{-0.52}$ & $-3.80^{+2.71}_{-2.35}$ & $-3.91^{+3.05}_{-3.80}$ \\[1pt]
                & $[4.0, 6.0]$  & $7.18^{+0.27}_{-0.34}$ & $-2.48^{+0.87}_{-0.56}$  & $-2.21^{+0.12}_{-0.10}$ & $-3.37^{+0.22}_{-0.35}$
                &  $1.19^{+0.25}_{-0.28}$ & $-0.67^{+0.64}_{-0.64}$ & $-2.74^{+4.14}_{-4.72}$ & $-6.12^{+9.00}_{-7.29}$ \\[1pt]
                & $[6.0, 8.0]$  & $6.07^{+0.31}_{-0.49}$ & $-4.06^{+1.44}_{-0.82}$  & -$2.57^{+0.11}_{-0.09}$ & $-4.05^{+0.34}_{-0.30}$
                &  $2.45^{+0.30}_{-0.46}$ & $-0.36^{+1.10}_{-1.58}$ & $-3.99^{+8.91}_{-4.65}$ & $14.2^{+41.1}_{-34.0}$ \\[1pt]
        \midrule
            \multirow{5}{*}{\rotatebox{90}{$\quad B_s\to\phi \mu^+\mu^-$}}
                & $[0.1, 0.98]$ & $3.30^{+0.48}_{-0.49}$ &  $1.26^{+0.55}_{-0.57}$  &  $0.94^{+0.04}_{-0.04}$ &  $2.46^{+0.09}_{-0.11}$
                & $-0.91^{+0.07}_{-0.07}$ & $-1.14^{+0.11}_{-0.12}$ & $-2.31^{+0.45}_{-0.62}$ & $-4.11^{+2.32}_{-2.30}$ \\[1pt]
                & $[1.1, 4.0]$  & $8.05^{+0.31}_{-0.34}$ & $-0.44^{+0.37}_{-0.38}$  & $-0.74^{+0.18}_{-0.17}$ & $-0.38^{+0.45}_{-0.45}$
                & $-0.87^{+0.21}_{-0.26}$ & $-1.38^{+0.47}_{-0.52}$ & $-3.59^{+2.48}_{-2.83}$ & $-4.46^{+3.92}_{-4.90}$ \\[1pt]
                & $[4.0, 6.0]$  & $7.54^{+0.39}_{-0.43}$ & $-3.01^{+1.17}_{-1.29}$  & $-2.18^{+0.20}_{-0.15}$ & $-3.00^{+0.61}_{-0.47}$
                &  $0.91^{+0.44}_{-0.43}$ & $-0.90^{+0.99}_{-0.95}$ & $-2.66^{+5.41}_{-6.55}$ &  $4.60^{+13.3}_{-14.9}$ \\[1pt]
                & $[6.0, 8.0]$  & $6.55^{+0.51}_{-0.56}$ & $-4.92^{+1.79}_{-1.88}$  & $-2.58^{+0.20}_{-0.15}$ & $-3.66^{+0.58}_{-0.49}$
                &  $1.98^{+0.66}_{-0.63}$ & $-0.63^{+1.67}_{-1.77}$ & $-2.89^{+11.8}_{-12.4}$ & $-9.41^{+53.3}_{-57.2}$ \\[1pt]
        \bottomrule
    \end{tabular}
    }
    \caption{
    \label{tab:AO_predictions}
    SM predictions for the angular observables for the $B^0\to K^{*0}\mu^+\mu^-$ and $B_s \to\phi\mu^+\mu^-$ decays.
    Note that we provide the central $68\%$ intervals and that the posterior-predictive
    distributions are not Gaussian.
    }
\end{table}

\begin{table}[t!]
    \centering
    \setlength{\tabcolsep}{10pt}
    \renewcommand{\arraystretch}{1.2}
    \begin{tabular}{@{}lccccc@{}}
        \toprule
            \multirow{2}{*}{Process} & \multirow{2}{*}{$q^2$ bin} & \multicolumn{2}{c}{$\hat{\BR}~[10^{-5}]$} & \multicolumn{2}{c}{\BR~$[10^{-8}]$} \\
            & & SM & LHCb & SM & LHCb \\
        \midrule
            \multirow{8}{*}{$B^+\to K^+\mu^+\mu^-$}
                & $[0.1, 0.98]$ & $4.28^{+0.42}_{-0.43}$ & $2.93 \pm 0.16$ & $4.18^{+0.40}_{-0.38}$ & $2.92 \pm 0.22$ \\[1pt]
                & $[1.1, 2.0]$  & $4.38^{+0.40}_{-0.41}$ & $2.10 \pm 0.14$ & $4.29^{+0.35}_{-0.37}$ & $2.10 \pm 0.17$ \\[1pt]
                & $[2.0, 3.0]$  & $4.78^{+0.39}_{-0.43}$ & $2.83 \pm 0.16$ & $4.69^{+0.36}_{-0.41}$ & $2.82 \pm 0.21$ \\[1pt]
                & $[3.0, 4.0]$  & $4.67^{+0.40}_{-0.39}$ & $2.54 \pm 0.15$ & $4.57^{+0.35}_{-0.36}$ & $2.54 \pm 0.20$ \\[1pt]
                & $[4.0, 5.0]$  & $4.57^{+0.40}_{-0.38}$ & $2.21 \pm 0.14$ & $4.48^{+0.33}_{-0.35}$ & $2.21 \pm 0.18$ \\[1pt]
                & $[5.0, 6.0]$  & $4.49^{+0.36}_{-0.39}$ & $2.31 \pm 0.14$ & $4.39^{+0.34}_{-0.38}$ & $2.31 \pm 0.18$ \\[1pt]
                & $[6.0, 7.0]$  & $4.38^{+0.46}_{-0.43}$ & $2.45 \pm 0.14$ & $4.28^{+0.43}_{-0.39}$ & $2.45 \pm 0.18$ \\[1pt]
                & $[7.0, 8.0]$  & $4.28^{+0.74}_{-0.54}$ & $2.31 \pm 0.14$ & $4.17^{+0.71}_{-0.49}$ & $2.31 \pm 0.18$ \\[1pt]
        \midrule
            \multirow{5}{*}{$B^0\to K^{*0}\mu^+\mu^-$}
                & $[0.1, 0.98]$ & $7.21^{+0.92}_{-0.75}$ & $7.51^{+0.53}_{-0.58}$ & $9.86^{+1.11}_{-1.25}$ & $8.94^{+0.88}_{-0.92}$ \\[1pt]
                & $[1.1, 2.5]$  & $4.55^{+0.74}_{-0.55}$ & $3.84^{+0.40}_{-0.38}$ & $6.24^{+0.98}_{-0.90}$ & $4.56^{+0.56}_{-0.55}$ \\[1pt]
                & $[2.5, 4.0]$  & $4.59^{+0.68}_{-0.62}$ & $4.21^{+0.41}_{-0.44}$ & $6.42^{+0.73}_{-1.00}$ & $5.01^{+0.59}_{-0.62}$ \\[1pt]
                & $[4.0, 6.0]$  & $6.81^{+0.97}_{-0.95}$ & $5.94^{+0.49}_{-0.47}$ & $9.74^{+0.98}_{-1.71}$ & $7.08^{+0.74}_{-0.73}$ \\[1pt]
                & $[6.0, 8.0]$  & $7.88^{+1.23}_{-1.01}$ & $7.22^{+0.47}_{-0.51}$ & $11.2^{+1.9 }_{-2.1 }$ & $8.58^{+0.83}_{-0.82}$ \\[1pt]
        \midrule
            \multirow{5}{*}{$B_s\to\phi \mu^+\mu^-$}
                & $[0.1, 0.98]$ & $9.89^{+1.91}_{-1.69}$ & $6.70 \pm 0.47$ & $11.1^{+2.1 }_{-1.7 }$ & $6.81 \pm 0.58$ \\[1pt]
                & $[1.1, 2.5]$  & $6.47^{+1.32}_{-0.98}$ & $4.33 \pm 0.42$ & $7.23^{+1.24}_{-1.16}$ & $4.41 \pm 0.47$ \\[1pt]
                & $[2.5, 4.0]$  & $6.56^{+1.32}_{-1.13}$ & $3.45 \pm 0.38$ & $7.30^{+1.26}_{-1.23}$ & $3.51 \pm 0.43$ \\[1pt]
                & $[4.0, 6.0]$  & $9.84^{+1.88}_{-1.70}$ & $6.10 \pm 0.49$ & $10.8^{+1.8 }_{-1.7 }$ & $6.22 \pm 0.58$ \\[1pt]
                & $[6.0, 8.0]$  & $11.4^{+2.3 }_{-1.9 }$ & $6.20 \pm 0.48$ & $12.6^{+2.3 }_{-1.9 }$ & $6.30 \pm 0.58$ \\[1pt]
        \bottomrule
    \end{tabular}
    \caption{
    \label{tab:BR_predictions}
    SM predictions and measurements by the LHCb experiment\cite{Aaij:2014pli,LHCb:2016ykl,LHCb:2021zwz} for the integrated branching ratios \BR\ and
    the integrated normalized branching ratios $\hat{\BR} \equiv \BR(B\to M \mu^+\mu^-) / \BR(B\to MJ/\psi)$.
    Our predictions for the integrated \BR~account for correlations between the normalized predictions and the normalization channel.
    Ignoring tiny differences due to the masses, the predictions for the
    modes $B^0\to K^0\mu^+\mu^-$ and $B^+\to K^{*+}\mu^+\mu^-$ can be obtained by rescaling with the
    ratio of $B$ lifetimes. Note that we provide the central $68\%$ intervals and that the posterior-predictive
    distributions are not Gaussian.
    }
\end{table}

\begin{figure}[t!]
    \centering
    \includegraphics[width=.32\textwidth]{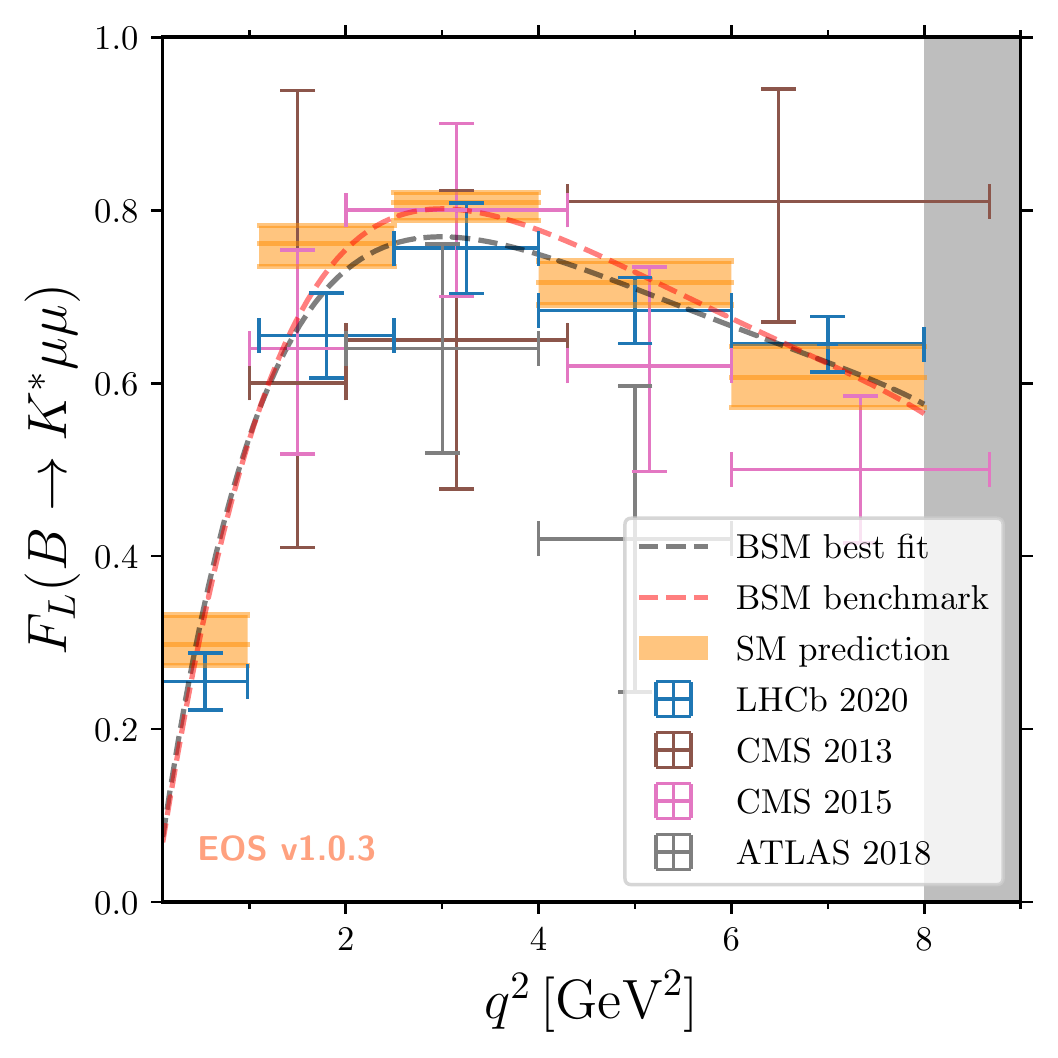}
    \includegraphics[width=.32\textwidth]{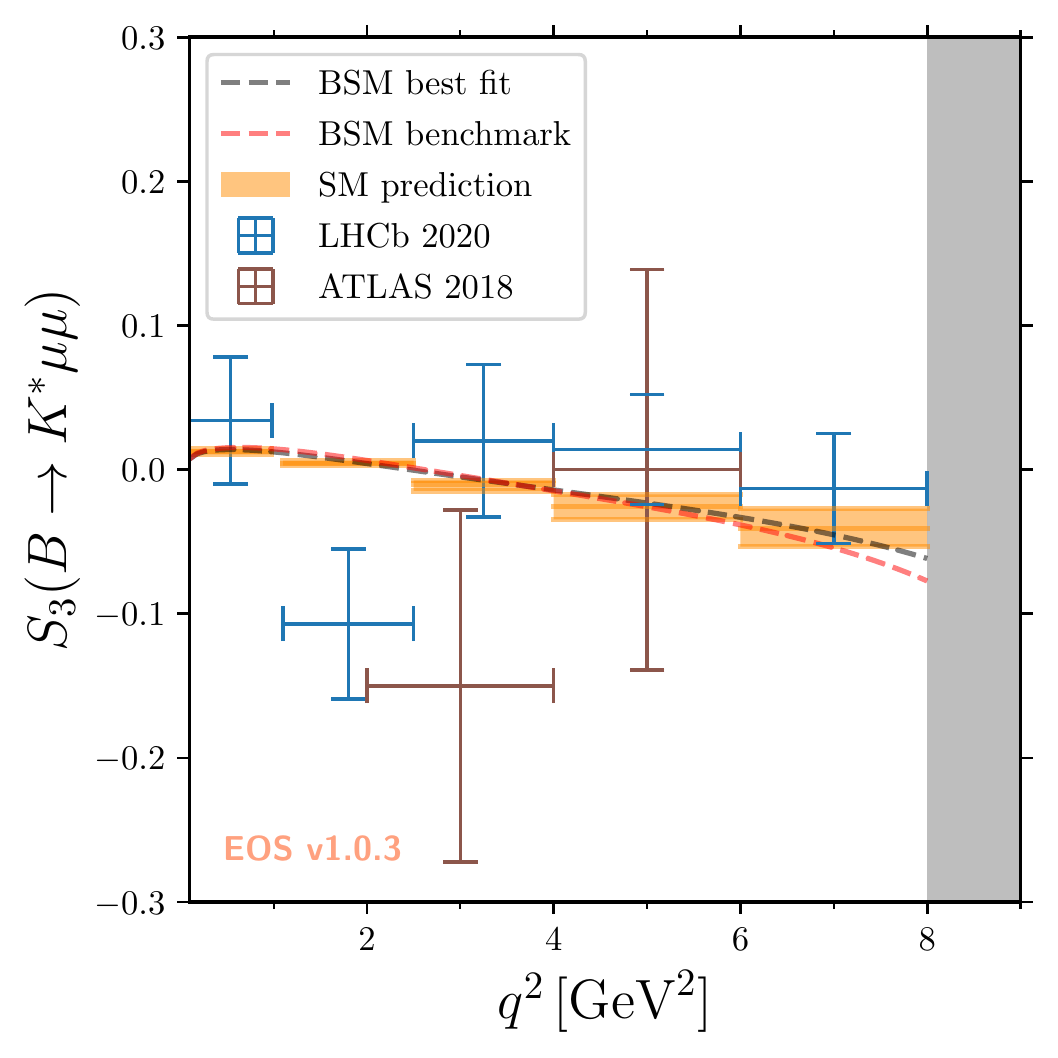}
    \includegraphics[width=.32\textwidth]{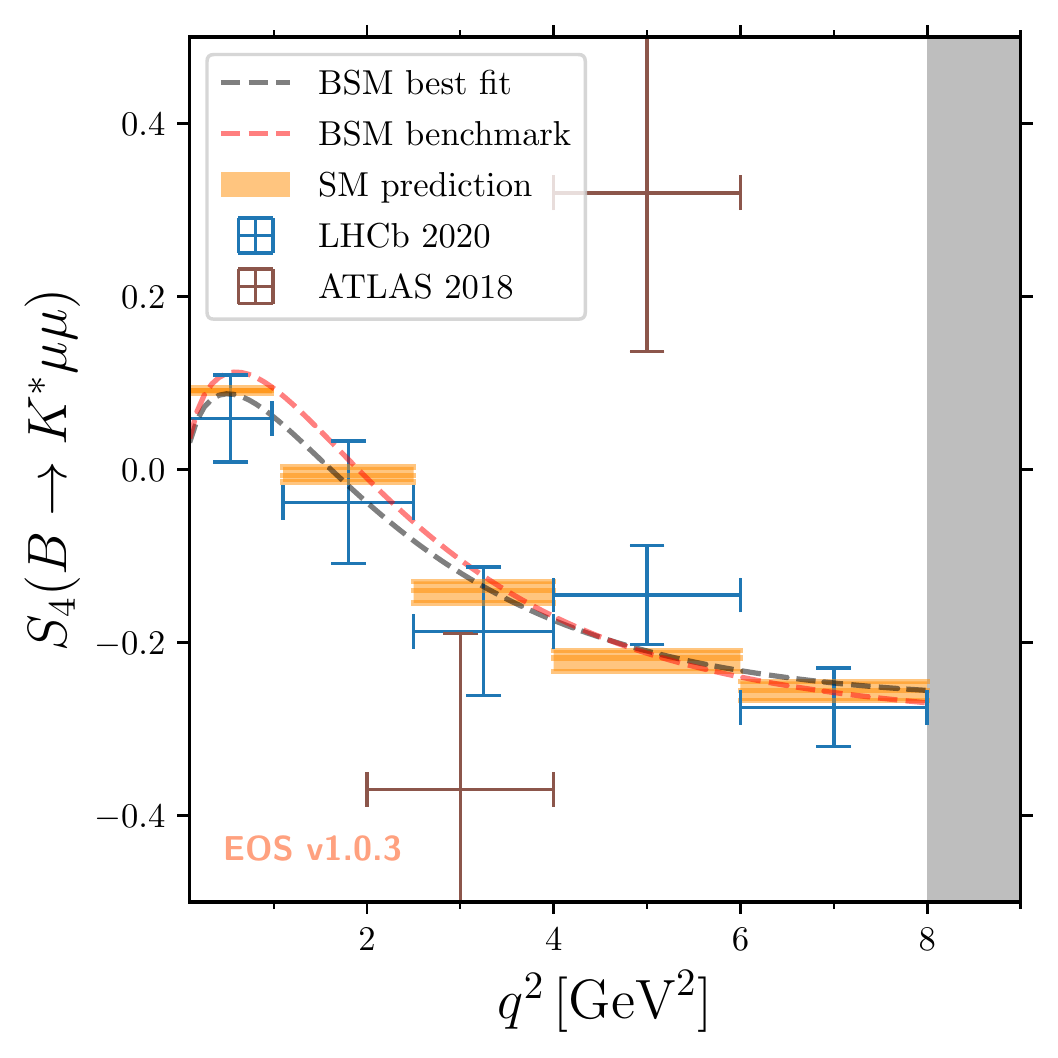} \\
    \includegraphics[width=.32\textwidth]{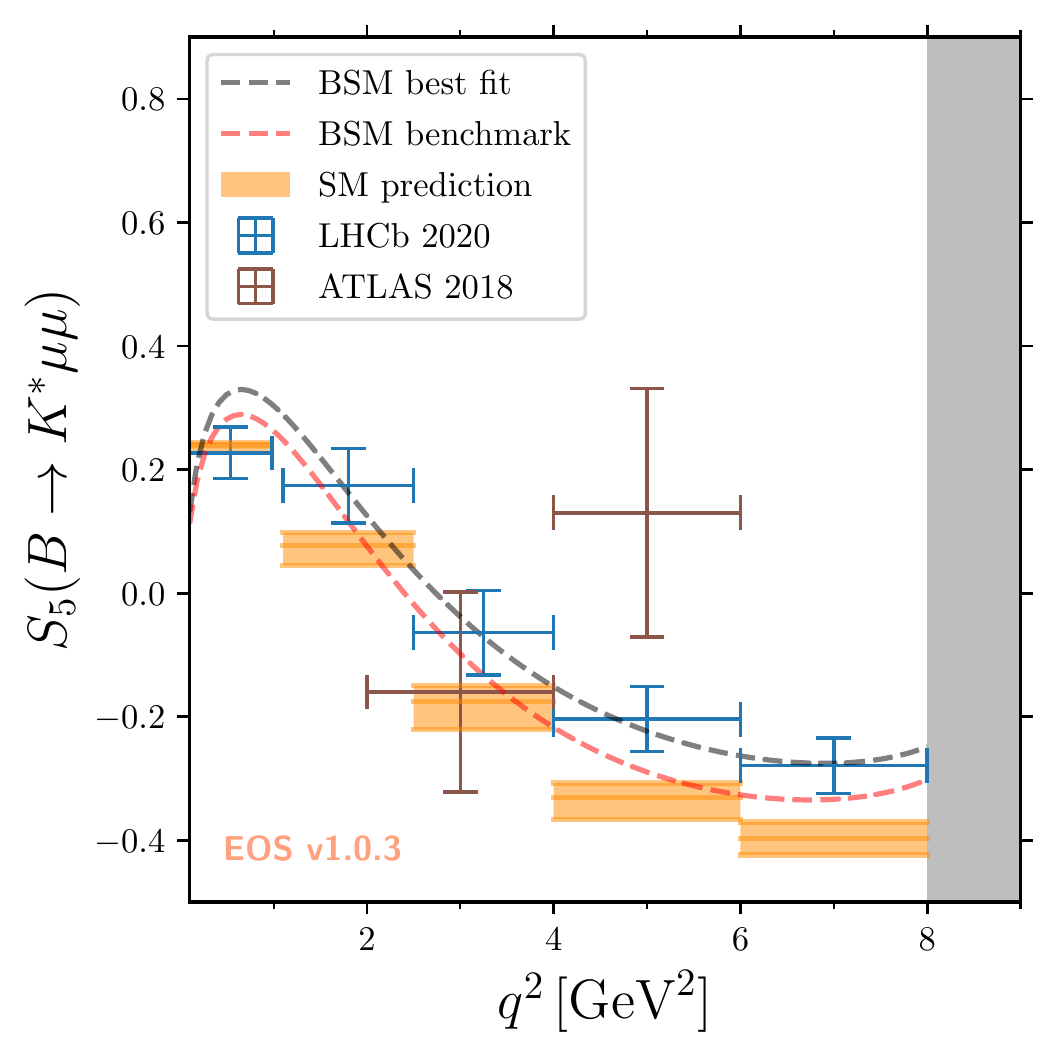}
    \includegraphics[width=.32\textwidth]{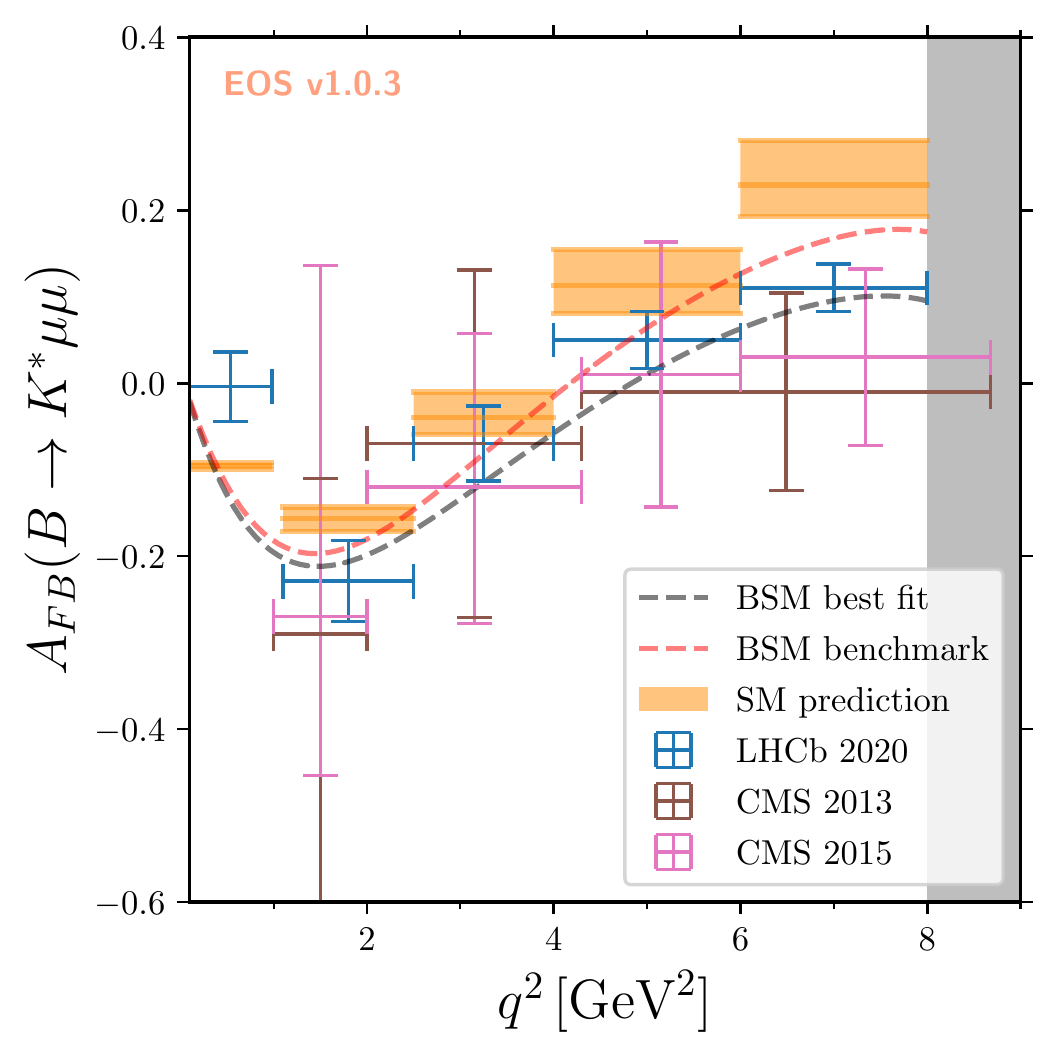}
    \includegraphics[width=.32\textwidth]{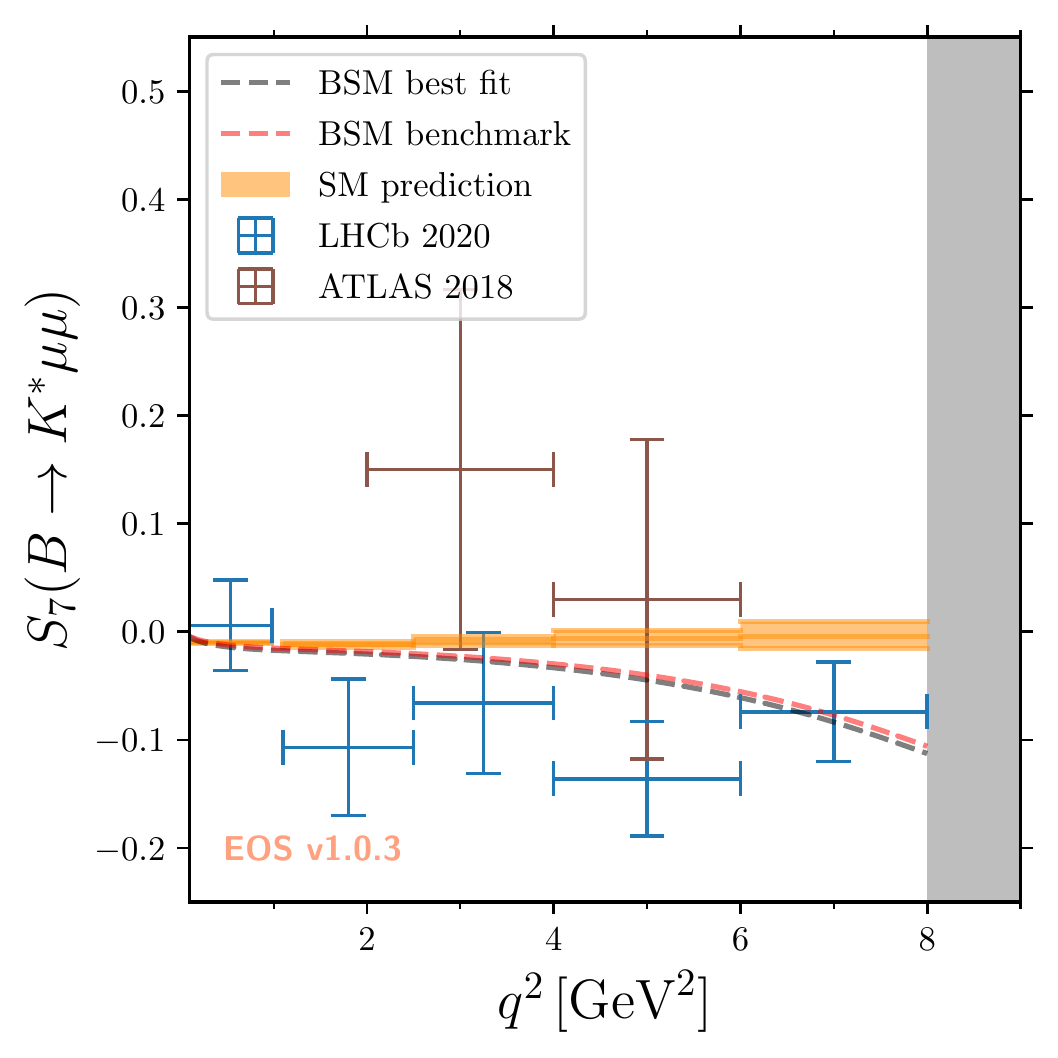} \\
    \includegraphics[width=.32\textwidth]{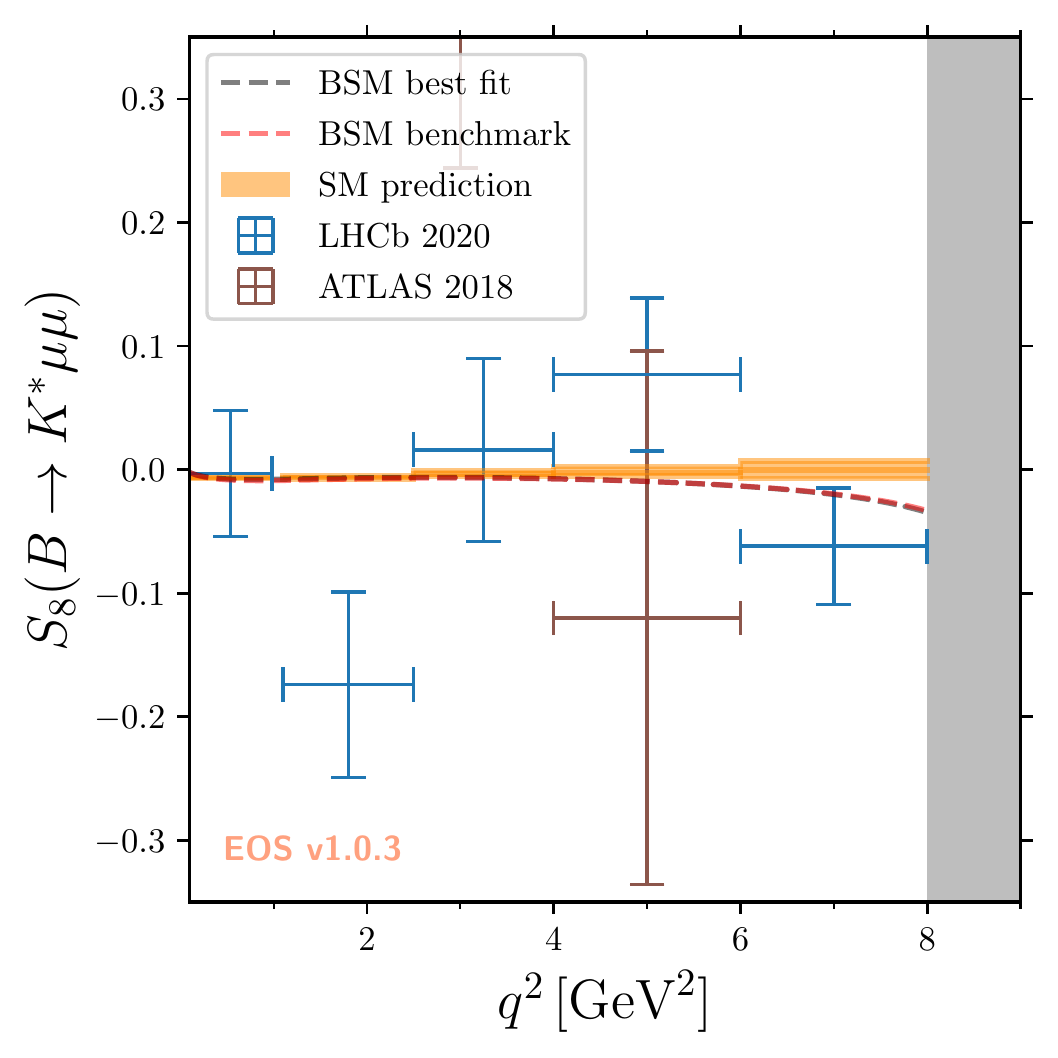}
    \includegraphics[width=.32\textwidth]{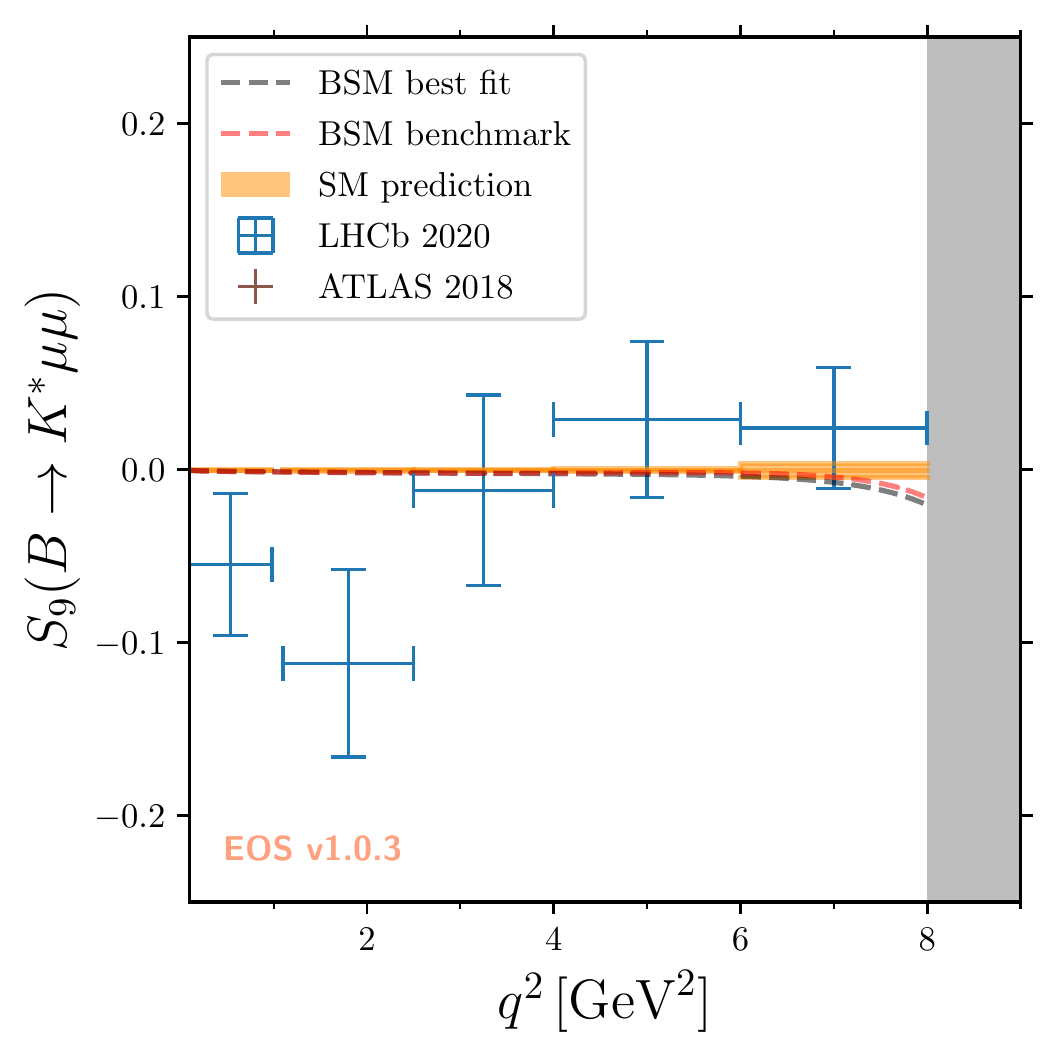}
    \caption{SM predictions for the angular observables of the $B\to K^*\mu^+\mu^-$ decays.
    Our SM predictions are also given in \Tab{tab:AO_predictions}.
    }
    \label{fig:BToKstar_AA}
\end{figure}

\begin{figure}[t!]
    \centering
    \includegraphics[width=.4\textwidth]{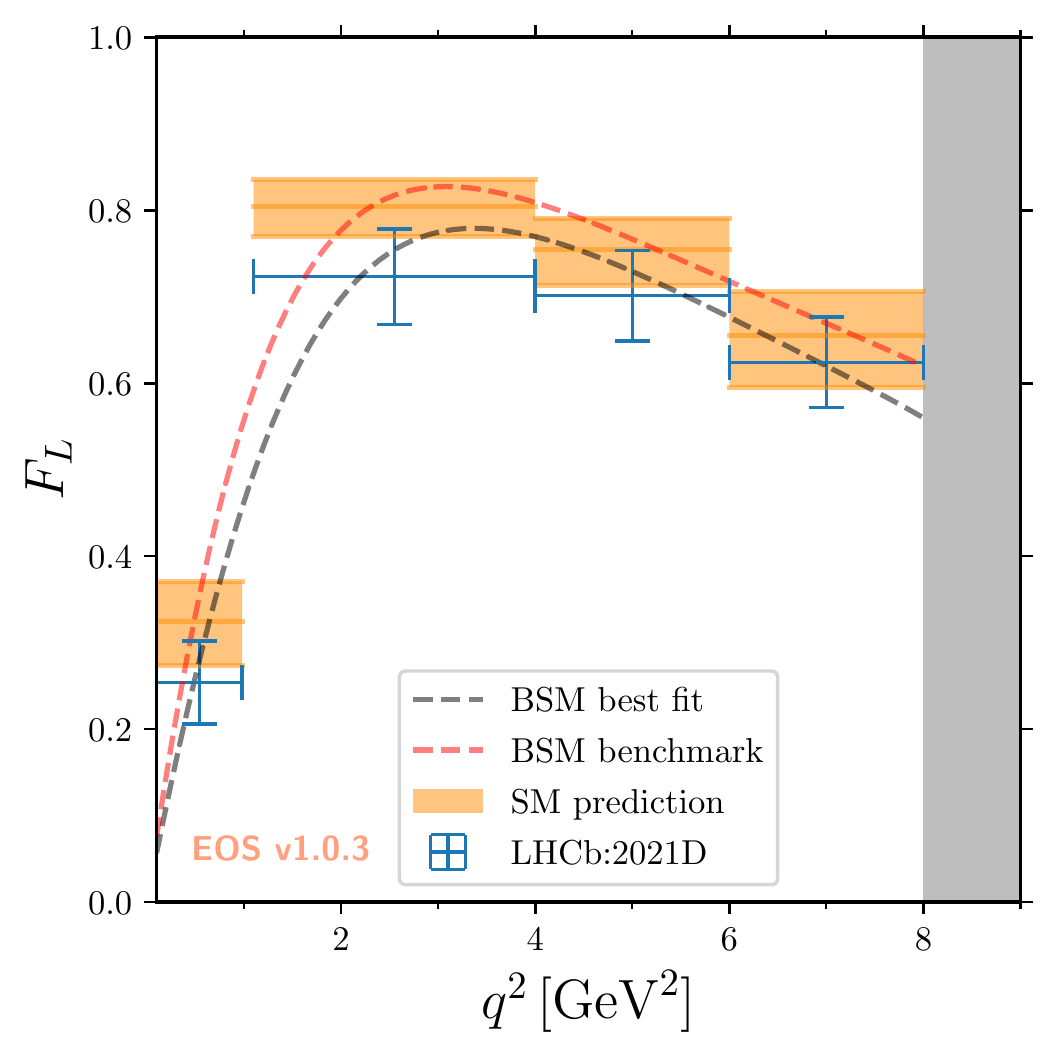}
    \includegraphics[width=.4\textwidth]{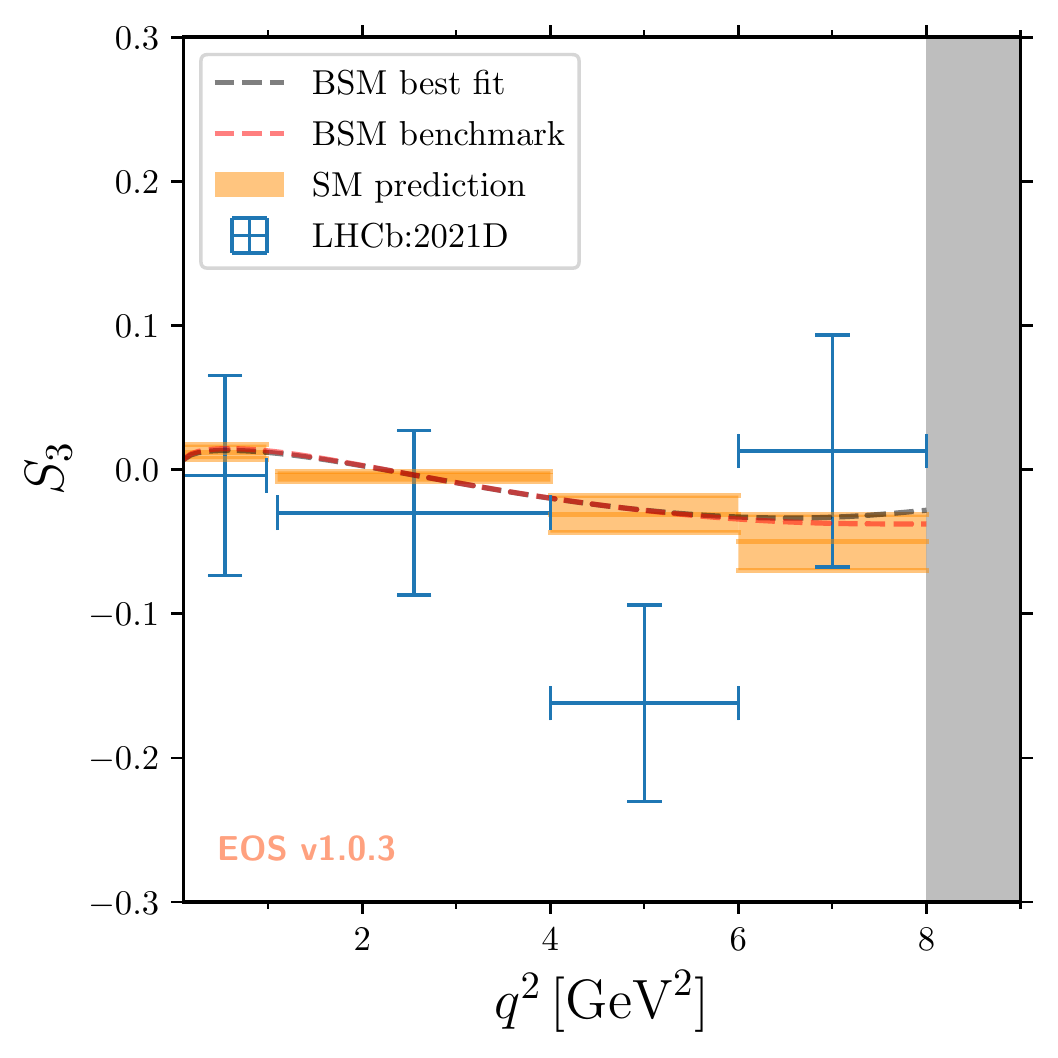} \\
    \includegraphics[width=.4\textwidth]{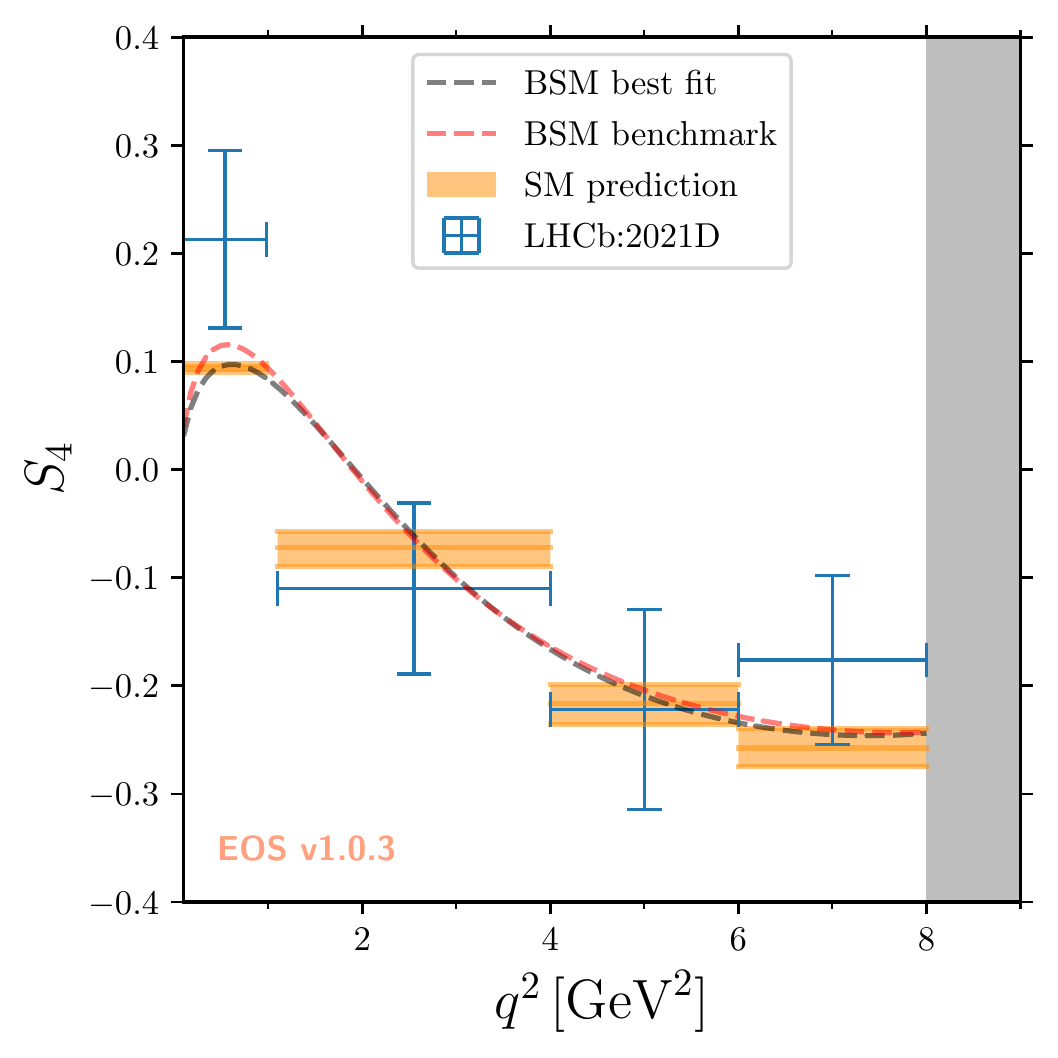}
    \includegraphics[width=.4\textwidth]{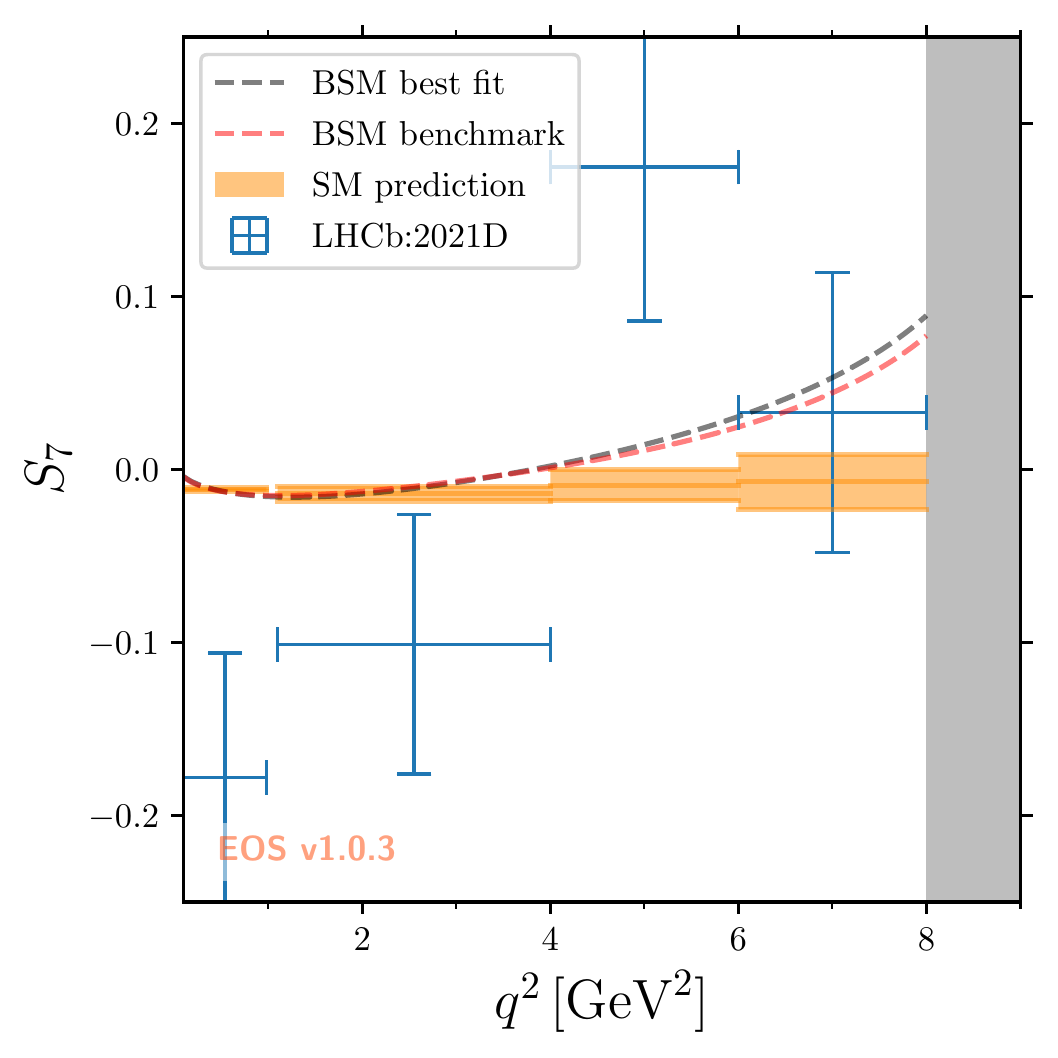}
    \caption{SM predictions for the angular observables of the $B_s\to \phi\mu^+\mu^-$ decay.
    Our SM predictions are also given in \Tab{tab:AO_predictions}.}
    \label{fig:BsToPhi_AA}
\end{figure}

\Tab{tab:BR_predictions} contains the median and the standard deviation of our SM predictions for the $B\to M\mu^+\mu^-$ branching ratios.
We compare these predictions to experimental measurements and two BSM scenarios in \Fig{fig:SM_Plots}.
\Tab{tab:AO_predictions} contains the  and the standard deviation of our SM predictions for the $B\to V\mu^+\mu^-$ angular observables.
We compare these predictions to experimental measurements and two BSM scenarios in Figures~\ref{fig:BToKstar_AA} and \ref{fig:BsToPhi_AA} for $B\to K^*\mu^+\mu^-$ and $B_s\to \phi\mu^+\mu^-$ decays, respectively (see \Fig{fig:SM_Plots} for the plot of $P_5'$). 
For the original definition of the angular observables see \Reff{Altmannshofer:2008dz}.
We use the conventions defined in the Appendices of \Reff{LHCb:2015svh}; see also the discussion of angular
conventions in \Reff{Gratrex:2015hna}.
Our SM predictions are also valid for $B\to M e^+e^-$ observables up to tiny effects due to the lepton masses.

\FloatBarrier

\bibliographystyle{JHEP}
\bibliography{references.bib}

\end{document}